# A Systematic Review of User-Centred Evaluation of Explainable AI in Healthcare

Ivania Donoso-Guzmán[a,b], Kristýna Sirka Kacafírková[c], Maxwell Szymanski[b], An Jacobs[c], Denis Parra[a], Katrien Verbert[b]

[a]*Department of Computer Science, Pontificia Universidad Católica de Chile, Chile*
[b]*Department of Computer Science, KU Leuven, Belgium*
[c]*imec-SMIT, Vrije Universiteit Brussel, Belgium*

**Abstract**

Despite promising developments in Explainable Artificial Intelligence, the practical value of XAI methods remains under-explored and insufficiently validated in real-world settings. Robust and context-aware evaluation is essential, not only to produce understandable explanations but also to ensure their trustworthiness and usability for intended users, but tends to be overlooked because of no clear guidelines on how to design an evaluation with users.

This study addresses this gap with two main goals: (1) to develop a framework of well-defined, atomic properties that characterise the user experience of XAI in healthcare; and (2) to provide clear, context-sensitive guidelines for defining evaluation strategies based on system characteristics.

We conducted a systematic review of 82 user studies, sourced from five databases, all situated within healthcare settings and focused on evaluating AI-generated explanations. The analysis was guided by a predefined coding scheme informed by an existing evaluation framework, complemented by inductive codes developed iteratively.

The review yields three key contributions: (1) a synthesis of current evaluation practices, highlighting a growing focus on human-centred approaches in healthcare XAI; (2) insights into the interrelations among explanation properties; and (3) an updated framework and a set of actionable guidelines to support interdisciplinary teams in designing and implementing effective evaluation strategies for XAI systems tailored to specific application contexts.

# 1. Introduction

In the healthcare domain, the decisions made by AI systems can impact the well-being or life of people [1]. In this context, it has been proposed that providing explanations about AI models or single predictions could potentially increase clinicians' appropriate trust [2–5] and ultimately boost adoption in healthcare settings [6–8]. This has led to increased research in the field of eXplainable AI (XAI), especially in the healthcare sector [9, 10].

Although this area shows promising results, the value of XAI methods still needs to be proven in practice [11]. Systematic and consistent evaluation is crucial: not only for generating understandable explanations but also for ensuring that these explanations are trustworthy and usable for their intended users [12]. Methodologies from the AI/ML community, such as evaluation with a Ground Truth Dataset, cannot be used in these scenarios: the success of an explanation depends on the user [13], its context, the AI model and the explanation [14].

In this discussion, user-centric evaluation in XAI has been recognised as crucial for designing systems that build trust and ensuring explanations are meaningful and actionable [15–19]. Involving users in the evaluation has been recognised as valuable because the explanations' value depends on how **real users** perceive, understand, and interact with them [15, 17]. Nevertheless, some XAI evaluations tend to overlook this aspect and focus rather on abstract measurements without users [20, 21].

Researchers have tried to tackle user-centric evaluation in two ways. First, they have tried to disentangle explanation's characteristics into simple, measurable properties such as completeness [15, 22–24], novelty [25–28], and interactivity [22, 23, 29]. Second, they have tried to provide structure to the properties by providing frameworks that cluster the properties in meaningful groups and provide general guidelines on how to measure the properties [15, 17, 22, 30, 31].

These efforts have tried to stress the importance of involving the users in the evaluation of XAI systems, but despite attempts to establish a unified framework for XAI evaluation [15, 22, 32], no consensus has been reached [32]. We believe this occurs for three reasons. First, there are multiple definitions of the aspects to be evaluated and their corresponding measurements [32]. Current efforts have focused on summarising or redefining the aspects based on current research, but do not use previously defined aspects in their frameworks. Second, there are no clear indications of what aspects are more important to be evaluated. Only [26] attempted to create such guidelines based on usage contexts, but still most papers focus on measuring *Understanding*, *Trust* and *Performance*. Finally, current frameworks do not provide clear guidelines on what aspects to measure based on the system's context. They cluster aspects into meaningful groups but do not state when to measure them.

This research aims to close this gap by analysing user studies to understand how the contextual aspects influence the evaluation and by providing clear guidelines on what aspects to measure based on the context of the application. This paper builds on previous work by Donoso-Guzmán et al. [15], which identified multiple granular properties that could be measured as a foundation for consistent evaluation across multiple layers. In this study, we examine empirical evidence from user studies regarding these properties. Specifically, we explore which properties are most commonly investigated and considered core to XAI evaluation in healthcare, as well as additional properties that frequently emerge in the literature. Based on our findings, we provide guidelines on how to conduct the evaluation of XAI systems, specifying what aspects to measure based on the system and the users.

The study goals can be summarised as:

(**RG1**) To provide a framework of well-defined and atomic properties that are part of the XAI user experience in the healthcare domain

(**RG2**) To provide clear guidelines on how to define the evaluation of XAI systems based on the system characteristics

To achieve these goals, we conducted a systematic review of the literature. We coded all the selected papers to identify explanation properties, users' characteristics, system conditions and connections between properties, and with the collected information, we provide an evaluation framework with guidelines on how to use it to design an evaluation of an XAI system.

The contributions of this paper can be summarised as follows: we present a comprehensive survey of XAI evaluation studies that involve users and are conducted within healthcare settings. Our overview not only identifies gaps in the current XAI healthcare literature but also serves as a source of inspiration for designing more effective explanations and evaluating them rigorously. Furthermore, we classify explanations based on their visual representation and level of interactivity, offering additional insights for future research and practical application. Finally, we propose a set of guidelines to follow during the evaluation of XAI systems. These guidelines aim to support researchers and practitioners in a consistent and informed selection of evaluation strategy, while encouraging a more comprehensive approach to assessing explanations.

## 2. Related Work

*2.1. Broad XAI surveys*

Much of the literature on Explainable Artificial Intelligence focuses on methods to explain complex black-box models. The first surveys of the area tried to standardise the conceptual frameworks to describe these systems and unify the diverse approaches and definitions across the field. Guidotti et al. [33] present a classification scheme for explanation methods pertinent to various black-box systems, targeting specific challenges related to interpretability. Similarly, Barredo Arrieta et al. [34] analyse XAI methods and delve into the context in which explanations are utilised and the objectives behind incorporating them into systems. Sahakyan et al. [35] investigate the interpretability of models that use tabular data, highlighting the necessity for clarity in understanding decisions made by opaque machine learning models.

More recently, the surveys have focused on examining the current challenges of the area. Saeed and Omlin [31] present a meta-survey that underscores the significance of transparency in creating trust and acceptance. The authors analyse the challenges faced in XAI development and implementation. They identify the varying user needs and the balance between accuracy and explanation complexity as the main barriers. Ali et al. [36] also identify several obstacles that must be addressed to advance towards trustworthy AI. They conduct a comprehensive literature review to identify the key concepts, frameworks, and challenges surrounding XAI and trust. They state that user diversity would be better addressed by creating user-centric explanations.

*2.2. Challenges and Applications in Healthcare*

The integration of XAI within healthcare has been subject to extensive exploration, especially concerning the interpretability of medical decision-making processes. Ooge et al. [37] offer a visual analytics review, focusing specifically on visual explanations in XAI, emphasising the importance of visual analytics in healthcare to facilitate understanding of AI outputs. The authors suggest that these visual tools can bridge the gap between opaque algorithmic processes and the interpretative needs of healthcare providers, thereby enabling more informed decision-making. Antoniadi et al. [38] explore challenges for clinical decision support systems (CDSS), highlighting that XAI augments clinical decision-making by improving transparency and trust in AI systems, which are critical for adoption in healthcare environments. They identify key challenges in integrating XAI into clinical workflows and state the need for XAI methodologies that enhance transparency while accommodating clinicians' needs. Similarly, Chaddad et al. [10] categorise different XAI techniques specific to healthcare applications, identifying challenges and proposing future directions for improving interpretability, particularly in medical imaging. They state that assessing how these methods align with clinicians' mental models will improve acceptance and operational efficiency in clinical environments. More recently, Mienye et al. [39] conducted a comprehensive survey of XAI applications in healthcare, identifying challenges such as the necessity for explainability, algorithm transparency, and ethical considerations in AI deployment. They state that the success of AI in healthcare depends on users' ability to understand and trust these technologies.

All these works emphasise the importance of the system's integration into the clinical workflow and the importance of appropriate trust to increase the adoption of these systems in clinical settings.

*2.3. XAI Evaluation Surveys*

Several studies have tried to establish frameworks to design and evaluate XAI systems. Table 1 presents a summary of the main characteristics of the proposals in the area. Mohseni et al. [40] surveys and organises the diverse research on XAI across multiple disciplines, including machine learning, visualisation, and human-computer interaction. The authors propose a step-by-step framework to guide multidisciplinary teams in designing and evaluating XAI systems, providing guidelines and evaluation methods. Vilone and Longo [23] present a comprehensive survey on XAI evaluation, focusing on empirical evaluation methods. They propose a taxonomy that categorises various evaluation approaches based on both theoretical and practical perspectives on explainability. Löfström et al. [27] conduct a semi-systematic meta-survey to identify and organise evaluation criteria for explanation methods in XAI. They present a taxonomy grouping properties into three aspects: model, explanation, and user. They identify four commonly accepted properties: performance, appropriate trust, explanation satisfaction, and fidelity, recommending these for more generalisable research in explanation quality. Nauta et al. [22] offer an extensive review of computational evaluation methods that do not involve user studies, providing a detailed taxonomy of computational techniques for assessing the quality of explanations.

Recent evaluation surveys have focused on human-centric evaluation. Lopes et al. [32] introduces a new taxonomy to organise XAI evaluation methods, aiming for clarity and intuitiveness by considering the multidisciplinary nature of XAI research. The taxonomy is divided into two families: Human-centred and Computer-centred methods. The taxonomy is designed to serve as a map for XAI evaluation methods during the development process, helping researchers from different disciplines systematically select and apply appropriate evaluation strategies. Kim et al. [17] presents a systematic review of 73 studies evaluating XAI systems with users, focusing on what makes explanations meaningful from a human-centred perspective. The authors identify 30 components of meaningful explanations and organise them into a taxonomy with three main dimensions: the contextualised quality of the explanation, its contribution to human-AI interaction, and its contribution to human-AI performance. Rong et al. [16] also presents a review of user studies in human-centred XAI. The authors categorise user studies based on measured characteristics such as trust, understanding, usability, and human-AI collaboration performance. The paper offers practical guidelines for designing and conducting user studies and identifies open research directions, especially the need to integrate psychological science into human-centred XAI research.

*2.4. Identified Gap*

Current evaluation frameworks emphasise the multidisciplinary nature of XAI. There is widespread recognition of the need for frameworks and methodologies that incorporate human factors into the evaluation process. These studies assert that to effectively evaluate XAI applications, the evaluation should be human-centred [16, 17, 32, 36, 40] and should consider the context of the system's deployment [10, 38]. It is also understood that the lack of standardisation in the field hinders the consistent evaluation of systems [17, 23, 32, 36, 40].

While previous surveys have effectively highlighted the importance of multidisciplinary collaboration and user-centric design in explanations, they have not contributed to standardising the field (see 1). These works have proposed various aspects, properties, components, and definitions for measuring or assessing evaluations, but they have not built upon previous definitions or connected them. Furthermore, although all these frameworks recognise the importance of context in the evaluation process, they tend to apply these evaluations broadly across different fields without explicitly considering the specific conditions of each area.

|  | Properties | Connection between properties | Context | Domain |
|---|---|---|---|---|
| Mohseni et al. [40] | Definitions Proposal | Taxonomy | Yes | General |
| Vilone and Longo [41] | Proposal | No | No | General |
| Löfström et al. [27] | Proposal | Taxonomy | No | General |
| Nauta et al. [22] | Proposal | No | Yes | General |
| Lopes et al. [32] | Proposal | Taxonomy | No | General |
| Kim et al. [17] | Proposal | Taxonomy | Yes | General |
| Rong et al. [16] | Proposal | No | Yes | General |
| This paper | Reuse from [15] | Yes | Yes | Healthcare |

Table 1: Summary table with main characteristics of Evaluation-related surveys. Kim et al. [17] presents the most similar evaluation survey, but the inclusion criteria are different; they include Wizard of Oz and exclude systems that make predictions for images.

In this work, we focus exclusively on the healthcare domain and employ the definitions presented by Donoso-Guzmán et al. [15] to analyse user studies. We aim to provide clear guidelines on what aspects to measure based on the characteristics of both the system and the users. Unlike Kim et al. [17], we do not restrict our analysis by data type; we include studies involving images and other modalities, whereas they limit their scope to certain types of data. Additionally, while their review includes Wizard-of-Oz-style studies where humans simulate AI behaviour, we only consider studies in which an actual AI system, regardless of its specific implementation, is responsible for making predictions.

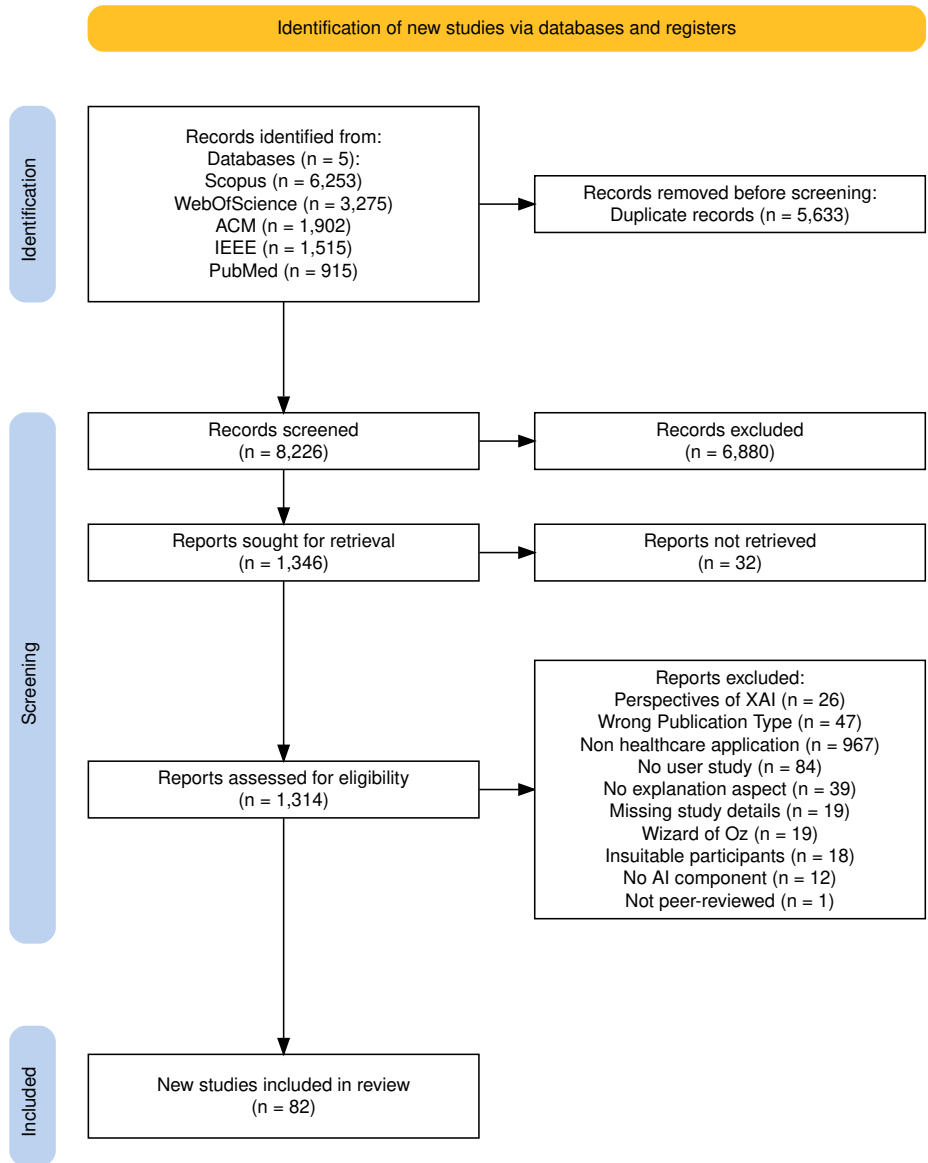

Figure 1: Prisma schema of the survey process

# 3. Methodology

## 3.1. Paper retrieval

We collected papers from 5 databases: Scopus, Web of Science, ACM, IEEE, and PubMed on 15$^{\text{th}}$ May 2024. The first two are general databases that include journals that publish in computer science and medicine. IEEE has several journals in computer science, and ACM is specialised in computer science. PubMed was chosen to also capture papers that are published in more specialised medical journals. The query for our systematic review was constructed using the PICO framework to ensure a comprehensive search across multiple academic databases. The search terms were categorised into two main components:

- **Population**: Keywords related to explanation, interpretability, and AI systems, including terms such as "Explainable AI (XAI)", "Interpretable Machine Learning (IML)", "Human-centred AI", "Decision Support Systems".

- **Intervention**: Terms focusing on human evaluation and empirical assessment, including "user", "empirical", "experiment", "evaluation", and "user-study".

At this stage, no restrictions were applied on the publication date, topic, journal or conference; the only enforced restriction was that the paper had to be peer-reviewed. We did not apply restrictions on the healthcare domain because, in the case of interdisciplinary research, abstracts need to capture the attention of multiple profiles and make compromises based on their most important target or publication venue. We left the healthcare domain filtering for a full-text screening phase. The query was applied to each database with some changes to comply with their different formats and restrictions [1]. We collected $13,860$ papers, out of which $5,633$ were duplicates.

## 3.2. Screening process

### 3.2.1. Title and Abstract screening

After removing duplicate records, $8,226$ proceeded to Title and Abstract Screening. The focus of this step was to keep papers that were user studies of XAI applications using real AI models. The inclusion criteria were:

---

[1] See Appendix A to see the full query for each database.

1. **The AI was a model that created predictions, not a Wizard of Oz**. In this review, we consider that AI models and XAI methods are designed materials with specific affordances and limitations [42]. In order to ensure that we would analyse explanations that took into consideration these restrictions, we decided to exclude studies that did not have an actual AI model, but a simulated AI, i.e. a Wizard of Oz.

2. **The XAI method is reproducible**, which means the explanations were not generated by people but with a replicable computational process. Just like for the previous criterion, the XAI method also has specific limitations and capabilities that the explanation design needs to take into account.

3. **The assessment was done by people**, not with simulations or automatic metrics. The goal is to provide guidelines for studies with users, so only those studies were considered.

4. The paper does not explicitly mention a user study, but the described results suggest that the study was conducted with users.

   - The authors reported results on user satisfaction, performance, and preference.
   - The authors claimed to increase user interpretability.

The exclusion criteria were:

1. People generated the AI predictions.

2. People generated the explanations.

3. The paper did not describe the assessment of an XAI system.

4. The XAI was used to make the model interpretable, but the interpretability was not evaluated.

To ensure the criteria were followed, four screeners reviewed papers and discussed disagreements until achieving a Fleiss Kappa score above 0.8. After this, one screener continued with the rest of the papers. In case of doubt, it was decided to include the paper to be checked in the full-text screening phase.

At the end of this stage, $6,880$ papers were excluded for not meeting the inclusion criteria, leaving $1,346$ papers eligible for full-text screening.

*3.2.2. Full-text Screening*

The full-text screening was conducted in two phases: first, we screened the papers to look for healthcare applications, and second, we screened the healthcare papers to check if they presented user studies of XAI applications.

*Finding healthcare applications.* In this stage, we filtered papers to look for studies that evaluated healthcare applications. We skimmed the full text to include papers that had:

1. Presented an evaluation with users.

2. AI models that predicted:

    (a) Health outcomes or risks

    (b) Diseases' outcomes or risk

    (c) Drug-related information

    (d) Information related to the general population health

3. Users that were:

    (a) Healthcare professionals

    (b) Patients or people with a certain condition

    (c) People who did not have any health condition, and the application's goal was to evaluate the risk of a specific disease or condition

    (d) People who did not have any health condition, and the application's goal was to teach people about specific risk factors.

The exclusion criteria in this part were:

1. The AI model predicted

    (a) Food recommendations to people who did not have a specific condition.

    (b) General behaviour changes to improve overall health.

2. Publication was not peer-reviewed

3. Publication was a Thesis, Workshop proposal, or Editorial

Of the 1,346 papers subjected to full-text retrieval,

- **967 papers** were excluded as they focused on non-healthcare applications
- **26 papers** focused on XAI perspectives rather than empirical studies
- **47 papers** were excluded due to being the wrong publication type
- **32 papers** were excluded due to lack of access

*Selecting papers for coding.* The previous process left **274 papers** related to **healthcare applications**. These papers were analysed to check whether they complied with the exclusion criteria. These are the following:

1. **No user study** ($n = 84$). The paper did not present an evaluation with users.

2. **Missing study details** ($n = 19$). The paper presented an evaluation with users, but the description of the study missed details that are part of the review.

3. **Lack of an AI component** ($n = 12$). The study did not include an AI component.

4. **Lack of explanation aspects** ($n = 39$). The paper did not evaluate the explanation that was created.

5. **Wizard of Oz** ($n = 19$). The paper presents an evaluation where the AI model is a Wizard of Oz, not a trained AI.

6. **Unsuitable participants** ($n = 18$). The participants had no relevant knowledge for the task, i.e. for a breast cancer prediction task, non-medical related users conducted the evaluation.

7. **Not being peer-reviewed** ($n = 1$).

After applying these criteria, 82 papers were included in the final review.

*3.3. Analysis Criteria*

Papers were coded based on the coding scheme developed by the research team, which is presented in the following sections. This deductive coding was conducted by three researchers in ATLAS.ti software and discussed in regular sessions to reach consensus and ensure consistency. During the coding process, new codes and criteria emerged to better describe the evaluation and the system. These are mentioned in this section and explained in the results.

*3.3.1. User description*

To distinguish between users' backgrounds, we adopted the classification proposed by Suresh et al. [43], which defines three types of user knowledge: formal, instrumental, and personal, as well as three contextual categories: machine learning, data domain, and milieu.

Formal knowledge refers to training acquired through formal education, such as a university degree. For example, in the data domain context, this could include medical students. Instrumental knowledge represents applied expertise gained through practical experience, such as doctors specialising in a particular field (e.g., diabetologists). Personal knowledge, on the other hand, refers to lived experience. In the context of machine learning, for instance, this could apply to a doctor who has neither formal nor instrumental training in ML but has developed an understanding of it through personal interest and self-learning.

*3.3.2. AI model and XAI method*

The AI model and XAI method were identified in each study. No previously defined categories were defined before starting the coding. Nevertheless, we conducted an inductive analysis of this data once all papers were coded to find the different models and methods categories.

*3.3.3. Data*

Data refers to the nature of the dataset for which the system is created. We used the same categories as [22], without the item-matrix type: tabular, text, images, time series, audio, video, and graphs.

*3.3.4. Usage Context*

Usage context is "a situation for which a user seeks explanations." [26]. We used the same usage contexts as defined by [26]. They state that the application scenario dictates the XAI usage context, and define six possible contexts which we use to classify papers:

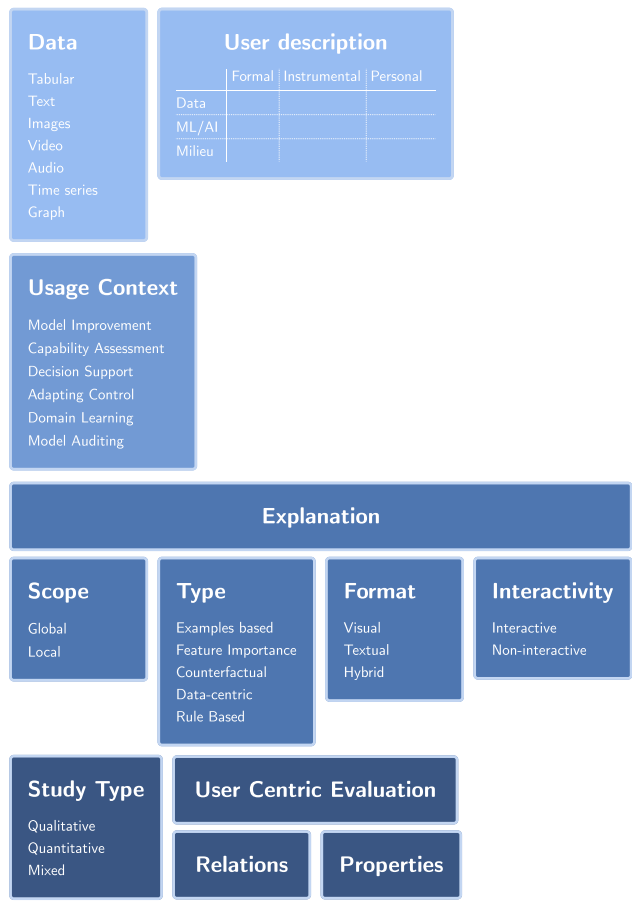
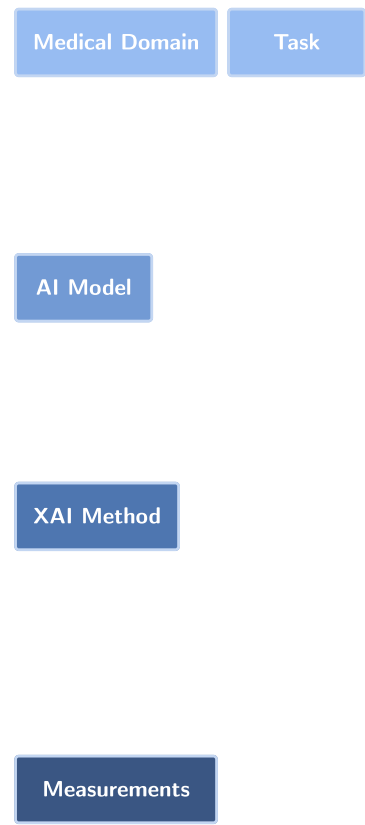

Figure 2: Summary of criteria used to code the selected papers.

- **Model Improvement**: Ensuring the AI model performs as intended by analysing and improving it during development and after deployment.

- **Capability Assessment**: Evaluating the AI system's capabilities and limitations to determine its reliability and appropriate usage.

- **Decision Support**: Understanding AI predictions to make informed decisions and take appropriate actions based on the causes of predictions.

- **Adapting Control**: Gaining insights into how the AI system processes data to better control and adjust system behaviours.

- **Domain Learning**: Identifying patterns and knowledge from AI-extracted historical data for improved prediction tasks.

- **Model Auditing**: Verifying AI compliance with fairness, security, and privacy standards.

*3.3.5. Explanation*

Unlike other surveys, where the explanation and the method used to generate it are analysed together [22, 34], we split the analysis to decouple how the explanation is generated from what information the user will see. To achieve this, we analysed the explanations according to explanation elements, as defined in previous work [15]. These elements are: Generation, the process of identifying and selecting causes; Abstraction, the content of the explanation provided by the XAI method; Format, design of the explanation; and Communication, the interaction of users with the explanation (see fig. 3).

For the generation level, we only decided to label the scope of the explanation as global or local. Other characteristics, like agnostic vs specific model type, were left for analysis of the method. For the abstraction level, we used types defined by [44]. These authors define different ways in which "people understand events or observations through explanations". They define that for *inquiry and reason*, users can reason in a deductive, inductive, or analogical way. For the *causal explanation and causal attribution*, they state that users seek counterfactual, contrastive and attribution types of explanations. Based on this, we defined four types of abstractions using the common names used in XAI literature: example-based (inductive, analogical), feature importance (attribution), counterfactual and rule-based (deductive). Additionally, we added the data-centric type, which focuses on providing explanations based on the data features. This type of explanation has proven to be relevant in healthcare domain settings [45, 46]. For the format level, we focused on what the user would interact with and what type of cognitive effort they would have to make. We choose three basic types: textual, visual and hybrid based on [47]. Lastly, for the communication level, we classified interfaces based on whether the user could interact with the explanation. We included all types of interaction that allowed the user to change what they saw. This could be filtering, having tooltips, selecting samples, etc. A summary of these criteria and their values can be found in table 2.

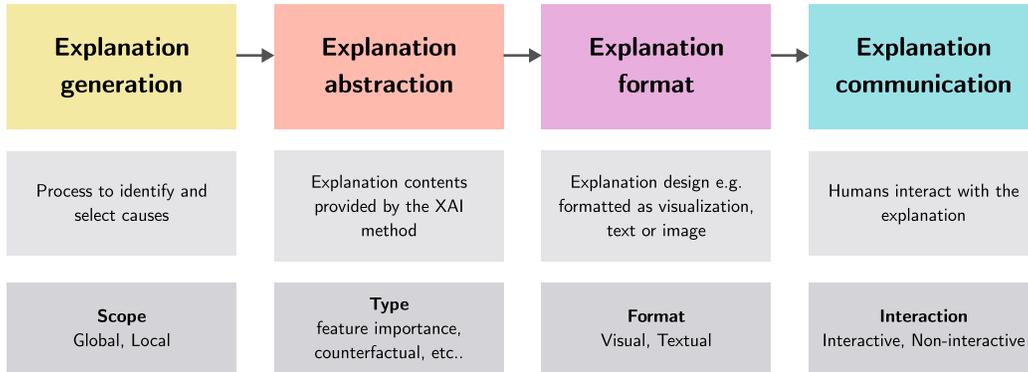

Figure 3: Explanation elements as defined by [15]. These elements help to identify characteristics of explanations that are defined at the different steps of the process and within the explanation product.

| Explanation element | Values |
| --- | --- |
| Generation | Local, Global |
| Abstraction | Example-based, Feature Importance, Counterfactual, Data-centric, Rule-based |
| Format | Visual, Textual, Hybrid |
| Communication | Interactive, Non-interactive |

Table 2: Explanation Categories and Types

*3.3.6. Study type*

All papers were classified based on the type of user evaluation they used. We distinguished between qualitative, quantitative and mixed-methods studies.

*3.3.7. Measurements*

While coding the studies, we also distinguished between particular measurements that were used. These categories were created during coding based on the evidence and are referred to in the results section. These measurements were associated with the elements of explanation described in fig. 3.

*3.3.8. Properties and their relations*

A **property**, in the XAI context, can be defined as a measurable characteristic of an explanation. The properties belonging to the same framework should have non-overlapping definitions, but can be related to each other by correlation or causation. We use the framework of properties defined in [15] to deductively code the properties measured in the studies. This framework describes 6 *Conceptual Components* based on the work by Knijnenburg and Willemsen [48], although they only describe properties in 4 of them. These components group properties that measure related concepts. The framework has a total of 30 properties. They present theoretical relations among properties according to the surveyed papers. This framework was chosen because it was strongly built on theory and previous work in the AI evaluation field. The properties definition was created by unifying multiple previous definitions in the field, and the properties were organised using a widely used evaluation framework in the recommendation systems field.

In this survey, a property was coded as part of a study when the definition of the measurement matched the definition of the chosen framework. In qualitative studies, we looked at the definitions or meanings the authors were communicating when discussing the concept. For instance, in [49] the authors say:

> Participants also used the highlighted sentences to verify whether the most and least influential sentences matched their expectations.

In this case, the whole sentence was marked as *Information Correctness* because the users performed the action because they wanted to check how the explanation complied with the information they expected to see.

In the case of quantitative studies, when the authors used User Behaviour or Metrics to measure the property, we made sure to understand the goal of the measurement and also to consistently use the same criteria for all papers using the same or similar measurement. When the study used Closed Questions and the questions were available, we made sure to read each question and code the property each one was measuring. In this way, we ensured that the properties were coded without overlaps between them. In all these cases, the name the authors used to describe it was not important. The only relevant information was the actual property that was measured.

A relation between any two properties was marked in qualitative and quantitative studies. In the case of quantitative studies, the study had to measure a correlation between two properties and the correlation was found to be significant. In the case of qualitative studies, the authors had to mention that a specific property caused or affected another property. For instance, in the quote

> In this case, the participants expressed that, as annotated regions from the system matched the important regions determined by the users, they were confidently able to continue to the next stages.[50]

we see that *Information Expectedness* influences the *Confidence* the users have on their decision.

## 4. Results

*4.1. General statistics*

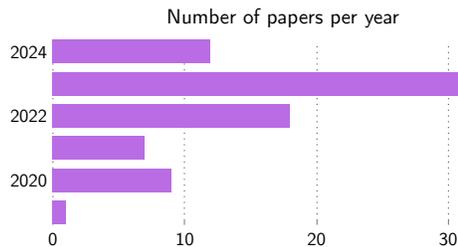

Figure 4: Publication year of selected papers. Most of them ($n = 80$) were published in 2019 or later. Only two papers from before 2019 complied with the inclusion criteria [51, 52] and are not shown in the chart.

We coded 82 papers, where 80 were published in 2019 or later (figure 4). Publication journals can be classified into three groups. The first group consists of Computer Science journals (44 papers), which focus on the technical aspects of implementation as well as the human factors involved in interactions with computer-based systems. The second group includes interdisciplinary journals (31 papers) that explicitly state their focus on applying Computer Science to the healthcare domain. Finally, healthcare journals (7 papers) primarily concentrate on healthcare topics.

Among the 82 publications, 56 were journal articles, while 26 were part of conference proceedings. The most frequently published journal was Artificial Intelligence in Medicine (Elsevier), with seven papers. It was followed by Medical Image Analysis and BMC Medical Informatics and Decision Making, each with three papers. The conferences with the highest number of papers were CHI Conference on Human Factors in Computing Systems, IUI Intelligent User Interfaces, and CD-MAKE International Cross-Domain Conference for Machine Learning and Knowledge Extraction, each featuring three papers.

*4.2. Participants*

In this section, we describe the characteristics of the participants involved in the user studies. Except for 7 studies [47, 49, 53–57], the selected studies primarily involved users with formal medical knowledge (n = 75), including medical students and residents. Fewer studies (n = 56), additionally included users with extensive practical knowledge, such as clinicians with a certain level of experience or specific expertise (e.g., oncologists, paediatricians, neurosurgeons). In cases where practical experience was not explicitly mentioned or the term "medical expert" was used without further specification, we classified participants solely within the formal data domain category. The personal data domain was rarely utilised; it appears in only 8 studies, as it primarily refers to patients who have personal experiences or connections with the AI's area of focus. Only three of them evaluated how people who experienced a specific healthcare condition interacted with explanations [46, 47, 54]. Some participants also had notable knowledge of machine learning or AI. Six studies [55, 57–61] included participants with formal machine learning knowledge, while nine [59–67] involved users with instrumental machine learning expertise.

Some studies state that they had participants with different knowledge levels. For instance, the studies [68, 69] had two user groups: residents and specialists. Both groups had *Formal Data Domain* knowledge, but only the specialists had *Instrumental Knowledge*. We identified 30 studies with more than one group in their evaluation. Most of them only described the knowledge these participants had in the domain, but did not assess their ML/AI knowledge. fig. 5 shows the different types of knowledge combinations that were present. Notably, the ML/AI knowledge was not specified for most groups. Out of the 116 groups of participants that were studied, only 9 had knowledge in both the *Data Domain* and *ML/AI* (figure 5).

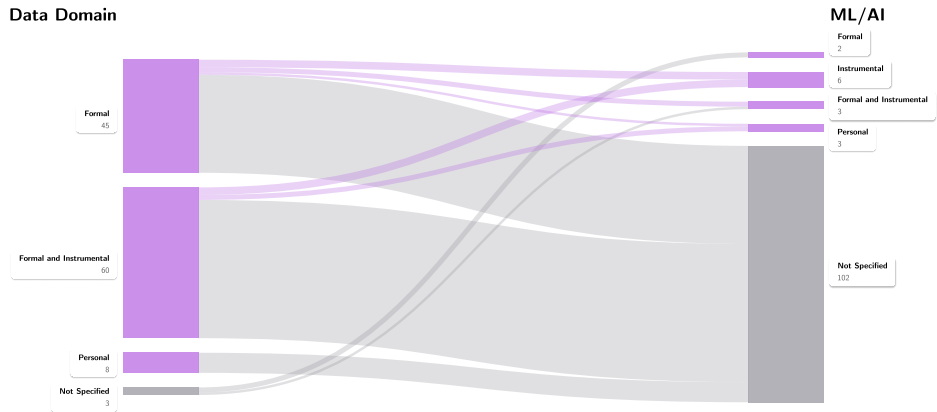

Figure 5: Groups of participants in the selected papers, according to their knowledge of the Domain and of ML/AI. Each paper can have more than one group. Most studies do not evaluate the ML/AI knowledge their participants have.

Regarding the number of participants, the studies had, on average, 34.5 users per study. Qualitative studies had between 2 and 21 users, mixed studies between 1 and 262, and quantitative studies between 3 and 223. In this last group, 21 studies (51%) had 16 or fewer participants, out of which 10 were able to provide confidence scores for their results, and 6 recognised that the sample size was a limitation of the research.

*4.3. Data*

The majority of papers use one type of data in their systems. We found that only two [54, 70] papers incorporate more than one type of data: [54] uses Graph, Tabular and Text while [70] uses Tabular and Video. Eight papers were marked as using both Tabular and Time Series datasets because they use raw Electronic Health Records. These papers show the user some kind of patient evolution over time and do not compress the records to one single time point when explaining. Overall, we find that the most used type of dataset is Tabular ($n = 35$), followed by Images ($n = 27$) and Time series datasets ($n = 12$). The other types are used in fewer than 10 papers each.

| Type | # | Papers |
|---|---|---|
| Audio | 1 | [71] |
| Graph | 7 | [53, 54, 72–76] |
| Images | 27 | [51, 52, 58, 59, 62, 64, 68, 69, 77–95] |
| Tabular | 35 | [46, 54–56, 60, 63, 65, 66, 70, 96–121] |
| Text | 4 | [47, 49, 54, 57] |
| Time series | 12 | [50, 61, 98, 99, 102, 104, 112, 115, 119, 121–123] |
| Video | 8 | [67, 70, 116, 124–128] |

Table 3: Studies per data type. Some studies used more than one type of data. For instance, [54] has graph, tabular and text data types.

*4.4. Usage Context*

Each study has one or more encodings of usage contexts. The most common context is *Decision Support*, which is defined as using XAI to support informed decision-making and enhance understanding of prediction causes [26]. This was the most frequent context, appearing in 50 out of 82 papers. Decision support applications were most common in oncologic radiology [51, 52, 67, 69, 86, 87, 92, 95, 116, 125], emergency medicine [66, 98, 101, 103, 104, 113, 115], and oncology [63, 72, 74, 79, 91, 110].

The second most common usage context was *Capability A*, present in 24 studies. Examples include using explanations to assess the practical validity of AI systems [57, 74] or engaging experts to assess whether the system is capable enough through evidence provided by explanations [62, 89].

The third most common was using explanations for *Domain Learning*, appearing 9 times. Examples include an interactive and explainable dashboard for drug repurposing [73, 76], a chatbot helping patient relatives understand causes of cancer [53], or a model explaining and detailing risk factors of coronary heart diseases [55].

The usage context of *Model Improvement* employs explanations to verify model behaviour and inspect how the model needs to improve. Only three studies (those by Lee et al. [124, 126, 128]) explore this capability, where physical therapists could provide feature-based feedback in a system for rehabilitation assessment.

The final context emerging from our studies is *Model Auditing*, in a study where health experts were involved to assess an explainable model and remove any problematic risk functions if needed [60].

Notably, the usage context of *Adapting Control*, which aims to understand how to achieve desired AI system behaviour, did not emerge in any of the included studies

*4.5. AI models*

Most of the studies control($n = 72$) consisted of evaluating a system using one AI model. Ten studies deviated from this: Four papers presented one system that used multiple models to generate information for its users [63, 119, 121, 127]; Three papers [59, 91, 125] present two evaluations, one for each model; and three papers [79, 116, 128] make an explicit comparison of their performance and explainability.

Most of the models were used for classification tasks ($n = 92$), followed by regression with 11, and finally, only one model was a ranking mechanism. The most used models were CNN-based: different DenseNet [129] ($n = 7$) and ResNet [130] ($n = 6$) architectures, as well as other architectures and custom implementations ($n = 14$). They are followed by Random Forest [131]($n = 8$) and Boosting Techniques [132]($n = 6$). These models were grouped in the following categories (see table 4):

- **Neural Networks**: Deep learning models, including CNNs, RNNs, and transformers.

- **Tree-Based Methods**: Decision trees, random forests, and boosting models.

- **Linear Models**: Regression and classification models such as logistic regression and support vector machines (SVMs).

- **Probabilistic Models**: Bayesian networks and Naive Bayes classifiers.

- **K-Nearest Neighbors (KNN)**: Distance-based models like KNN and weighted KNN.

- **Ensemble Methods**: Techniques combining multiple models.

- **Ensembles with Knowledge**: Hybrid approaches that integrate knowledge-based reasoning with trained models.

- **Knowledge Graphs**: Models that used knowledge graphs for classification.

- **Reinforcement Learning (RL)**: Models using reinforcement learning for decision-making and classification.

- **Rule-Based Systems**: Models using predefined rule-based logic.

- **Private Models**: Proprietary models that are not explicitly described.

- **Other Methods**: Miscellaneous techniques that do not fit in the above categories.

| Category | # | Papers |
| --- | --- | --- |
| Ensemble | 4 | [97, 101, 104, 107] |
| Ensemble with Knowledge | 5 | [80, 103, 124, 126, 128] |
| KNN | 2 | [63, 77] |
| Knowledge Graphs | 4 | [47, 54, 74, 75] |
| Linear Models | 9 | [46, 49, 93, 105, 116, 117, 119, 120, 128] |
| Neural Networks | 37 | [50, 58, 59, 61, 62, 64, 67–70, 76, 78, 79, 81–92, 94, 99, 100, 102, 112, 116, 118, 121, 123, 125, 127, 128] |
| Other | 2 | [57, 128] |
| Private model | 3 | [52, 65, 95] |
| Probabilistic | 4 | [53, 72, 108, 119] |
| RL | 3 | [73, 115, 128] |
| Rule-Based | 5 | [51, 63, 66, 113, 128] |
| Tree-Based | 16 | [55, 56, 60, 71, 96, 98, 106, 109–111, 114, 116, 119, 121, 122, 128] |

Table 4: Studies per AI model type. The model category can be used multiple times per paper because we included comparative studies.

*4.6. XAI methods*

In this section, we present methods used in the studies to generate XAI explanations. Regarding the usage of XAI methods in the studies, we analysed whether the systems used one or multiple XAI methods to create the explanations (see table 5). Most of the studies ($n = 55$) used only one XAI method in their systems. Twenty-three papers used multiple methods to generate their explanations in one system. Four papers presented an explicit comparison of various XAI methods, which they subsequently tested and compared.

| Use of XAI | # | Papers |
| --- | --- | --- |
| Comparison using multiple methods | 4 | [59, 79, 125, 127] |
| One system using multiple methods | 23 | [46, 50, 55, 61–63, 70, 87, 91, 92, 97, 101–104, 109–111, 115, 118–120, 124] |
| One system using one method | 55 | [47, 49, 51–54, 56–58, 60, 64–69, 71–78, 80–86, 88–90, 93–96, 98–100, 105–108, 112–114, 116, 117, 121–123, 126, 128] |

Table 5: Studies organised by how they used the XAI methods. Most of the studies only use one method.

As explained in the methods section, the XAI method was marked in each paper without making any categorisation to simplify the coding process. Once the coding phase was over, we analysed the different descriptions and looked for an up-to-date taxonomy that could help us categorise the methods. We decided to use the taxonomy defined by Ali et al. [36] (see table 6). In this taxonomy, the authors classify the methods into three big categories: data, model, and post-hoc explainability, and there are several subcategories within each. According to [36], these categories can be defined as:

- **Data** explainability methods focus on helping users understand the underlying data

- **Model** explainability aims to use the AI model to create explanations

- **Post-Hoc** explainability methods group techniques that are applied after the training and prediction to create explanations for users

Most methods used were classified as Post-hoc ($n = 51$), while Model and Data methods had 23 and 20 papers each. The most used method is SHAP values [133] ($n = 17$), followed by GradCam [134] ($n = 7$).

| | | | |
|---|---|---|---|
| Data explainability | Dataset summarizing methodologies | | [46, 81, 101, 103, 118, 119] |
| | Explainable feature engineering | | [58, 61, 63, 68, 80, 86, 115, 124, 126, 128] |
| | Exploratory Data Analysis | | [61] |
| | Knowledge graphs | | [47, 73, 99, 123] |
| Model explainability | Explainability through architectural adjustments | | [50, 54, 102] |
| | Explainability through regularization | | [119] |
| | Family of inherently interpretable models | | [49, 53, 55, 57, 60, 71, 79, 85, 103, 105, 109, 122] |
| | Joint prediction and explanation | | [66, 113] |
| | Other | | [72, 76, 104, 108, 116] |
| Post-hoc | Attribution | Backpropagation | [50, 64, 67, 69, 82, 83, 87, 88, 90–92, 94, 118, 125, 127] |
| | | Deep Taylor Decomposition | [59] |
| | | DeepLift | [125, 127] |
| | | Perturbation Methods | [62, 93, 97, 110, 125, 127] |
| | Example-based | Adversarial Example | [78, 89] |
| | | Counterfactuals | [46, 70, 109, 111, 112] |
| | | Prototypes and Criticisms | [59, 62, 63, 74, 77, 79, 84, 87, 102, 120] |
| | Game Theory Methods | SHAP | [46, 55, 56, 59, 65, 70, 96–98, 100, 101, 106, 107, 109, 110, 114, 115, 117, 119–121, 125, 127] |
| | Knowledge extraction methods | Rule extraction | [51, 75] |
| | Neural Methods | Influence Methods | [59] |
| | Other | | [46] |
| | Visualization Methods | | [104, 111] |
| Private model | | | [52, 92, 95] |

Table 6: XAI methods for each paper classified by the Taxonomy by Ali et al. [36]

*4.7. Explanations*

Feature importance is the most commonly used type of explanation abstraction. A total of 58 studies presented explanations of this type, with 36 relying exclusively on it. The second most common type is data-centric explanations, which focus on highlighting aspects of the input dataset to provide context. This approach appeared in 20 studies, with 9 using it as the sole explanation type. Its most frequent pairing was with feature importance. The systems by [101, 119] are prominent examples of the integration of both approaches. They combined multiple visualisations using different explanation styles to provide a holistic view of the model predictions they were explaining

Example-based explanations were used in 14 studies, with three employing them as the only explanation type. Similar to data-centric explanations, example-based explanations were often paired with feature importance, appearing together in seven studies. The study by Röhrl et al. [62] is a good example of the integration of example-based explanation and feature importance. Here, the user can identify parts of the cell that were relevant to the classification, and they can also see examples of similar classes to compare the cell image.

Rule-based explanations appeared in 10 studies, while counterfactual explanations were present in 8. These less common explanation types were usually combined with others; both were paired with additional explanation types in six studies each.

Regarding the scope of explanations, 69 studies used local explanations, while only four relied solely on global explanations. Nine studies incorporated both global and local explanations.

Most studies ($n = 68$) used a single explanation format. The visual-non-interactive format was the most prevalent ($n = 43$), followed by visual-interactive ($n = 15$) and purely textual ($n = 10$). Fourteen studies employed hybrid formats, combining text with visuals. Among these, the most common hybrid approach combined text with visual-non-interactive elements ($n = 12$), while only two studies combined text with visual-interactive explanations.

*4.8. Study Type*

As shown in section 4.8, the majority of the papers ($n = 41$) used purely quantitative evaluation methods, such as questionnaires using scales. Ten studies used purely qualitative methods, such as interviews, and 31 had a mixed methods design, for example, a combination of an interview with a Likert scale questionnaire.

| Category | # | Papers |
|---|---|---|
| Mixed | 31 | [46, 47, 49, 50, 60, 63, 67, 72, 74, 76, 80, 81, 84, 86, 99, 100, 102–104, 108, 113, 117, 120–128] |
| Qualitative | 10 | [58, 61, 66, 88, 98, 106, 110, 114, 118, 119] |
| Quantitative | 41 | [51–57, 59, 62, 64, 65, 68–71, 73, 75, 77–79, 82, 83, 85, 87, 89–97, 101, 105, 107, 109, 111, 112, 115, 116] |

Table 7: Papers per study type. The majority of papers conducted a quantitative study.

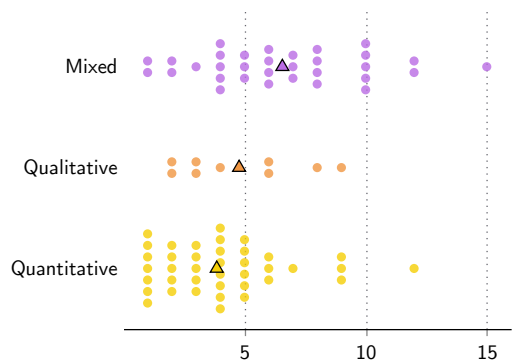

Figure 6: Number of properties studied by papers and type of study. Triangles represent the average number of properties per type. Quantitative studies measured the least number of properties.

The type of study highly determined the number of properties that were evaluated or considered. Figure 6 shows that quantitative studies usually measured around 4 properties, qualitative studies close to five properties and mixed studies measured almost 7 properties on average.

*4.9. Type of methods used for evaluation*

To closely examine the methods used, we first analysed how the property was measured. Here, we used inductive coding and identified five distinct ways, corresponding to the explanation elements Abstraction and Communication (see fig. 3):

- Closed Questions: questions with a limited set of answers. For instance, Likert-type questions or yes/no questions.

- Open Questions: questions with no predetermined answer. The participant is free to use the words and expressions she wants.

- User Behaviour: questions or metrics that measure the user's actions or knowledge. For instance, user answers to questionnaires that measure the objective understanding and performance.

- Interview Analysis: The property appears as part of qualitative data analysis. The analysis could have been deductive or inductive. It is possible that no open questions are marked in the paper because the authors did not disclose the interview protocol.

- Metrics: standardised measurement using a mathematical formula, which focus on assessing the XAI system's competencies without gathering direct user feedback.

Figure 7 presents a chart with the use of these methods according to the study type (Quantitative, Qualitative, Mixed) and the property they measure. Mixed-studies use closed questions as much as Quantitative studies, but for properties like *Relevance to the Task* and *Information Expectedness*, they used interview analysis much more. Quantitative studies do not use measurements at the Abstract level, and rely on closed questions and user behaviour to conduct their analysis.

To describe the methodology of qualitative and the qualitative section of mixed studies, we analysed the study descriptions to understand the authors' qualitative method. We identified seven distinct qualitative methods, as seen in fig. 8: most studies employed interviews, either before or after using the system. These were followed by open-ended questions in questionnaires and the think-aloud method during system interaction, including constructive interaction. A few studies utilised ethnography, focus groups and observation as continuous evaluation methods.

*4.10. Medical Domain*

We now give an overview of the medical domains we identified where XAI is applied. The most common medical domain XAI was used for is **internal medicine** (25/82 papers), with specialisations including *oncology* (i.e., cervical cancer diagnosis using tabular data [110], or predicting the recurrence of breast cancer [107] or lung cancer [74]), *cardiology* (i.e., risk assessment or diagnosis of coronary heart diseases [55, 56, 105, 109], or classifying pulmonary heart diseases [99] or cardiomegaly cases [84]), endocrinology (mainly focussing on diabetes - either diabetes monitoring [46] or risk prediction [97, 117, 120] with only one study being non-diabetes related, being thyroid tumours classification [69]), *pulmonology* (systems for classifying asthma and bronchitis [65], paediatric pneumonia [94] or the previously mentioned system for classifying pulmonary heart diseases [99]) or neurology (i.e., systems predicting stroke likelihood due to obstruction or rupture in brain [111], sleep staging predictions using EEGs [50] or predicting rehabilitation of comatose patients [61]), and haematology (assessing risk for coagulopathy [72, 108]).

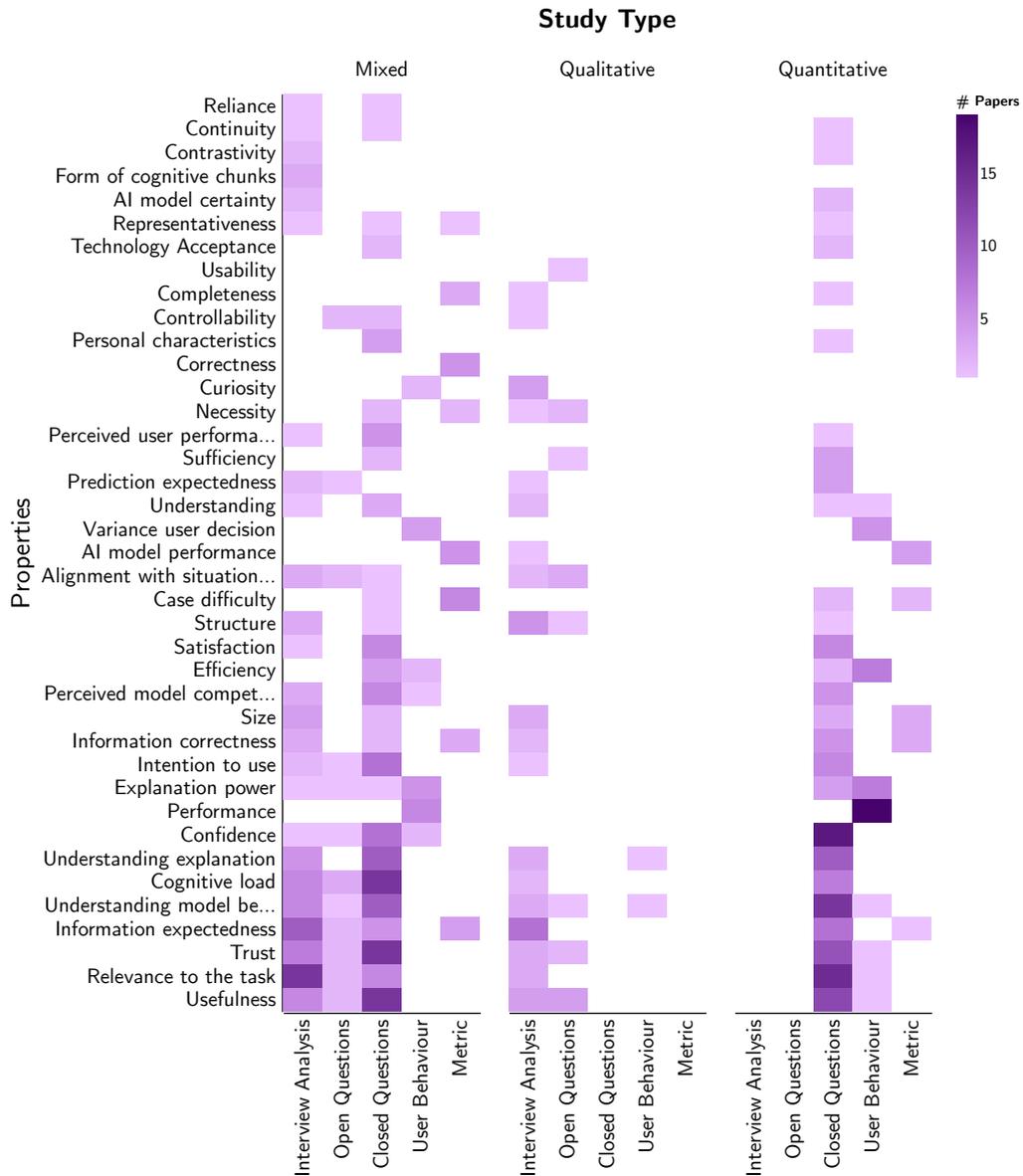

Figure 7: Measurement method for each property depending on the study type. Quantitative studies rely heavily on *Closed questions* to conduct their studies, and Mixed-studies use *Interview Analysis* to capture feedback on specific properties.

| Medical Domain | Feature Importance | Data-Centric | Example Based | Rule Based | Counterfactual |
|---|---|---|---|---|---|
| Oncologic Radiology | [59, 67, 69, 83, 86, 87, 95, 116, 125, 127] | [52, 58, 81] | [52, 87, 116] | [51, 116] | [116] |
| Emergency Medicine | [60, 66, 93, 98, 101, 103, 104, 113, 115, 119, 121] | [101, 103, 119] | [104] | | [112] |
| Oncology | [53, 72, 91, 107, 110] | [68, 79] | [63, 74] | [72, 75] | |
| Physiotherapy | [47, 70, 96, 124] | [118, 122, 124, 126, 128] | | [122] | [70] |
| Endocrinology | [46, 69, 97, 117, 120] | [46] | [120] | | [46] |
| Cardiology | [55, 56, 99] | [105] | | [55, 105, 109] | [84, 109] |
| Psychology | [47, 71, 100, 106, 114] | [114] | | | |
| Pulmonology | [65, 94, 99, 101] | [101] | | | |
| General Medicine | [49, 102, 123] | [57] | [102] | | |
| Radiology | [90, 92] | [80] | [92] | | |
| Pharmacology | [76] | | | [54, 73] | |
| Orthopedagogy | [82, 92] | | [77, 92] | | |
| Pathology | [62] | [79] | [62] | | |
| Neurology | [50, 111] | [50, 61] | | | [111] |
| Dermatology | [91] | | [78, 89] | | [78] |
| Hematology | [72, 108] | | | [72] | |
| Ophthalmology | [64, 88] | | | | |
| Dentistry | | [85] | | | |

Table 8: Studies organised by medical domain and explanation type.

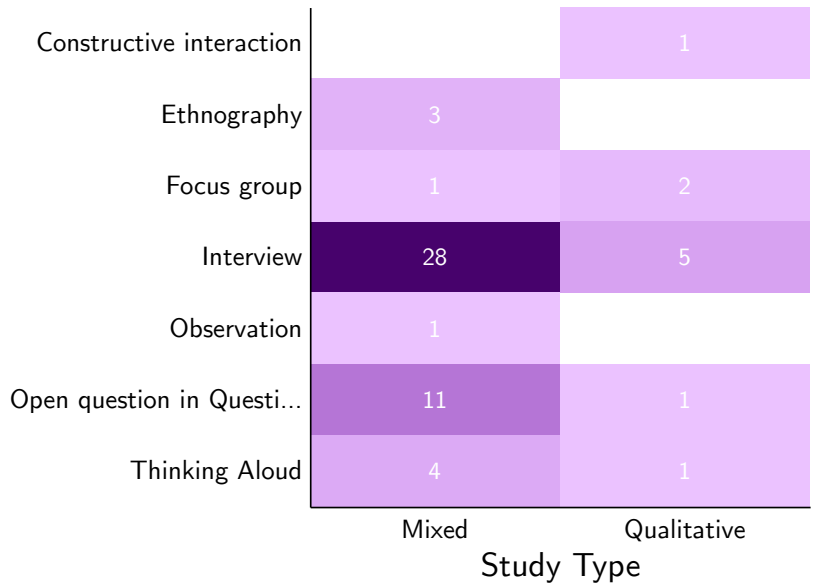

Figure 8: Qualitative methods for mixed and quantitative studies. By far, the most used method is interview.

The second most common was the use of XAI for **support diagnostics** (16/82 papers), with most applications centring around *oncologic radiology*. Examples include systems - and by extension explanations - for glioma or other tumour classifications [52, 58, 69, 83, 95, 125], of which most use feature importance explanations. Support diagnostics also include systems used within *radiology* that do not involve oncology, used for i.e., assessing spleen injuries [80] or bone fractures [90, 92] through X-ray images.

The third most frequent domain was **emergency medicine** (11/82 papers), with applications such as patient triaging [66, 101] and predicting ICU stay duration [121].

Applications within *general medicine* (4/82 papers) include broader topics such as predicting diagnoses based on electronic health records [102] and clinical history [123], classifying medical articles [49], or supporting consumer health search [57].

Finally, we have specialisations that fall outside the aforementioned categories, grouped under other medical domains (19/ 82 papers). These include applications in physiotherapy (8 papers), where explanations supported systems for detecting wandering patterns [122] and evaluating rehabilitation progress [124, 126]. Studies within the field of *psychology* have a slightly broader focus, ranging from assessing anxiety levels from speech [71], stress levels from wearable sensor data [114] and classifying a person's mental state [100], to systems for automating "Grief Inquiries Following Tragedies" [106] or mental health recommendations for people with chronic pain [47]. Pharmacological studies centred around either drug repurposing for treatments [54, 73] or assessing drug-disease treatment pairs [76]. Studies within orthopedagogy focused on lesion or fracture classification through X-rays [77, 82, 92], whereas studies within pathology focused on cytological image analysis [62]. For *dermatology*, two studies focused on assessing skin lesions [78, 89], whereas one study pertained to melanoma classification [91] (also falling under the oncology domain). Two studies centred around *ophthalmology*, both using feature importance explanations for assessing glaucoma cases [64, 88]. Finally, only one study focused on *dentistry*, identifying furcation involvement lesions on a series of dental radiographs and explaining through highlighting image regions [85].

### *4.11. Relations between properties*

We found 85 relations between properties. As seen in figure 9, most of these relations occur between *Subjective System Aspects* and *User Experience*.

These relations were found in 34 papers. These studies are split equally between quantitative and mixed ($n = 15$ each), and 4 qualitative studies also describe relations. The average number of users does not differ between the relations supported by 1, 2 or 3 papers.

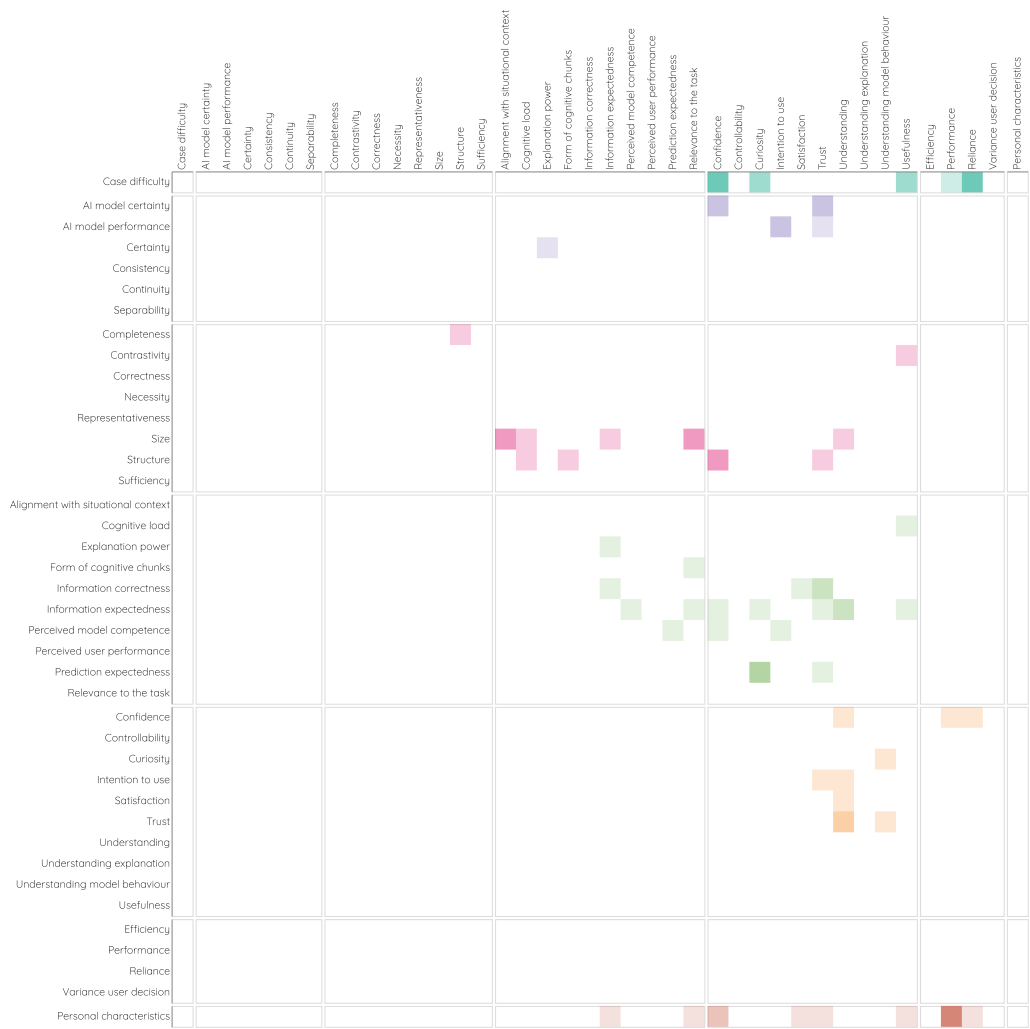

Figure 9: **Relations between the properties**. Each square represents a relation between two properties. Darker squares mean that more papers found that relation. Most of the relations are found between the Subjective System Aspects and User experience aspects.

*4.12. Properties*

In the following sections, we present observations for each property identified in the studies.

Table 9: Table of all explanation properties and their definitions based on the reviewed literature.

| Property | Definition | References |
| --- | --- | --- |
| Personal Characteristics | | |
| Personal characteristics | Individual traits that could impact the user experience | [47, 71, 99, 123] |
| Technology Acceptance | Level of inclination users have about using AI in their work | [65, 71, 103, 120] |
| Situational Characteristics | | |
| Case difficulty | Measures the level of complexity of the case users have to assess | [49, 50, 63, 64, 77, 85, 86, 88, 92, 104, 117] |
| Objective System Aspects | | |
| AI model certainty | Measures the confidence level the AI model has in its prediction. | [74, 99, 101, 111] |
| AI model performance | Measures the accomplishment level the AI model has with respect to the task for which it was trained. | [59, 79, 91, 98, 116, 121, 125, 128] |
| Continuity | Measures the extent to which the function provides similar explanations for similar instances. | [64, 100, 122] |
| Explanation Aspects | | |
| Completeness | Measures if all the causes that the model used to generate the prediction are present in the explanation | [110, 111, 125] |
| Contrastivity | Measures whether the explanation contains reasons that highlight differences with respect to other possible outcomes. Low contrastivity will provide the same reasons for instances in which the model predicts different classes. | [82, 84, 124] |

Table 9 – continued from previous page

| Property | Definition | References |
|---|---|---|
| Correctness | Measures the extent to which the selected causes are correct with respect to the model reasoning. | [125] |
| Necessity | Measures to which extent only the reasons that cause the decision were selected. Are the reasons provided responsible for the decision? | [72, 84, 100, 110, 114] |
| Representativeness | Measures similarity of explanation of similar but distinct instances | [64, 100, 122] |
| Size | Measures the amount of information present in the explanation. | [55, 57, 58, 74, 90, 98–100, 104, 111, 119, 122, 128] |
| Structure | Measures the extent to which the information is displayed in a way that allows the users to navigate the information. | [74, 97, 98, 100, 104, 110, 114, 119] |
| Sufficiency | Measures to which extent all the reasons that cause the decision were selected. Do the reasons provided are enough on their own to make a decision? Do we need more information? | [57, 97, 100, 110, 111, 122] |
| Subjective System Aspects | | |
| Alignment with situational context | Level of appropriateness of the explanation to the usage context. | [50, 58, 66, 80, 104, 110, 118] |
| Cognitive load | Measures the cognitive effort the user has to do to achieve the task. | [47, 50, 64, 66, 70, 74, 80, 81, 86, 87, 98, 100, 101, 103, 104, 117, 122, 124–126] |
| Explanation power | Measures whether the selected causes are capable of making the user accept the prediction. | [49, 55, 56, 67, 70, 71, 75, 77, 82, 99–101, 117, 120, 123] |
| Form of cognitive chunks | Measures the semantics and structure of the pieces of information the user receives | [67, 80] |
| Information correctness | Measures the level of the truth contained in the explanation | [57, 59, 61, 68, 73, 76, 84, 100, 110, 112, 121] |

Table 9 – continued from previous page

| Property | Definition | References |
|---|---|---|
| Information expectedness | Measures how much information presented in the explanation was anticipated by the user. | [49, 50, 59, 60, 67, 68, 70, 73, 76, 82, 84, 94, 97–99, 104, 108, 117–119, 121, 124–127] |
| Perceived model competence | Measures the user's impression of the model competence for the task at hand | [49, 50, 81, 83, 89, 99, 103, 113, 121, 123] |
| Perceived user performance | Measures the user's impression of its own competence for the task at hand | [49, 70, 80, 81, 122] |
| Prediction expectedness | Measures how much the AI model prediction was anticipated by the user. | [50, 65, 97, 101, 108, 119] |
| Relevance to the task | Level of explanation usefulness to the user's task. This refers to the task in the context of the data domain | [46, 60, 62, 70, 72–74, 76, 78, 80, 83, 84, 86, 96, 98, 100–102, 106, 112, 117, 121, 124–126] |
| User experience | | |
| Confidence | Measures the user's subjective belief in the correctness of a decision | [49, 54, 63, 65, 68, 75–78, 80, 85, 89, 90, 92, 94, 97, 99, 101–103, 108, 123] |
| Controllability | Measures whether the user perceives she has some level of control over the system. This could manifest as the ability to reverse actions, correct the system, filter or zoom the explanation, or ask questions to clarify the explanation or prediction. | [80, 100, 103, 119] |
| Curiosity | Measures whether the user is intrinsically motivated to understand the explanation. If the user is curious, she will be more attentive to the task and, therefore, more engaged with the system | [66, 104, 119] |
| Intention to use | Measures an individual's willingness to use an explanation in the future [135, 136] | [62, 67, 70, 74, 80, 81, 83, 94, 106, 107, 115, 117, 123, 124, 126] |
| Satisfaction | Measures the level of fulfilment the user gets while interacting with the system. | [47, 59, 74, 75, 81, 99, 101, 103, 111, 116, 122] |

Table 9 – continued from previous page

| Property | Definition | References |
|---|---|---|
| Trust | We use Tintarev et al. definition: "Perceived confidence in a system's competence". | [46, 47, 49, 50, 62, 66–68, 70–72, 74, 78, 80, 83, 88, 89, 96, 97, 102, 104, 110, 111, 117–119, 122, 124, 126] |
| Understanding | Measures the user's ability to correctly interpret the system's outputs | [46, 47, 53, 66, 70, 74] |
| Understanding explanation | Measures the user's ability to correctly interpret the explanation generated by the system | [46, 47, 61, 62, 64, 72, 83, 84, 97, 100, 101, 104, 109–111, 117, 122, 125] |
| Understanding model behaviour | Measures the user's ability to correctly interpret the model behaviour | [46, 47, 49, 55, 57, 60, 62, 70, 81, 83, 84, 87, 96, 97, 100, 101, 103, 108, 110, 111, 117, 119, 122, 126] |
| Usefulness | Measures whether the explanation helps the user to achieve a certain goal. This refers to the fact that it makes the user experience better | [46, 47, 50, 57, 58, 66, 74, 77, 78, 80, 84, 85, 88–90, 92, 96, 100, 107, 109, 111, 114, 115, 117, 122, 123, 128] |
| Interaction | | |
| Efficiency | Measures the speed at which a task can be performed | [49, 56, 64, 70, 75, 76, 80, 83, 85, 91, 100, 122] |
| Performance | Measures how well the user can do the task while using the system (prediction+explanations) | [49–52, 56, 60, 65, 69–71, 75–78, 82, 85, 86, 90–92, 95, 116, 128] |
| Reliance | Measures whether the user is willing to provide control to the machine for the given task | [88, 104] |
| Variance user decision | Measures users' agreement in a decision-making scenario. | [50, 65, 70, 71, 75, 80, 81, 94, 126] |

*4.12.1. Personal Characteristics*

*Domain Experience* refers to the level of knowledge and practice users have in the data domain and it is the personal characteristic that has been studied the most in the area. Twenty-five papers had users with different knowledge and experience levels. Only 5 of them analysed how these differences impacted other properties. Another 4 papers analysed how the experience level impacted other properties. Among these 9 studies, the most analysed connection was to *Performance* (4 papers). Interestingly, not all papers found the same result: [77] found that residents performed than specialists when they did not have the XAI support; [69] found that senior performed better than junior physicians; [65] found no significant differences between groups; and [68] found that people with more experience obtained higher gains when using the XAI system that when using AI alone and that the people with the least experience had the largest decline when using the XAI system. These mixed results could be due to the fact that all the scenarios where in a Decision Support context with a Diagnostic Decision Support task and less-experienced-participants might have used the opportunity of participating to also test their skills in the problem [77]. The other properties that are affected by *Domain Experience* are: *Confidence* [85, 92]; *Information Expectedness*, *Relevance to the Task* and *Trust* [104] ; *Size* and *Structure* [98]; *Explanation Power* [71] and *Usefulness* [92].

*Attitude Towards AI* refers to the inclination users have to use AI in their work. This differs from their skill level for this type of technology, as it aims to measure users' likeness of AI tools. This property was explicitly measured in four studies [65, 71, 103, 120] using questionnaires with closed answers. One study assessed the relation between this characteristic and *Reliance*, but it did not find evidence supporting it [120]. Another study had the same result with *Explanation Power* [71]. Additionally, one study evaluated how *Performance* was affected by this property, and no evidence was found [65].

*Personal Characteristics* is a broad term used to describe any kind of user trait that could impact the result. In the survey, four papers presented an assessment of these traits. Papers [99, 123] used Need for cognition [137], [71] used the Big Five Inventory [138], and [47] used Need for cognition and Ease of *Satisfaction* [139]. Two studies evaluated the relation of *Personal Characteristics* with other properties of the XAI user experience: [71] evaluated its influence on *Explanation Power* without finding significant results, and [47] found that higher ease-of-satisfaction led to higher *Satisfaction* with the explanations.

*4.12.2. Situational Characteristics*

*Case Difficulty* is the only aspect of situational context mentioned in the papers. It refers to the complexity of the case that users face. Papers measured it in two ways: with closed-questions, by asking participants about the difficulty of the case ($n = 3$) and at the abstract level by measuring it while selecting cases ($n = 8$). For this, the papers used difficulty scores provided by other health care providers in a previous stage [92], difficulty based on domain knowledge [50, 63, 104], severity of the case [86, 88], and scores provided by the AI model or another model [85, 117]. This aspect of the situational context influences the user's *Confidence* in their decision [49, 50, 63, 77] and *Reliance* on the system [60, 85, 88]. *Case Difficulty* also affects *Usefulness*, but the results were mixed: [49] found a negative correlation and [77, 117] found a positive correlation. *Performance* and *Curiosity* were also influenced by the *Case Difficulty*: *Performance* decreased when the *Case Difficulty* was high [77], and *Curiosity* increased when the *Case Difficulty* was high [104].

*4.12.3. Objective System Aspects*

Objective systems aspects (OSAs) are "the aspects of the system that are currently being evaluated" [48]. The previous analysis conducted in [15] yielded six properties: *AI Model Performance*, *AI Model Certainty*, *Certainty*, *Continuity*, *Separability* and *Consistency*.

*AI Model Performance* aims to quantify the model competence. In all these studies, an AI model was trained with a specific dataset, and therefore, all these papers measure this property. The majority of papers ($n = 76$) only test one unified system, and therefore, they cannot compare how the performance of the AI model can affect other properties. However, there are 6 papers where the *AI Model Performance* is used to compare different systems: [116, 128] compare multiple AI models for one system and [59, 79, 91, 125] compare different AI models for different datasets. In all these papers, the *AI Model Performance* is measured at the abstract level using traditional Machine Learning metrics like accuracy, precision or recall. Additionally, two papers measure or mention *AI Model Performance* without making an explicit comparison: He et al. [121] presents a system that uses multiple models at the same time. In their study, the goal of the user is to understand the performance of these various models using a visualization; and in Barda et al. [98], the *AI Model Performance* is mentioned in the qualitative analysis as an aspect that was missing in the system. In regards to its relations with other aspects, He et al. [121] evaluated how it related to the *Perceived Model Competence* and it found a small correlation between the two. The two studies that compare multiple AI models for the same system also studied the relations of this property. Herm et al. [116] found that higher *AI Model Performance* is associated with higher *Satisfaction*, and [128] found that when using higher-performance models, users had better *Performance*.

*AI Model Certainty* was part of the evaluation in 4 papers, two mixed studies and two quantitative. In these mixed studies, it was mentioned in the qualitative analysis as a missing aspect of the explanation. To capture this feedback, Panigutti et al. [99] used an open question in a questionnaire, and Anjara et al. [74] used a Think Aloud protocol.

*Certainty*, *Consistency* and *Separability* were not part of the evaluation of the selected papers. The first one was mentioned in one qualitative study as an aspect that influences the *Explanation Power* [49], but it was not analysed as an individual aspect of the experience. The second and third were not even mentioned in the studies. We believe this happens because these specific properties are related to the XAI method's mathematical function and should be evaluated before conducting any user studies.

*4.12.4. Explanation Aspects*

This component groups the properties that measure the quality of the generated explanation. The papers surveyed presented evaluations that covered all these properties.

*Size* and *Structure* were the ones that appeared the most. *Size* has been mostly mentioned in qualitative data analysis ($n = 7$). The property was evaluated using closed questions in four studies, and in two others, it was measured at the abstract level. This property is related to *Cognitive Load*, *Relevance to the Task*, and *Alignment with Situational Context*. The relation between these two properties is well illustrated in this quote:

> [Size] The full explanation with all the details of significant evidence is accessed only if desired, and [Relevance to the task] it is more suitable to retrospectively analysing the prediction or the decision in [Aligment with situational context] the user's own time, or in retrospective clinical meetings. [108]

*Structure* was also primarily mentioned in qualitative research: it is part of the qualitative analysis of 6 papers, and it is measured by closed questions in only two. Its relations to other properties are deeply explored in Pisirir et al. [108]. This study presents a comparison of two different explanations, one narrative and one with bullet points, and compares their effects on the users. They found that the *Structure* affected *Cognitive Load*, *Confidence*, and *Trust*.

*Representativeness* was defined as "An explanation is representative if it holds for many distinct but similar instances" and *Continuity*, as part of Objective System Aspects, as "The function should provide similar explanations for similar instances". By analysing the use of this concept in the user studies and by re-analysing the literature on these topics [11, 23, 25, 28, 29, 140, 141], it was decided to join these two properties in one only concept. The reason for this is that both focus on the fact that one explanation or style of explanation can be used to explain the prediction of similar instances, but the *Continuity* was focused on measuring the ability of the model to achieve it, and the *Representativeness* was focused on how the user would perceive this effect. By joining one concept that can be measured at different levels, we can reflect how these properties are used in user studies. The new concept is still called *Representativeness*, and its definition is *measures similarity of explanations of similar but distinct instances*. We found that *Representativeness* was evaluated using closed questions in two studies [64, 100], mentioned in one qualitative analysis [122], and also measured the abstract level using a metric in [100].

*Completeness* and *Correctness* were mentioned in three studies of with *Decision Support* as usage context. All of them were measured with different measurement types. In [125], a mixed study, both properties were measured at the abstract level using metrics proposed in the paper. In [110], a qualitative study, *Completeness* is mentioned in the qualitative analysis of the interview. Finally, [111] uses a closed question to evaluate *Completeness*.

*Necessity* and *Sufficiency* refer to the appropriateness of the information that is present in the explanation. The first one was evaluated with closed questions [100] and metrics [72, 84]. Jaber et al. [114] asked about this property in an open question as part of a questionnaire, and it appeared as a topic in their qualitative analysis. Morais et al. [110] also asked about it as part of their interview script, but it was not part of the aspects users mentioned. The second property was investigated by explicitly asking questions about it in one study [110], and quantitatively, using questionnaires, in five studies [57, 97, 100, 111, 122].

The last property of this group, *Contrastivity*, was measured quantitatively in one study [82] that also found it is correlated with *Usefulness*. In [124], it was mentioned by the participants as an aspect that was desired, especially when the patient was on the "edge of two classes". Lastly, the qualitative analysis in [84] revealed that poor *Contrastivity* leads to poor *Understanding*.

*4.12.5. Subjective System Aspects*

Subjective System Aspects (SSA) are "users' perceptions of the Objective System Aspects" [48]. This component experienced the most changes of all the components. We found 3 new properties that reflect the nuances in what users perceive in the XAI scenario.

*Alignment with Situational Context* context was evaluated in three mixed studies [50, 80, 104] and in three qualitative studies [58, 66, 110, 118]. Users mentioned that time pressure and *Case Difficulty* were factors considered when requesting explanations and when pondering whether the explanation was useful or not. Only one study, [50], specifically asked questions that related to understanding the integration of the explanations into the users' workflow.

*Cognitive Load* is an aspect commonly evaluated in HCI research. XAI research is not an exception. Twenty papers evaluated this Property, out of which only five ([70, 81, 86, 124, 126]) used the standardized NASA-TLX questionnaire [142]. Nine other papers evaluated this property with tailored questionnaires. Only two papers [117, 124] evaluated this property, both with a quantitative questionnaire and also in a qualitative evaluation. The relations of *Cognitive Load* with other properties have not been as prominent as expected. *Size* [58] and *Structure* [108] were found to influence *Cognitive Load*. Only one paper [98] explored how *Cognitive Load* can affect the perceived *Usefulness* of the explanations.

In the previous version of the framework, we identified *Information Expectedness* as an important property. During this review, we also found another property that was close but different: *Information Correctness*. On the one hand, *Information Expectedness* refers to whether the information provided in the explanation was anticipated by the users based on the input information and their knowledge. *Information Correctness*, on the other hand, refers to whether the information that is provided is accurate in terms of the domain knowledge. For instance, when explaining a diagnosis based on an X-ray image, if the anatomical information is tagged correctly, then the information that is shown is correct. At the same time, the explanation uses the anatomical information to say "The diagnosis is X based on A and B", and that explanation is expected by the user because of the knowledge she has on the domain. If the anatomical information was tagged incorrectly, then the *Information Correctness* would be zero. In the case, the explanation does not use the information the user would expect (it uses partial information or other information), then the *Information Expectedness* would be low.

*Information Correctness* was part of the evaluation of 11 studies. Four studies evaluated it using more than one type of measurement: [100, 112] used close questions and abstract level metrics, [73] used close questions and it was mentioned in their qualitative data analysis, finally [84] used abstract level metrics, close questions and it also appear in their qualitative analysis. The other eight papers measured this property only with one type of measurement: in 4, it was part of the qualitative data analysis, 3 used close questions, and one used only abstract-level metrics. In regards to its relations to other properties, it was found that it correlated with *Trust* [68], and *Information Expectedness* and *Satisfaction* [84]. Even though this aspect is an important part of the user experience, it has not yet been included in the evaluation of many papers. We believe that, given the explosion of generative-AI-based explanations, this aspect might become more important. For instance, when explaining images using AI-generated examples, their correctness should be considered as part of the evaluation.

*Information Expectedness* was part of the evaluation of 25 studies. We found it relates to 11 properties: *Domain Experience*, *Information Correctness*, *Confidence*, *Explanation Power*, *Intention to Use*, *Perceived Model Competence*, *Relevance to the Task*, *Understanding*, *Trust*, and *Usefulness*. These relations were supported by 8 different papers that showcase the importance of this aspect in the evaluation: the difference between what users anticipate and what users see in the explanation can determine a big part of their experience.

Another new property we found in this survey is *Prediction Expectedness*. While *Information Expectedness* is related to how much information of the explanation can be anticipated, this property relates to the anticipation of AI prediction. *Prediction Expectedness* affects user *Curiosity*, reflecting more engagement with the system. For instance, in [50], a user reported that when her prediction did not match the system predictions she " [Curiosity] tried to re-investigate the recordings based on the AI explanations to find out whether my reasoning on predictions was strong enough to modify the AI prediction". This connection was found in 4 papers [50, 66, 67, 101], which reflects the importance of this aspect in the experience. One paper [50] also found that this property affects *Perceived Model Competence* and *Trust*.

*Explanation Power* is a property that was identified in the previous version as "Measures whether the selected causes make the user understand the reasons the model considered when making a decision". During the revision, we noticed that this definition was focused more on *Understanding Explanation* than on the capability of the reasons to make the user accept the prediction. We noticed that many papers discuss the convincingness of the reasons and how much users change their decisions based on the information they received in the explanation. Based on all this, we decided to change the focus of this property by redefining it to "Measures whether the selected causes are capable of making the user accept the prediction". This property is mainly measured with implicit measurements. Six studies measure this aspect by evaluating how much a user prediction changes after seeing the explanation [70, 71, 77, 99, 120, 123], while other 4 [49, 55, 56, 75] measure how many times the user agrees with the system or how much their perception of the model outcomes changes after seeing the explanation. With respect to its relations, it was found that it relates to 5 Properties: *Certainty*, *Confidence*, *Usefulness*, *Information Expectedness* and *Structure*. In total, four papers support these relations.

*Perceived Model Competence* refers to how the user perceives the model's performance. This property can affect user *Trust* and *Confidence* as stated in this quote from [49]:

> When participants found errors in the highlighted sentences, `Confidence` they felt "unsure and insecure". For example, P27 said: "I was confused and stopped using the system after `Perceived Model Competence` I found it made obvious mistakes."

In total, 10 papers evaluated this property, out of which 7 evaluated this property using close questions.

Related to this property is the *Perceived User Performance*. Users tend to have an idea of how much their performance improves when using the system, and this idea can affect how likely they are to accept the system. This property was measured by questionnaires in four studies [49, 70, 80, 122], and it was mentioned in the interviews in one qualitative study [81].

*Relevance to the Task* measures how well-adjusted the explanation is to the task the user has to perform. This property has the same spirit as alignment to the situational context, but instead of measuring with respect to the context, it is measured with respect to the task. The property is widely evaluated: it was measured with quantitative questionnaires in 14 studies [62, 70, 72, 73, 78, 83, 84, 96, 100, 101, 112, 117, 124, 126] and appears in the qualitative results of 11 papers [46, 60, 74, 76, 80, 86, 98, 102, 106, 121, 125]. Only one paper [112] studied an implicit metric, an abstract metric and used close questions to measure the medical relevance of specific parts of the explanation.

Doshi-Velez and Kim defined Cognitive chunks for XAI as "basic units of explanation" [143]. They present it as part of method-related dimensions of interpretability that "may correspond to different explanation needs" [143]. In [15] is presented as one property called *Form of Cognitive Chunks*. Unexpectedly, this aspect is only mentioned in two mixed studies [67, 80] as part of its qualitative section. During the analysis, we saw this aspect incorporated into the design choices of a few papers [98, 101, 119], but its impact on the user experience was not measured or evaluated.

*4.12.6. User experience*

We found 9 properties in this component, two more than in the previous framework definition.

*Curiosity* aims to measure how much the user can potentially engage with the system. This property appears in two qualitative studies and one mixed study. In this last study, only this property is mentioned in the qualitative interview and also measured with implicit feedback by collecting user interaction with the system [104]. In the other two studies [66, 119], it is part of the qualitative data analysis. Although this property is not part of the evaluation of many papers, it does appear to be related to other properties. Cases with high difficulty tend to increase user *Curiosity* [104], *Information* and *Prediction Expectedness* affect user *Curiosity* [50, 66, 67, 101] and by doing so, they can increase *Understanding* [50, 66]. Users "expressed a *Curiosity* in why the agent would make such a decision to be able to better understand the system." [66].

*Usefulness* aims to measure the utility of the explanation for the user. This property is widely used: 25 papers explicitly measure it in the studies via close questions ($n = 19$), open questions ($n = 6$) and implicit measures ($n = 1$), and in the other two, it is spontaneously mentioned by users. Regarding its relations with other properties, we found that *Case Difficulty* tends to increase the *Usefulness* of explanations: the harder the case, the more useful the explanation [49, 77, 117]. We also found that it has relations with *Cognitive Load*, *Contrastivity*, *Explanation Power*, and *Information Expectedness*.

*Understanding* is a common property that is widely used in the area and appears in survey papers, for instance [17, 40, 141], as part of their proposed evaluation. The conducted analysis found, as expected, 33 papers (40% of the papers) that evaluated this property covering all Usage Contexts present in the survey. Twelve papers included more than one dimension of *Understanding*. For instance, [101] has these two sentences to evaluate *Understanding*: "I understand when and why CORONET may provide the wrong recommendation in some cases", that refers to model understanding, and "The scatterplot with all patients is easy to interpret", which refers to understanding the shown explanation. This pattern was repeated in several papers [46, 47, 62, 83, 84, 97, 100, 101, 110, 111, 117, 122], which supports the idea that *Understanding* is composed of at least two factors: explanation understanding, that refers to apprehend what the explanation is informing, and model understanding, that refers to apprehend how the model works. The first type of understanding, called *Understanding Explanation*, evaluates whether the user can grasp the ideas that the explanation is trying to convey. Usually, this type of understanding needs to happen in order to acquire model understanding. *Understanding Model Behaviour* is a more complex achievement; not only does the user have to understand the information that she is seeing, but she also has to infer how the model created that information. The achievement of *Understanding Explanation* cannot be used as a proxy for *Understanding Model Behaviour*, as shown in [110]:

> As mentioned at the beginning of this section, (Understanding Model not achieved) the most salient aspect of the analysis is related to the explainability of the XAI methods, which is primarily evidenced in the XAI is not explanatory code.
>
> Despite the issue regarding explainability, (Understanding Explanation achieved) most domain experts acknowledged that the visual elements are easy to interpret and were able to perform the identification of major/minor influencing features.

Considering that people tend to overestimate how well they understand [141, 144], it is surprising that this property is most usually measured only with self-reported feedback from users ($n = 17$) and not via objective understanding measurement or techniques. Only two papers tested whether users understood using questionnaires to elicit the user mental model [53, 55] and [110] evaluated this understanding by asking qualitative questions for model and explanation understanding.

*Trust* is one of the most important properties of explanations [31, 36, 38, 145, 146]. It is measured in 31 studies: 14 mixed, 10 quantitative and 5 qualitative. Twenty papers measure *Trust* using closed questions, and three papers only with open questions. In one of this last set of papers, it does not appear as part of the interview analysis because users discuss the AI context regarding *Trust* and not their *Trust* towards the system [110]. In six papers, it appears as part of qualitative analysis, even without asking about it explicitly. One study [71], in addition to closed questions, implicitly measured the evolution of *Trust* over samples according to self-reported *Trust* and *Correctness* of AI model prediction. Just like *Understanding*, these studies cover all Usage Contexts. Five papers found relations between *Trust* and other properties [50, 67, 68, 104, 108]. These are: *AI Model Certainty*, *AI Model Performance*, *Information Correctness*, *Information Expectedness*, *Prediction Expectedness*, *Structure*, *Intention to Use*, *Understanding Model Behaviour*.

*Satisfaction*, defined as the level of fulfilment the user gets while interacting with the system, is measured in 11 papers, five quantitative and 6 mixed studies. In all these studies, it is measured by closed questions, and in one [47], it additionally appears as part of the interview analysis.

*Intention to Use* measures the willingness of users to use the technology. It appears in 8 mixed studies, 6 quantitative and 1 qualitative study. This property is measured only by closed questions in 12 papers and by open and closed questions in one paper. In two papers, it appears as part of the interview analysis. According to qualitative evidence, this property depends on users' *Perceived Model Competence* [49, 67] and *Perceived User Performance* [81]. Using a Mutual Information Analysis, [117] found that *Intention to Use* depended on *Understanding Model Behaviour* and [67] found a correlation between *Trust* and *Intention to Use*.

*Confidence* is a property that is not part of the original version of the evaluation framework. It is defined as a measure of the user's subjective belief in the correctness of a decision. This property appears in a total of 22 studies: 13 quantitative studies and 9 mixed. However, only in two of these mixed studies it is evaluated in a qualitative way: [102] it appears as part of the qualitative data analysis and [103] asks about it in their open questions. Its importance in the area is reflected also in the number of relations it has with other properties: 11 papers have evidence of relations with other properties. The properties that are most mentioned are: *Case Difficulty* ($n = 4$), *AI Model Certainty* ($n = 2$) and *Domain Experience* ($n = 2$). The other properties that relate to *Confidence* are: *AI Model Certainty*, *Case Difficulty*, *Case Difficulty*, *Domain Experience*, *Explanation Power*, *Information Expectedness*, *Perceived*

*Model Competence*, *Performance*, *Reliance*, *Structure* and *Understanding Model Behaviour*.

*Controllability* is a property widely used in recommender systems. Several papers have shown that the higher the *Domain Experience*, the higher the desired control. In this survey, we found that it is part of the evaluation of only 4 papers: 3 mixed studies and one qualitative. Half of these papers used closed questions to measure it, two used open questions, and in one study, it appeared in the qualitative analysis. It is interesting to note that only 2 [80, 119] out of the 18 papers that had some interactive elements in their systems measured this property.

*4.12.7. Interaction*

*Efficiency* was measured in 7 purely quantitative and 5 mixed studies. Although this property can be easily measured implicitly by measuring the time it takes for users to complete the task, it is less used than other more complex to acquire properties. In the mixed studies, it was measured implicitly only in two of them [49, 76]. In 5 of 7 of the quantitative studies, it was measured implicitly [56, 70, 75, 85, 91], and one study measured it both implicitly and self-reported [64], but they did not compare the contrast between the measurements. This property is measured mostly in *Diagnostic Decision Support* tasks (9/12). The datasets used in these studies follow a different pattern from the average: five studies explain the prediction of images, three of tabular data, and two of graphs. Text, time series and video have one each.

*Performance* measures the user and how well the user can perform the action using the system, including predictions and explanations. This property is measured in 17 quantitative studies and 6 mixed studies. It is only measured implicitly by checking the user's predictions against the ground truth. The dataset used in these studies does not follow the typical trend: 12 studies used images, followed by 5 tabular, with the other types having fewer than 3 occurrences. Most of the studies are set in a *Decision Support* usage context ($n = 17$).

*Reliance* measures how willing the user is to provide decision control to the machine. It is an important aspect of the adoption of healthcare systems [147]. However, only two papers measured it or discussed it with users. Chen et al. [88] measured it with a likert-type questions by asking users directly whether they would be willing to provide decision power to the system, and in [104], users said that they should not rely on the system 100% and they were "concerned with liability and responsibility if [they] followed the model and the patient had a bad outcome". This put emphasis on the legal aspect of using AI systems in healthcare settings. We found some papers that claimed they were measuring *Reliance* [71, 77], but they measure how much the user decision changes after seeing the prediction and/or explanation. In this survey, we considered that to be a measure of *Explanation Power* because it does not reflect whether the user will give decision power to the machine but whether she can change her decision based on the information the machine displays.

The last property that we coded is *Variance User Decision*. This a new property that measures how much different users agree on a specific decision. This does not relate to agreeing with the machine but among users. This measure is relevant in the healthcare scenario because one of the goals of these systems is to achieve standardised care, which only happens if all healthcare providers share the same criteria and knowledge. Five quantitative and 4 mixed studies measured this property, all of them using an implicit metric to measure it and it was mostly used in Decision Support Contexts ($n = 6$).

## 5. Updated User Centric Evaluation framework

During the coding process, we found seven new properties related to SSA, UX and Interaction. These are: *Information Correctness*, *Perceived User Performance*, *Prediction Expectedness*, *Confidence*, *Intention to Use*, and *Variance User Decision*. Additionally, we found 2 Personal Characteristics and 1 Situational Characteristic properties that were repeatedly used by several studies papers. These are: *Domain Experience*, *Attitude Towards AI* and *Case Difficulty*.

We combined the properties *Continuity* with *Representativeness* and based on the evidence we found, we split *Understanding* into *Understanding Explanation* and *Understanding Model Behaviour*.

Additionally, we redefined *Explanation Power* to make it clear that it refers to how much the user changes their mind based on the explanation they receive.

Based on these changes and the evidence we gathered, we also redefine how the framework components are related. In the original work, it was not clear how the *Personal Characteristics* and the *Situational Characteristics* were related to the *Explanation Aspects*. We found one paper [98] that found that *Domain Experience* affects *Size* and *Structure*, so we decided to add the possible influence of personal characteristics to the explanation aspects.

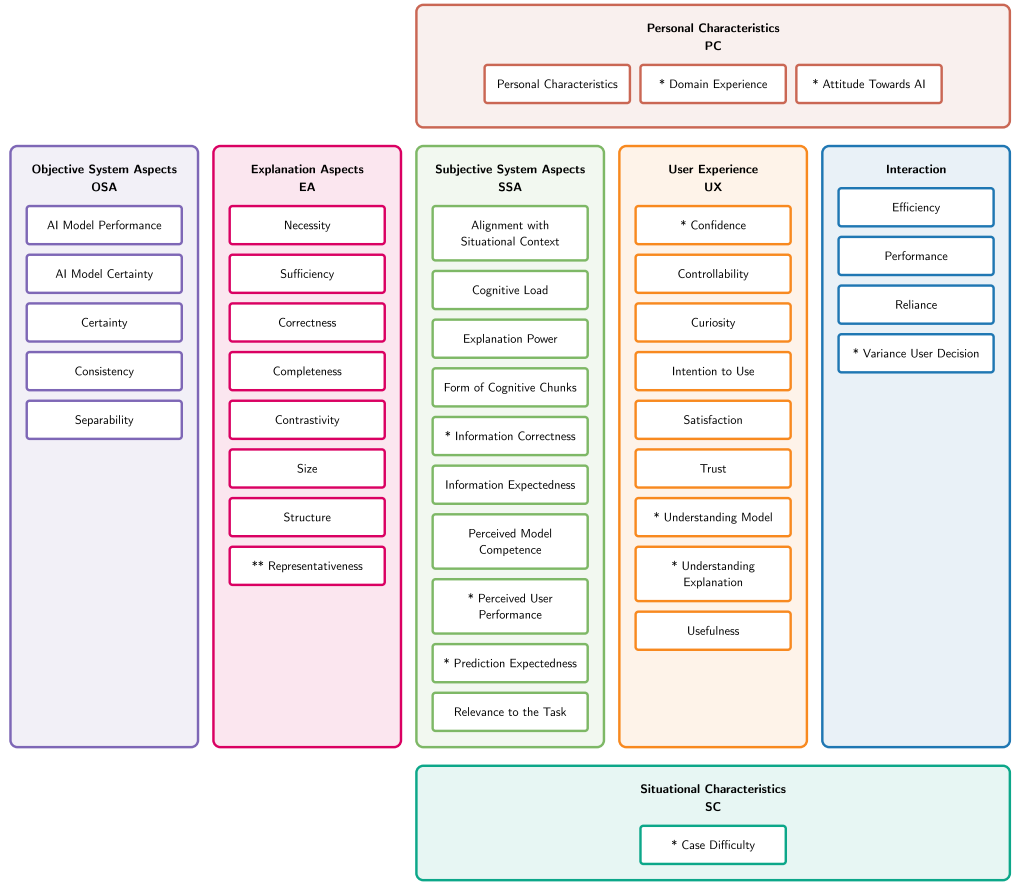

Figure 10: User Centric Evaluation framework updated. New properties are marked with * and modified properties are marked with **.

## 6. How to use the framework

To understand the relations of the different elements that were encoded in this survey and the way they define the evaluation procedure, we propose a layered approach to connect them made based on the observations during coding of the user studies. This approach can serve as a guide to design XAI systems in the early stage and the evaluation that is more appropriate for it. As shown in fig. 11, each layer contains elements that are interdependent, and each layer is informed by all previously encompassing layers. Sections 6.1, 6.2 and 6.3, will elaborate on the layers' aspects and how the design choices are limited in each layer, and section 6.4 explains how to use them to define which properties to evaluate. The last section, provides general recommendations for reporting the results.

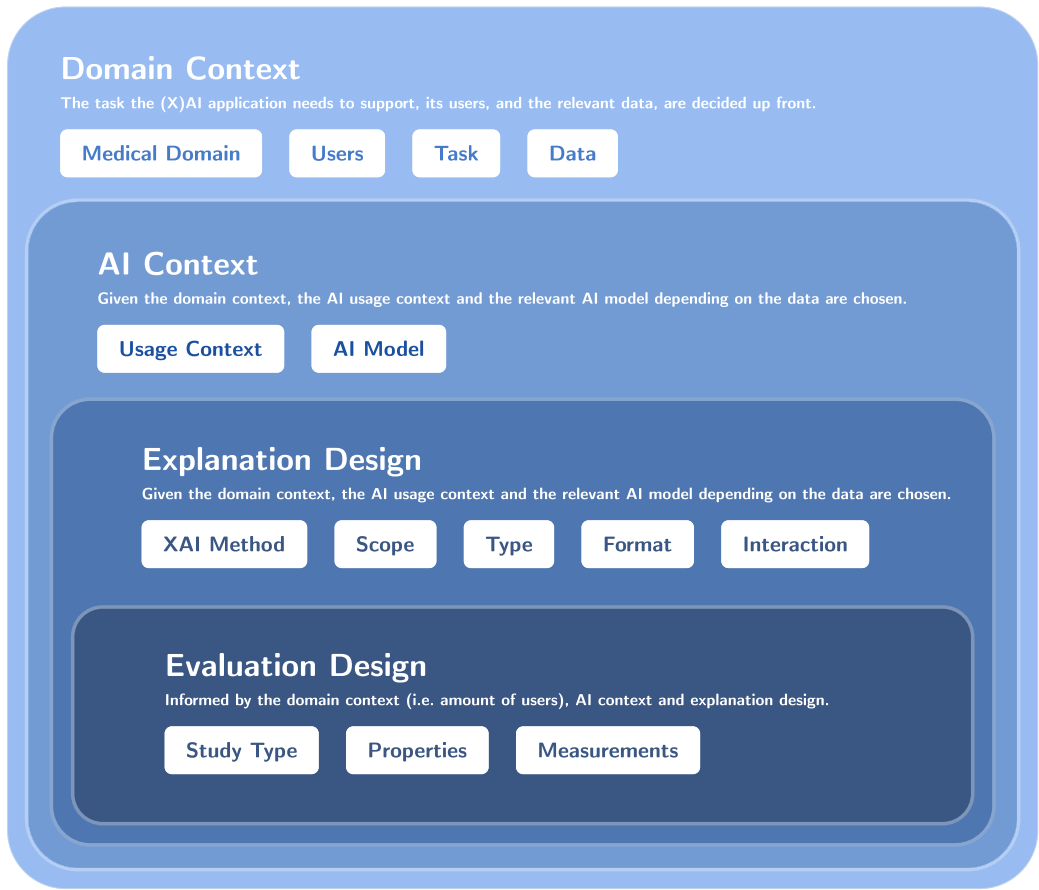

Figure 11: Layered model to design evaluations. During a *Designing explanation* stage, the layers *Domain Context* and *AI Context* will strongly influence the *Explanation Design*. All layers influence the *Evaluation Design*.

*6.1. Domain Context*

The foundational layer aims to build a contextual ground that consists of what is known about the deployment space. It consists of the *medical domain* in which the task is situated, the *users* performing the task (i.e., medical expert, patient, etc.), the *medical task* performed by the users and assisted by AI and the *data* used for the task.

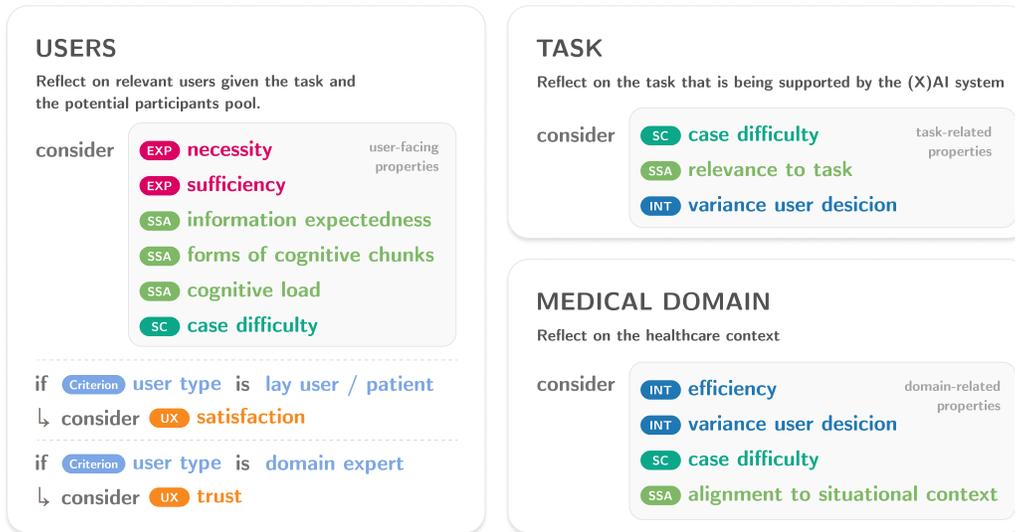

Figure 12: Recommendation and suggestion to select properties according to the *Domain Context* Layer

## 6.2. AI Context

The second stage involves selecting an appropriate *AI model* to support the medical task. This decision is influenced by the nature of the task and the available data. Additionally, the intended usage context (e.g., decision support or model auditing, following the taxonomy of [26]) further refines model selection. For example, if the model's performance in the given task is not yet established, usage scenarios like auditing or capability assessment should be prioritised over assessing decision support.

## 6.3. Explanation Design

The previous layers strongly inform this stage. Here, the *XAI method* options are limited by the Task, the Data and the AI model. The usage context might strongly influence the design choices along the four axes we have identified: *Scope*, *Type*, *Format* and *Interaction*. These choices might shape the selection of the XAI method as well. The technological limitations of XAI methods and the nature *data* might also influence the design choices along the four axes. It is possible that some combination of them finds no XAI method that works for the context, so these axes would be revisited to adapt them to the technological limitations.

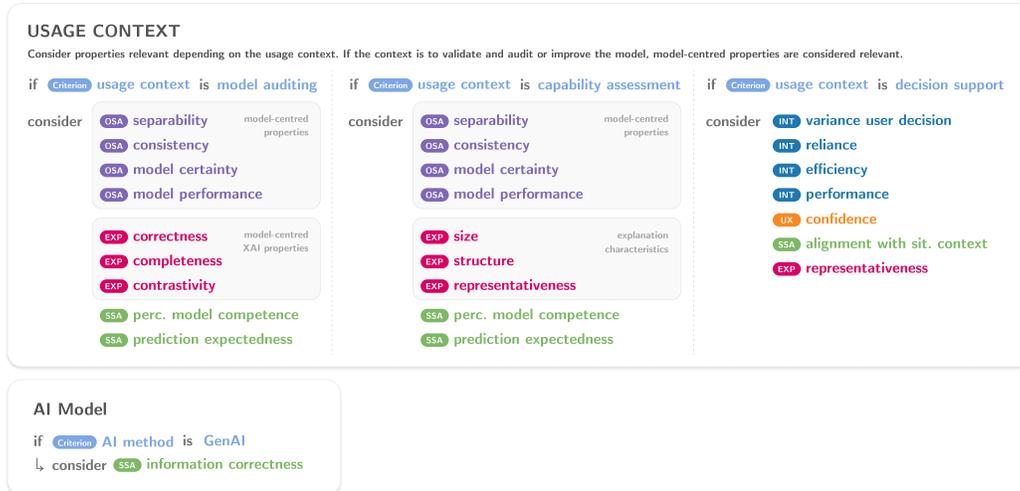

Figure 13: Recommendation and suggestion to select properties according to the *AI Context* Layer

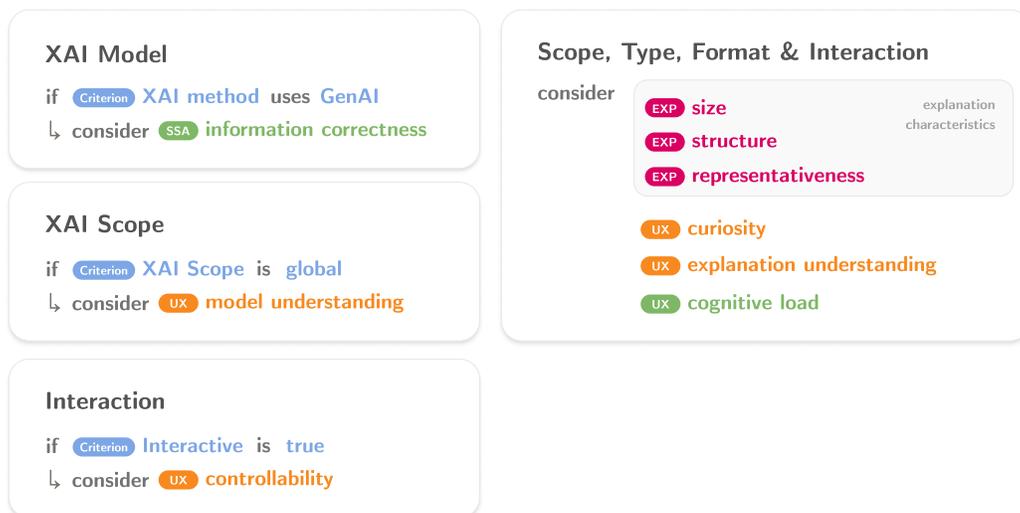

Figure 14: Recommendation and suggestion to select properties according to the *Explanation Design* Layer

*6.4. Evaluation design*

The final stage consists of the evaluation itself: the *study type*, the choice of *properties* and their *measurements* To be able to decide on the evaluation design in XAI studies, all layers should be taken into account to reflect on the evaluation appropriately and effectively. The following five-step guide outlines a structured process for this decision-making.

*Step 1: Select relevant properties*

Choosing which properties to evaluate can be a complex task. Our analysis of user studies revealed recurring patterns that emerge when properties are applied in specific contexts. Each of the criteria (such as medical domain, users' characteristics, data used, etc.) can potentially influence which properties are relevant. These connections are demonstrated in figures 13, 12 and 14, which show how to make an informed decision with the properties selected.

One of the most crucial criteria we observed is a usage context, which also aligns with Liao et al.'s study [26]. Usage context, among other criteria, could serve as a good starting point when uncertain which properties should be selected.

Furthermore, an overview of relations between properties (see section 4.11 and fig. 9) can help identify how the properties being measured relate to those already selected. This can be facilitated through the use of the *Relations Between Properties*. These relations can help to identify possible mediating factors that can influence results, and also properties that can influence results but were not identified before.

*Step 2: Identify measurements available at for all elements*

Once all the potential properties have been identified, the process of filtering and selecting appropriate measurements and study design starts. Not all properties can be measured at all levels, and some automatic metrics may not be compatible with the available data or specific XAI techniques. For these reasons, it is essential to check whether suitable measurement already exist.

If no existing measurement is found for a given property, new survey items or metrics may ned to be developed and validated. Alternatively, the evaluation design may need to shift toward qualitative or mixed-methods approaches to gather meaningful insights.

*Step 3: Identify properties that can be measured without user explicit feedback*

At certain stages of XAI evaluation, abstract level measurements may be sufficient and do not require user feedback. For example, abstract-level properties typically do not depend on user feedback. In contrast, measurements at the communication level require implicit user feedback.

Identifying which properties can be evaluated without involving users directly helps simplify the evaluation process and allocate user engagement more efficiently.

*Step 4: Prioritise properties and select*

The next step involves prioritising the selected properties. Properties that can only be evaluated through direct user input should be ranked from high to low in order of importance based on the research goals. The decision on which properties to measure with users should consider both the time users have available and the effort required to provide feedback. For example, if users can only participate for five minutes, a 20-question survey would be too long. Therefore, it is crucial to test the feedback methods in advance to ensure they fit within the available time.

*Step 5: Revise selection*

After selecting the properties based on the previous steps, a final review is necessary to ensure alignment with both the research objectives and the overall study design. This step is especially important for quantitative studies. If only a limited number of properties can be assessed using closed-ended questions or predefined metrics, the study design may need to shift to a mixed-methods or qualitative approach to maintain relevance.

This process can be iterative and repeated to align the research goals with the conditions of the evaluation. In addition, if the researchers are designing the system and the evaluation procedure at the same time, this process can help to design the system according to what aspects can be evaluated.

*6.5. Recommendations for reporting results*

The properties and their definitions are not random: the names have been chosen to minimise possible overlap between concepts, and the words chosen for the definitions try to avoid ambiguity. When reporting results, these names should always be kept and reused, even if they get repetitive. This helps in understanding the paper's results, and it makes it easier to compare with other studies.

When users can be grouped by their experience level, the results should always be reported at the group level. As explained in Section 4.2, there are differences between experience levels, so making sure these differences are presented makes it easier to compare studies.

## 7. Discussion of Research Goals

In this section, we discuss aspects related to the two research goals.

*7.1.* **RG1** *To provide a framework of well-defined and atomic properties that are part of the XAI user experience in the healthcare domain*

*7.1.1. Disentanglement of properties helps to better understand the evaluation*

We find that *Satisfaction* is mentioned in surveys as an aspect equally important as *Trust* and *Understanding* [27, 32]. However, our research shows that this is not the case in this domain. The most studied property is *Understanding* ($n = 33$), followed by *Trust* ($n = 30$) and *Satisfaction* is measured only by 11 studies. This can be explained by the different meanings assigned to this *Satisfaction*. Rong et al. [16] considers *Satisfaction* as part of a bigger aspect called *Usability* that comprises other properties: *Usefulness*, *Cognitive Load*, ease of use and detecting undesired behaviours. Löfström et al. [27] define *Satisfaction* as "The degree of how much the users **feel they understand** the system, the explanations, and the user interface", which in our framework is defined as *Understanding Explanation* and *Understanding Model Behaviour*. Mohseni et al. [40] also presents *Satisfaction* and *Usefulness* as part of the same construct. Lopes et al. [32] presents *Satisfaction* and *Usefulness* as part of the same category. These different definitions make it harder to compare the studies. Our framework proposes to disentangle these concepts into more granular properties. By doing this, we can better understand the connections between the concepts and not simply assume that they are related to each other. Our definition is closer to the definition of satisfaction by the Merriam-Webster dictionary ([148] sense 2).

*7.1.2. The healthcare domain has specific characteristics that make the evaluation different from a general domain*

Most evaluation frameworks have presented evaluation guidelines without any specific domain. In this study, we focused on the healthcare domain and discovered small differences that have not been considered in these general domain frameworks.

First, as mentioned previously, *Satisfaction* is not as important as other frameworks have indicated. In this context, *Understanding* and *Trust* are more relevant than *Satisfaction*. They are measured in 3 times more papers by using all types of measurements. The relations of *Understanding* and *Trust* with other properties are also studied.

Second, *Reliance*, defined as the user's willingness to provide control to the machine, is part of the evaluation of only 2 studies; however, *Trust* is measured in 30 studies. This difference can be explained by two factors: First, most of these studies present prototypes that are only evaluated at a single point but *Reliance* is a consequence of repeatedly using a system and developing appropriate *Trust*; Second, as stated in [104], it is not yet clear who holds responsibility when an AI recommendation is followed, and in the healthcare settings, this accountability of actions is an essential part of interacting with patients.

Third, the personal characteristics such as *Domain Experience* and *Attitude Towards AI*, and the *Case Difficulty* are aspects that influence the property outcomes and are never mentioned in these general frameworks. These aspects of the context in which the user's activity takes place are essential to first decide what should be measured, and then understand the results.

Finally, *Confidence* is a new property that is measured in several studies, particularly in decision-making scenarios. This aspect, influenced by *Case Difficulty* and *Certainty*, appears in 22 papers, more than *Intention to Use* or *Cognitive Load*, is not mentioned in general evaluation frameworks, and, as the numbers stress, it is a relevant aspect to consider in evaluations in this domain.

All these aspects considered, we can see how this domain has specific properties and relations that are not mentioned in other frameworks, and affect how the evaluation should be conducted.

*7.1.3. Current Standard measurements are not convenient for evaluation*

Several papers that we studied used standard measurements in the evaluation. Among them we find System Usability Scale [149], Hoffman's Satisfaction Scale [141], TAM [150], and UTAUT [151]. This review did not specifically analyse which standard measurements were used in the studies. This decision was made based on the fact that we analysed the studies by properties, and these instruments measure more than one property at the same time and provide a unified score. We noticed that some studies did not provide the unified score and simply reported the values question by question [86, 104], which reflects that these instruments may not be convenient in this context.

Most studies created their own questionnaires to evaluate the properties, and almost none of them repeated the questions from a previous study. This is an issue already reported by previous studies [18] and it could improve by conducting studies that specifically aim to validate and create constructs.

## 7.2. **RG2** *To provide clear guidelines on how to design the evaluation of XAI systems based on the system characteristics*

### 7.2.1. *Quantitative studies should comply with a minimum number of users*

The studies had, on average, 34.5 users per study. Purely quantitative studies had between 3 and 223. In this group, 21 studies had 16 or fewer participants, out of which 10 were able to provide confidence scores for their results, and only 6 recognised that the sample size was a limitation of the research. These sample sizes are too small to draw meaningful conclusions.

Based on the authors' experience, we understand this is a common problem in the healthcare domain: finding users that can participate is difficult and relying on services like Prolific or Mechanical Turk is expensive or not necessarily trustworthy [152].

For this reason, we encourage researchers in the area to carefully consider their research context when setting up these studies. Quantitative studies should be used to test hypotheses [18], and small samples do not allow for achieving this goal. If it is known or anticipated that finding users will be hard, it is recommended to conduct a thorough qualitative study with a few users. The results of these studies are more informative and legitimate than quantitative studies of a handful of users.

### 7.2.2. *Layered model and iterative design of evaluation*

In this study, we propose a process to select the properties that should be measured in user studies, taking into account the context of the system, the users and AI components. The layered model for evaluation design section 6 helps to understand the different dependencies that exist between the system's components in order to compile a set of appropriate aspects that should be used to measure the system's effects on the users.

Previous work had focused on identifying the aspects that needed to be taken into account when designing and/or evaluating an XAI-based system, but none of them specifically stated when these aspects should be measured. Lopes et al. [32] and Kim et al. [17] organised previously defined aspects in new taxonomies, but do not provide guidelines on which aspect to measure depending on the system characteristics. More recently, Rong et al. [16] provides a guide on how to conduct quantitative user studies, but it does not provide any specific details on how to decide what to measure. Furthermore, their guideline could be applied to any kind of user study, not just XAI user studies.

Our proposal closes this gap by explicitly stating what aspect to measure based on the system's context, the users and the AI and XAI components. This guideline will help other researchers design evaluations that are well justified and aligned with the research goals.

## 8. Limitations

The first limitation of this review is that it does not propose new measurement instruments or provide recommendations for them. There are very few studies in the Human-Computer-Interaction domain that evaluate whether a question really measures the construct, in this case, whether a question or metric measures a property. The works of Knijnenburg and Willemsen [48], Pu et al. [153] and Jin et al. [154] are well-known frameworks that work towards that direction. New studies need to be conducted to establish the most appropriate measurements for the properties. However, we do provide the list of questions and instruments that were present in the papers, with the specific property they measure. This list can be found in Appendix B.

The decision of not including Wizard of Oz studies was made to ensure we would analyse explanations that were created taking into account the AI model and XAI method affordances (see section 3.2). This led to leaving out of the study several works that are usually in the early stages of development. It is not clear the extent to which this decision impacted the usage context we were able to find in the studies.

Additionally, user studies are expensive, in time and monetary resources, so usually they are made when the systems are in late prototyping stages. We believe this decision also affected the number of usage contexts we were able to find.

## 9. Conclusion and Future work

In this study, we compiled and analysed evidence from user studies applying XAI systems within the healthcare domain that use human-centred approaches. Using a set of predefined criteria, we assessed how various properties of XAI systems are evaluated in practice and summarised how these are applied in healthcare contexts. Based on our insights, we developed practical guidelines to support the selection of appropriate and context-sensitive evaluation strategies.

Our analysis also highlights the extent to which human-centred measurement frameworks are currently adopted in XAI user studies in healthcare, providing insight into both common practices and existing gaps. These findings can inspire and inform future research by pointing to established good practices.

However, the proposed framework should be tested in other critical domains such as education, industrial settings, and human resources. If the framework does not generalise well across domains, domain-specific analyses will be necessary to identify distinct evaluation needs and property characteristics. Furthermore, exploring how the stage of system development influences the relevance and importance of specific evaluation properties could offer valuable guidance for applying human-centred evaluation approaches.

Additionally, our analysis revealed gaps in the availability and appropriateness of existing measurement metrics, indicating a need to develop new metrics tailored to particular properties or domain-specific frameworks. Lastly, while our guidelines represent an initial step towards identifying suitable evaluation strategies, further empirical validation is needed to assess the usability and effectiveness of these recommendations in real-world settings.

**References**

[1] A. Čartolovni, A. Tomičić, E. Lazić Mosler, Ethical, legal, and social considerations of AI-based medical decision-support tools: A scoping review, International Journal of Medical Informatics 161 (2022) 104738. URL: https://www.sciencedirect.com/science/article/pii/S1386505622000521?via=ihub. doi:10.1016/J.IJMEDINF.2022.104738.

[2] S. Tonekaboni, S. Joshi, M. D. McCradden, A. Goldenberg, What Clinicians Want: Contextualizing Explainable Machine Learning for Clinical End Use, Proceedings of Machine Learning Research (2019). URL: http://arxiv.org/abs/1905.05134.

[3] A. Bussone, S. Stumpf, D. O'Sullivan, The Role of Explanations on Trust and Reliance in Clinical Decision Support Systems, in: 2015 International Conference on Healthcare Informatics, IEEE, 2015, pp. 160–169. URL: http://ieeexplore.ieee.org/document/7349687/. doi:10.1109/ICHI.2015.26.

[4] M. Reyes, R. Meier, S. Pereira, C. A. Silva, F.-M. Dahlweid, H. v. Tengg-Kobligk, R. M. Summers, R. Wiest, On the Interpretability of Artificial Intelligence in Radiology: Challenges and Opportunities, Radiology: Artificial Intelligence 2 (2020) e190043. URL: http://pubs.rsna.org/doi/10.1148/ryai.2020190043. doi:10.1148/ryai.2020190043.

[5] J. Amann, D. Vetter, S. N. Blomberg, H. C. Christensen, M. Coffee, S. Gerke, T. K. Gilbert, T. Hagendorff, S. Holm, M. Livne, A. Spezzatti, I. Strümke, R. V. Zicari, V. I. Madai, o. b. o. t. Z.-I. initiative, To explain or not to explain?—Artificial intelligence explainability in clinical decision support systems, PLOS Digital Health 1 (2022) e0000016. URL: https://journals.plos.org/digitalhealth/article?id=10.1371/journal.pdig.0000016. doi:10.1371/JOURNAL.PDIG.0000016.

[6] C. M. Cutillo, K. R. Sharma, L. Foschini, S. Kundu, M. Mackintosh, K. D. Mandl, T. Beck, E. Collier, C. Colvis, K. Gersing, V. Gordon, R. Jensen, B. Shabestari, N. Southall, Machine intelligence in healthcare—perspectives on trustworthiness, explainability, usability, and transparency, 2020. doi:10.1038/s41746-020-0254-2.

[7] J. Wiens, S. Saria, M. Sendak, M. Ghassemi, V. X. Liu, F. Doshi-Velez, K. Jung, K. Heller, D. Kale, M. Saeed, P. N. Ossorio, S. Thadaney-Israni, A. Goldenberg, Do no harm: a roadmap for responsible machine learning for health care, Nature Medicine 25 (2019) 1337–1340. doi:10.1038/s41591-019-0548-6.

[8] European Institute of Innovation and Technology Health, McKinsey & Company, Transforming Healthcare with AI: The Impact on the Workforce and Organisations, Technical Report, European Institute of Innovation and Technology Health, 2020.

[9] H. W. Loh, C. P. Ooi, S. Seoni, P. D. Barua, F. Molinari, U. R. Acharya, Application of explainable artificial intelligence for healthcare: A systematic review of the last decade (2011–2022), Computer Methods

and Programs in Biomedicine 226 (2022) 107161. doi:`10.1016/J.CMPB.2022.107161`.

[10] A. Chaddad, J. Peng, J. Xu, A. Bouridane, Survey of Explainable AI Techniques in Healthcare, Sensors (Basel, Switzerland) 23 (2023). doi:`10.3390/s23020634`.

[11] A. F. Markus, J. A. Kors, P. R. Rijnbeek, The role of explainability in creating trustworthy artificial intelligence for health care: a comprehensive survey of the terminology, design choices, and evaluation strategies, Journal of Biomedical Informatics 113 (2020) 103655. URL: `http://arxiv.org/abs/2007.15911http://dx.doi.org/10.1016/j.jbi.2020.103655`. doi:`10.1016/j.jbi.2020.103655`.

[12] K. S. Kacafírková, S. Polak, M. S. Smitt, S. Elprama, A. Jacobs, Trustworthy enough? Evaluation of an AI decision support system for healthcare professionals, in: L. Longo (Ed.), Joint Proceedings of the xAI-2023 Late-breaking Work, Demos and Doctoral Consortium co-located with the 1st World Conference on eXplainable Artificial Intelligence (xAI-2023), CEUR Workshop Proceedings, 2023, pp. 7–11. URL: `https://ceur-ws.org/Vol-3554/paper2.pdf`.

[13] T. Miller, Explanation in artificial intelligence: Insights from the social sciences, Artificial Intelligence 267 (2019) 1–38. URL: `https://linkinghub.elsevier.com/retrieve/pii/S0004370218305988`. doi:`10.1016/j.artint.2018.07.007`.

[14] M. Velmurugan, C. Ouyang, Y. Xu, R. Sindhgatta, B. Wickramanayake, C. Moreira, Developing guidelines for functionally-grounded evaluation of explainable artificial intelligence using tabular data, Engineering Applications of Artificial Intelligence 141 (2025) 109772. URL: `https://www.sciencedirect.com/science/article/pii/S0952197624019316?via%3Dihub#sec11`. doi:`10.1016/J.ENGAPPAI.2024.109772`.

[15] I. Donoso-Guzmán, J. Ooge, D. Parra, K. Verbert, Towards a Comprehensive Human-Centred Evaluation Framework for Explainable AI, in: Communications in Computer and Information Science, volume 1903 CCIS, Springer Science and Business Media Deutschland GmbH, 2023, pp. 183–204. doi:`10.1007/978-3-031-44070-0{\_}10`.

[16] Y. Rong, T. Leemann, T. T. Nguyen, L. Fiedler, P. Qian, V. Unhelkar, T. Seidel, G. Kasneci, E. Kasneci, Towards Human-Centered Explainable AI: A Survey of User Studies for Model Explanations, IEEE Transactions on Pattern Analysis and Machine Intelligence 46 (2024) 2104–2122. doi:`10.1109/TPAMI.2023.3331846`.

[17] J. Kim, H. Maathuis, D. Sent, Human-centered evaluation of explainable AI applications: a systematic review, Frontiers in Artificial Intelligence 7 (2024) 1456486. doi:`10.3389/FRAI.2024.1456486/PDF`.

[18] S. Naveed, G. Stevens, D. Robin-Kern, An Overview of the Empirical Evaluation of Explainable AI (XAI): A Comprehensive Guideline for User-Centered Evaluation in XAI, Applied Sciences (Switzerland) 14 (2024). doi:`10.3390/APP142311288`.

[19] B. Y. Lim, Q. Yang, A. Abdul, D. Wang, Why these Explanations? Selecting Intelligibility Types for Explanation Goals, IUI Workshops (2019).

[20] S. S. Y. Kim, E. A. Watkins, O. Russakovsky, R. Fong, A. Monroy-Hernández, "Help Me Help the AI": Understanding How Explainability Can Support Human-AI Interaction, in: Proceedings of the 2023 CHI Conference on Human Factors in Computing Systems, 2023, pp. 1–17. URL: `http://arxiv.org/abs/2210.03735http://dx.doi.org/10.1145/3544548.3581001`. doi:`10.1145/3544548.3581001`.

[21] A. Suh, I. Hurley, N. Smith, H. C. Siu, Fewer Than 1% of Explainable AI Papers Validate Explainability with Humans, in: Proceedings of the Extended Abstracts of the CHI Conference on Human Factors in Computing Systems, ACM, New York, NY, USA, 2025, pp. 1–7. URL: `https://dl.acm.org/doi/10.1145/3706599.3719964`. doi:`10.1145/3706599.3719964`.

[22] M. Nauta, J. Trienes, S. Pathak, E. Nguyen, M. Peters, Y. Schmitt, J. Schlötterer, M. Van Keulen, C. Seifert, From Anecdotal Evidence to Quantitative Evaluation Methods: A Systematic Review on Evaluating Explainable AI, ACM Computing Surveys 55 (2023). doi:`10.1145/3583558`.

[23] G. Vilone, L. Longo, Notions of explainability and evaluation approaches for explainable artificial intelligence, Information Fusion 76 (2021) 89–106. doi:10.1016/J.INFFUS.2021.05.009.

[24] K. Beckh, S. Müller, S. Rüping, A Quantitative Human-Grounded Evaluation Process for Explainable Machine Learning, in: LWDA'22: Lernen, Wissen, Daten, Analysen, 2022. URL: http://ceur-ws.org.

[25] K. Sokol, P. Flach, Explainability fact sheets: A framework for systematic assessment of explainable approaches, in: FAT* 2020 - Proceedings of the 2020 Conference on Fairness, Accountability, and Transparency, Association for Computing Machinery, Inc, 2020, pp. 56–67. doi:10.1145/3351095.3372870.

[26] Q. V. Liao, Y. Zhang, R. Luss, F. Doshi-Velez, A. Dhurandhar, Connecting Algorithmic Research and Usage Contexts: A Perspective of Contextualized Evaluation for Explainable AI, Proceedings of the AAAI Conference on Human Computation and Crowdsourcing 10 (2022) 147–159. URL: https://ojs.aaai.org/index.php/HCOMP/article/view/21995. doi:10.1609/hcomp.v10i1.21995.

[27] H. Löfström, K. Hammar, U. Johansson, A Meta Survey of Quality Evaluation Criteria in Explanation Methods (2022). URL: http://arxiv.org/abs/2203.13929.

[28] D. V. Carvalho, E. M. Pereira, J. S. Cardoso, Machine Learning Interpretability: A Survey on Methods and Metrics, Electronics 8 (2019) 832. URL: https://www.mdpi.com/2079-9292/8/8/832. doi:10.3390/electronics8080832.

[29] J. H.-w. Hsiao, H. H. T. Ngai, L. Qiu, Y. Yang, C. C. Cao, Roadmap of Designing Cognitive Metrics for Explainable Artificial Intelligence (XAI) (2021). URL: https://arxiv.org/abs/2108.01737v1. doi:10.48550/arxiv.2108.01737.

[30] S. Mohseni, J. E. Block, E. Ragan, Quantitative Evaluation of Machine Learning Explanations: A Human-Grounded Benchmark, International Conference on Intelligent User Interfaces, Proceedings IUI (2021) 22–31. doi:10.1145/3397481.3450689.

[31] W. Saeed, C. Omlin, Explainable AI (XAI): A systematic meta-survey of current challenges and future opportunities, Knowledge-Based Systems 263 (2023) 110273. doi:10.1016/j.knosys.2023.110273.

[32] P. Lopes, E. Silva, C. Braga, T. Oliveira, L. Rosado, XAI Systems Evaluation: A Review of Human and Computer-Centred Methods, Applied Sciences 12 (2022) 9423. URL: https://www.mdpi.com/2076-3417/12/19/9423. doi:10.3390/app12199423.

[33] R. Guidotti, A. Monreale, S. Ruggieri, F. Turini, F. Giannotti, D. Pedreschi, A survey of methods for explaining black box models, ACM Computing Surveys 51 (2018) 1–45. doi:10.1145/3236009.

[34] A. Barredo Arrieta, N. Díaz-Rodríguez, J. Del Ser, A. Bennetot, S. Tabik, A. Barbado, S. Garcia, S. Gil-Lopez, D. Molina, R. Benjamins, R. Chatila, F. Herrera, Explainable Artificial Intelligence (XAI): Concepts, taxonomies, opportunities and challenges toward responsible AI, Information Fusion 58 (2020) 82–115. URL: https://doi.org/10.1016/j.inffus.2019.12.012. doi:10.1016/j.inffus.2019.12.012.

[35] M. Sahakyan, Z. Aung, T. Rahwan, Explainable Artificial Intelligence for Tabular Data: A Survey, IEEE Access 9 (2021) 135392–135422. doi:10.1109/ACCESS.2021.3116481.

[36] S. Ali, T. Abuhmed, S. El-Sappagh, K. Muhammad, J. M. Alonso-Moral, R. Confalonieri, R. Guidotti, J. Del Ser, N. Díaz-Rodríguez, F. Herrera, Explainable Artificial Intelligence (XAI): What we know and what is left to attain Trustworthy Artificial Intelligence, Information Fusion 99 (2023) 101805. doi:10.1016/J.INFFUS.2023.101805.

[37] J. Ooge, G. Stiglic, K. Verbert, Explaining artificial intelligence with visual analytics in healthcare, WIREs Data Mining and Knowledge Discovery 12 (2022). doi:10.1002/widm.1427.

[38] A. M. Antoniadi, Y. Du, Y. Guendouz, L. Wei, C. Mazo, B. A. Becker, C. Mooney, Current Challenges and Future Opportunities for XAI in Machine Learning-Based Clinical Decision Support Systems: A Systematic Review, Applied Sciences 2021, Vol. 11, Page 5088 11 (2021) 5088. URL: https://www.mdpi.com/2076-3417/11/11/5088/htmhttps://www.mdpi.com/2076-3417/11/11/5088. doi:10.3390/APP11115088.

[39] I. D. Mienye, G. Obaido, N. Jere, E. Mienye, K. Aruleba, I. D. Emmanuel, B. Ogbuokiri, A survey of explainable artificial intelligence in healthcare: Concepts, applications, and challenges, Informatics in Medicine Unlocked 51 (2024) 101587. URL: https://linkinghub.elsevier.com/retrieve/pii/S2352914824001448. doi:10.1016/J.IMU.2024.101587.

[40] S. Mohseni, N. Zarei, E. D. Ragan, A Multidisciplinary Survey and Framework for Design and Evaluation of Explainable AI Systems, ACM Transactions on Interactive Intelligent Systems 1 (2021) 1–45. URL: http://arxiv.org/abs/1811.11839. doi:10.1145/3387166.

[41] G. Vilone, L. Longo, Classification of Explainable Artificial Intelligence Methods through Their Output Formats, Machine Learning and Knowledge Extraction 2021, Vol. 3, Pages 615-661 3 (2021) 615–661. URL: https://www.mdpi.com/2504-4990/3/3/32/htmhttps://www.mdpi.com/2504-4990/3/3/32. doi:10.3390/MAKE3030032.

[42] L. E. Holmquist, Intelligence on Tap: Artificial Intelligence as a New Design Material, Interactions 24 (2017) 28–33. URL: https://dl.acm.org/doi/10.1145/3085571. doi:10.1145/3085571/ASSET/7D59872F-1989-49EA-8E74-B7228D403DB5/ASSETS/3085571.FP.PNG.

[43] H. Suresh, S. R. Gomez, K. K. Nam, A. Satyanarayan, Beyond Expertise and Roles: A Framework to Characterize the Stakeholders of Interpretable Machine Learning and their Needs, in: Proceedings of the 2021 CHI Conference on Human Factors in Computing Systems, volume 16, ACM, New York, NY, USA, 2021, pp. 1–16. URL: https://dl.acm.org/doi/10.1145/3411764.3445088. doi:10.1145/3411764.3445088.

[44] D. Wang, Q. Yang, A. Abdul, B. Y. Lim, Designing Theory-Driven User-Centric Explainable AI, in: Proceedings of the 2019 CHI Conference on Human Factors in Computing Systems, ACM, New York, NY, USA, 2019, pp. 1–15. URL: https://dl.acm.org/doi/10.1145/3290605.3300831. doi:10.1145/3290605.3300831.

[45] M. Szymanski, V. Vanden Abeele, K. Verbert, Designing and Evaluating Explanations for a Predictive Health Dashboard: A User-

Centred Case Study, Conference on Human Factors in Computing Systems - Proceedings (2024). URL: https://dl.acm.org/doi/10.1145/3613905.3637140. doi:10.1145/3613905.3637140/SUPPL{\_}FILE/3613905.3637140-TALK-VIDEO-SUBTITLES.VTT.

[46] A. Bhattacharya, J. Ooge, G. Stiglic, K. Verbert, Directive explanations for monitoring the risk of diabetes onset: Introducing directive data-centric explanations and combinations to support what-if explorations, in: International Conference on Intelligent User Interfaces, Proceedings IUI, Association for Computing Machinery, 2023, pp. 204–219. doi:10.1145/3581641.3584075.

[47] M. Szymanski, C. Conati, V. V. Abeele, K. Verbert, Designing and personalising hybrid multi-modal health explanations for lay users, in: Proceedings of the 10th Joint Workshop on Interfaces and Human Decision Making for Recommender Systems (IntRS 2023), volume 3534, CEUR-WS.org, 2023, pp. 35–52. URL: https://ceur-ws.org/Vol-3534/paper3.pdf.

[48] B. P. Knijnenburg, M. C. Willemsen, Evaluating Recommender Systems with User Experiments, in: Recommender Systems Handbook, Springer US, Boston, MA, 2015, pp. 309–352. URL: https://link.springer.com/10.1007/978-1-4899-7637-6_9. doi:10.1007/978-1-4899-7637-6{\_}9.

[49] J. Qu, J. Arguello, Y. Wang, Understanding the cognitive influences of interpretability features on how users scrutinize machine-predicted categories, in: CHIIR 2023 - Proceedings of the 2023 Conference on Human Information Interaction and Retrieval, Association for Computing Machinery, Inc, 2023, pp. 247–257. doi:10.1145/3576840.3578315.

[50] J. Hwang, T. Lee, H. Lee, S. Byun, A clinical decision support system for sleep staging tasks with explanations from artificial intelligence: User-centered design and evaluation study, 2022. doi:10.2196/28659.

[51] P. Taylor, J. Fox, A. T. Pokropek, The development and evaluation of cadmium: A prototype system to assist in the interpretation of mammograms, Medical Image Analysis 3 (1999) 321–337. doi:10.1016/S1361-8415(99)80027-9.

[52] M. Julià-Sapé, C. Majós, Àngels Camins, A. Samitier, M. Baquero, M. Serrallonga, S. Doménech, E. Grivé, F. A. Howe, K. Opstad, J. Calvar, C. Aguilera, C. Arús, Multicentre evaluation of the interpret decision support system 2.0 for brain tumour classification, NMR in Biomedicine 27 (2014) 1009–1018. doi:`10.1002/nbm.3144`.

[53] S. Srivastava, M. Theune, A. Catala, The role of lexical alignment in human understanding of explanations by conversational agents, in: International Conference on Intelligent User Interfaces, Proceedings IUI, Association for Computing Machinery, 2023, pp. 423–435. doi:`10.1145/3581641.3584086`.

[54] F. Liu, J. Zhou, M. Zuo, Y. Li, Dual-process theory-driven transparent approach for seniors to accept health misinformation detection results, Information Processing and Management 61 (2024). doi:`10.1016/j.ipm.2024.103751`.

[55] A. Sivaprasad, E. Reiter, N. Tintarev, N. Oren, Evaluation of human-understandability of global model explanations using decision tree, in: Communications in Computer and Information Science, volume 1947, Springer Science and Business Media Deutschland GmbH, 2024, pp. 43–65. doi:`10.1007/978-3-031-50396-2_3`.

[56] F. Sovrano, F. Vitali, An objective metric for explainable ai: How and why to estimate the degree of explainability, Knowledge-Based Systems 278 (2023). doi:`10.1016/j.knosys.2023.110866`.

[57] R. Upadhyay, P. Knoth, G. Pasi, M. Viviani, Explainable online health information truthfulness in consumer health search, Frontiers in Artificial Intelligence 6 (2023). doi:`10.3389/frai.2023.1184851`.

[58] L. Oberste, F. Rüffer, O. Aydingül, J. Rink, A. Heinzl, Designing user-centric explanations for medical imaging with informed machine learning, in: Lecture Notes in Computer Science (including subseries Lecture Notes in Artificial Intelligence and Lecture Notes in Bioinformatics), volume 13873 LNCS, Springer Science and Business Media Deutschland GmbH, 2023, pp. 470–484. doi:`10.1007/978-3-031-32808-4_29`.

[59] A. Katzmann, O. Taubmann, S. Ahmad, A. Mühlberg, M. Sühling, H. M. Groß, Explaining clinical decision support systems in medical imaging

using cycle-consistent activation maximization, Neurocomputing 458 (2021) 141–156. doi:10.1016/j.neucom.2021.05.081.

[60] S. Hegselmann, C. Ertmer, T. Volkert, A. Gottschalk, M. Dugas, J. Varghese, Development and validation of an interpretable 3 day intensive care unit readmission prediction model using explainable boosting machines, Frontiers in Medicine 9 (2022). doi:10.3389/fmed.2022.960296.

[61] M. Nauta, M. J. van Putten, M. C. Tjepkema-Cloostermans, J. P. Bos, M. van Keulen, C. Seifert, Interactive explanations of internal representations of neural network layers: An exploratory study on outcome prediction of comatose patients, 2020. URL: https://research.utwente.nl/en/publications/interactive-explanations-of-internal-representations-of-neural-ne.

[62] S. Röhrl, H. Maier, M. Lengl, C. Klenk, D. Heim, M. Knopp, S. Schumann, O. Hayden, K. Diepold, Explainable artificial intelligence for cytological image analysis, in: Lecture Notes in Computer Science (including subseries Lecture Notes in Artificial Intelligence and Lecture Notes in Bioinformatics), volume 13897 LNAI, Springer Science and Business Media Deutschland GmbH, 2023, pp. 75–85. doi:10.1007/978-3-031-34344-5_10.

[63] J. B. Lamy, B. Sekar, G. Guezennec, J. Bouaud, B. Séroussi, Explainable artificial intelligence for breast cancer: A visual case-based reasoning approach, Artificial Intelligence in Medicine 94 (2019) 42–53. doi:10.1016/j.artmed.2019.01.001.

[64] O. Deperlioglu, U. Kose, D. Gupta, A. Khanna, F. Giampaolo, G. Fortino, Explainable framework for glaucoma diagnosis by image processing and convolutional neural network synergy: Analysis with doctor evaluation, Future Generation Computer Systems 129 (2022) 152–169. doi:10.1016/j.future.2021.11.018.

[65] N. Das, S. Happaerts, I. Gyselinck, M. Staes, E. Derom, G. Brusselle, F. Burgos, M. Contoli, A. T. Dinh-Xuan, F. M. Franssen, S. Gonem, N. Greening, C. Haenebalcke, W. D. Man, J. Moisés, R. Peché, V. Poberezhets, J. K. Quint, M. C. Steiner, E. Vanderhelst, M. Abdo, M. Topalovic, W. Janssens, Collaboration between explainable artificial

intelligence and pulmonologists improves the accuracy of pulmonary function test interpretation, European Respiratory Journal 61 (2023). doi:`10.1183/13993003.01720-2022`.

[66] J. van der Waa, S. Verdult, K. van den Bosch, J. van Diggelen, T. Haije, B. van der Stigchel, I. Cocu, Moral decision making in human-agent teams: Human control and the role of explanations, Frontiers in Robotics and AI 8 (2021). doi:`10.3389/frobt.2021.640647`.

[67] W. Jin, M. Fatehi, R. Guo, G. Hamarneh, Evaluating the clinical utility of artificial intelligence assistance and its explanation on the glioma grading task, Artificial Intelligence in Medicine 148 (2024). doi:`10.1016/j.artmed.2023.102751`.

[68] T. Chanda, K. Hauser, S. Hobelsberger, T. C. Bucher, C. N. Garcia, C. Wies, H. Kittler, P. Tschandl, C. Navarrete-Dechent, S. Podlipnik, E. Chousakos, I. Crnaric, J. Majstorovic, L. Alhajwan, T. Foreman, S. Peternel, S. Sarap, İrem Özdemir, R. L. Barnhill, M. Llamas-Velasco, G. Poch, S. Korsing, W. Sondermann, F. F. Gellrich, M. V. Heppt, M. Erdmann, S. Haferkamp, K. Drexler, M. Goebeler, B. Schilling, J. S. Utikal, K. Ghoreschi, S. Fröhling, E. Krieghoff-Henning, A. Salava, A. Thiem, A. Dimitrios, A. M. Ammar, A. S. Vučemilović, A. M. Yoshimura, A. Ilieva, A. Gesierich, A. Reimer-Taschenbrecker, A. G. Kolios, A. Kalva, A. Ferhatosmanoğlu, A. Beyens, C. Pföhler, D. I. Erdil, D. Jovanovic, E. Racz, F. G. Bechara, F. Vaccaro, F. Dimitriou, G. Rasulova, H. Cenk, I. Yanatma, I. Kolm, I. Hoorens, I. P. Sheshova, I. Jocic, J. Knuever, J. Fleißner, J. R. Thamm, J. Dahlberg, J. J. Lluch-Galcerá, J. S. A. Figueroa, J. Holzgruber, J. Welzel, K. Damevska, K. E. Mayer, L. V. Maul, L. Garzona-Navas, L. I. Bley, L. Schmitt, L. Reipen, L. Shafik, L. Petrovska, L. Golle, L. Jopen, M. Gogilidze, M. R. Burg, M. A. Morales-Sánchez, M. Sławińska, M. Mengoni, M. Dragolov, N. Iglesias-Pena, N. Booken, N. A. Enechukwu, O. D. Persa, O. A. Oninla, P. Theofilogiannakou, P. Kage, R. R. O. Neto, R. Peralta, R. Afiouni, S. Schuh, S. Schnabl-Scheu, S. Vural, S. Hudson, S. R. Saa, S. Hartmann, S. Damevska, S. Finck, S. A. Braun, T. Hartmann, T. Welponer, T. Sotirovski, V. Bondare-Ansberga, V. Ahlgrimm-Siess, V. G. Frings, V. Simeonovski, Z. Zafirovik, J. T. Maul, S. Lehr, M. Wobser, D. Debus, H. Riad, M. P. Pereira, Z. Lengyel, A. Balcere, A. Tsakiri, R. P. Braun, T. J. Brinker, Dermatologist-like explainable ai enhances

trust and confidence in diagnosing melanoma, Nature Communications 15 (2024). doi:10.1038/s41467-023-43095-4.

[69] D. Song, J. Yao, Y. Jiang, S. Shi, C. Cui, L. Wang, L. Wang, H. Wu, H. Tian, X. Ye, D. Ou, W. Li, N. Feng, W. Pan, M. Song, J. Xu, D. Xu, L. Wu, F. Dong, A new xai framework with feature explainability for tumors decision-making in ultrasound data: comparing with grad-cam, Computer Methods and Programs in Biomedicine 235 (2023). doi:10.1016/j.cmpb.2023.107527.

[70] M. H. Lee, C. J. Chew, Understanding the effect of counterfactual explanations on trust and reliance on ai for human-ai collaborative clinical decision making, Proceedings of the ACM on Human-Computer Interaction 7 (2023). doi:10.1145/3610218.

[71] A. A. Tutul, E. H. Nirjhar, T. Chaspari, Investigating trust in human-ai collaboration for a speech-based data analytics task, International Journal of Human-Computer Interaction (2024). doi:10.1080/10447318.2024.2328910.

[72] E. Kyrimi, S. Mossadegh, N. Tai, W. Marsh, An incremental explanation of inference in bayesian networks for increasing model trustworthiness and supporting clinical decision making, Artificial Intelligence in Medicine 103 (2020). doi:10.1016/j.artmed.2020.101812.

[73] L. Stork, I. Tiddi, R. Spijker, A. ten Teije, Explainable drug repurposing in context via deep reinforcement learning, in: Lecture Notes in Computer Science (including subseries Lecture Notes in Artificial Intelligence and Lecture Notes in Bioinformatics), volume 13870 LNCS, Springer Science and Business Media Deutschland GmbH, 2023, pp. 3–20. doi:10.1007/978-3-031-33455-9_1.

[74] S. G. Anjara, A. Janik, A. Dunford-Stenger, K. M. Kenzie, A. Collazo-Lorduy, M. Torrente, L. Costabello, M. Provencio, Examining explainable clinical decision support systems with think aloud protocols, PLoS ONE 18 (2023). doi:10.1371/journal.pone.0291443.

[75] L. Xiao, H. Zhou, J. Fox, Towards a systematic approach for argumentation, recommendation, and explanation in clinical decision support, Mathematical Biosciences and Engineering 19 (2022) 10445–10473. doi:10.3934/mbe.2022489.

[76] Q. Wang, K. Huang, P. Chandak, M. Zitnik, N. Gehlenborg, Extending the nested model for user-centric xai: A design study on gnn-based drug repurposing, IEEE Transactions on Visualization and Computer Graphics 29 (2023) 1266–1276. doi:10.1109/TVCG.2022.3209435.

[77] F. Cabitza, C. Natali, L. Famiglini, A. Campagner, V. Caccavella, E. Gallazzi, Never tell me the odds: Investigating pro-hoc explanations in medical decision making, Artificial Intelligence in Medicine 150 (2024). doi:10.1016/j.artmed.2024.102819.

[78] C. Metta, A. Beretta, R. Guidotti, Y. Yin, P. Gallinari, S. Rinzivillo, F. Giannotti, Improving trust and confidence in medical skin lesion diagnosis through explainable deep learning, International Journal of Data Science and Analytics (2023). doi:10.1007/s41060-023-00401-z.

[79] P. Sabol, P. Sinčák, P. Hartono, P. Kočan, Z. Benetinová, A. Blichárová, Ľudmila Verbóová, E. Štammová, A. Sabolová-Fabianová, A. Jašková, Explainable classifier for improving the accountability in decision-making for colorectal cancer diagnosis from histopathological images, Journal of Biomedical Informatics 109 (2020). doi:10.1016/j.jbi.2020.103523.

[80] N. Sarkar, M. Kumagai, S. Meyr, S. Pothapragada, M. Unberath, G. Li, S. R. Ahmed, E. B. Smith, M. A. Davis, G. D. Khatri, A. Agrawal, Z. S. Delproposto, H. Chen, C. G. Caballero, D. Dreizin, An aser ai/ml expert panel formative user research study for an interpretable interactive splenic aast grading graphical user interface prototype, Emergency Radiology 31 (2024) 167–178. doi:10.1007/s10140-024-02202-8.

[81] F. M. Calisto, C. Santiago, N. Nunes, J. C. Nascimento, Breastscreening-ai: Evaluating medical intelligent agents for human-ai interactions, Artificial Intelligence in Medicine 127 (2022). doi:10.1016/j.artmed.2022.102285.

[82] F. Cabitza, A. Campagner, L. Famiglini, E. Gallazzi, G. A. L. Maida, Color shadows (part i): Exploratory usability evaluation of activation maps in radiological machine learning, Lecture Notes in Computer Science (including subseries Lecture Notes in Artificial Intelligence and Lecture Notes in Bioinformatics)

13480 LNCS (2022) 31–50. URL: https://dl.acm.org/doi/10.1007/978-3-031-14463-9_3. doi:10.1007/978-3-031-14463-9_3.

[83] A. Kumar, R. Manikandan, U. Kose, D. Gupta, S. C. Satapathy, Doctor's dilemma: Evaluating an explainable subtractive spatial lightweight convolutional neural network for brain tumor diagnosis, ACM Transactions on Multimedia Computing, Communications and Applications 17 (2021). doi:10.1145/3457187.

[84] S. Singla, M. Eslami, B. Pollack, S. Wallace, K. Batmanghelich, Explaining the black-box smoothly—a counterfactual approach, Medical Image Analysis 84 (2023). doi:10.1016/j.media.2022.102721.

[85] A. Glick, M. Clayton, N. Angelov, J. Chang, Impact of explainable artificial intelligence assistance on clinical decision-making of novice dental clinicians, JAMIA Open 5 (2022). doi:10.1093/jamiaopen/ooac031.

[86] F. M. Calisto, C. Santiago, N. Nunes, J. C. Nascimento, Introduction of human-centric ai assistant to aid radiologists for multimodal breast image classification, International Journal of Human Computer Studies 150 (2021). doi:10.1016/j.ijhcs.2021.102607.

[87] L. Pumplun, F. Peters, J. F. Gawlitza, P. Buxmann, Bringing machine learning systems into clinical practice: A design science approach to explainable machine learning-based clinical decision support systems, Journal of the Association for Information Systems 24 (2023) 953–979. doi:10.17705/1jais.00820.

[88] J. S. Chen, S. L. Baxter, A. V. D. Brandt, A. Lieu, A. S. Camp, J. L. Do, D. S. Welsbie, S. Moghimi, M. Christopher, R. N. Weinreb, L. M. Zangwill, Usability and clinician acceptance of a deep learning-based clinical decision support tool for predicting glaucomatous visual field progression, Journal of Glaucoma 32 (2023) 151–158. doi:10.1097/IJG.0000000000002163.

[89] C. Metta, A. Beretta, R. Guidotti, Y. Yin, P. Gallinari, S. Rinzivillo, F. Giannotti, Advancing dermatological diagnostics: Interpretable ai for enhanced skin lesion classification, Diagnostics 14 (2024). doi:10.3390/diagnostics14070753.

[90] C. Natali, L. Famiglini, A. Campagner, G. A. L. Maida, E. Gallazzi, F. Cabitza, Color shadows 2: Assessing the impact of xai on diagnostic decision-making, in: Communications in Computer and Information Science, volume 1901 CCIS, Springer Science and Business Media Deutschland GmbH, 2023, pp. 618–629. doi:10.1007/978-3-031-44064-9_33.

[91] L. Gamage, U. Isuranga, D. Meedeniya, S. D. Silva, P. Yogarajah, Melanoma skin cancer identification with explainability utilizing mask guided technique, Electronics (Switzerland) 13 (2024). doi:10.3390/electronics13040680.

[92] F. Cabitza, A. Campagner, L. Famiglini, C. Natali, V. Caccavella, E. Gallazzi, Let me think! investigating the effect of explanations feeding doubts about the ai advice, in: Lecture Notes in Computer Science (including subseries Lecture Notes in Artificial Intelligence and Lecture Notes in Bioinformatics), volume 14065 LNCS, Springer Science and Business Media Deutschland GmbH, 2023, pp. 155–169. doi:10.1007/978-3-031-40837-3_10.

[93] A. Derathé, F. Reche, P. Jannin, A. Moreau-Gaudry, B. Gibaud, S. Voros, Explaining a model predicting quality of surgical practice: a first presentation to and review by clinical experts, International Journal of Computer Assisted Radiology and Surgery 16 (2021) 2009–2019. doi:10.1007/s11548-021-02422-0.

[94] S. Domínguez-Rodríguez, H. Liz-López, A. Panizo-LLedot, Álvaro Ballesteros, R. Dagan, D. Greenberg, L. Gutiérrez, P. Rojo, E. Otheo, J. C. Galán, S. Villanueva, S. García, P. Mosquera, A. Tagarro, C. Moraleda, D. Camacho, Testing the performance, adequacy, and applicability of an artificial intelligence model for pediatric pneumonia diagnosis, Computer Methods and Programs in Biomedicine 242 (2023). doi:10.1016/j.cmpb.2023.107765.

[95] J. E. Huh, J. H. Lee, E. J. Hwang, C. M. Park, Effects of expert-determined reference standards in evaluating the diagnostic performance of a deep learning model: A malignant lung nodule detection task on chest radiographs, Korean Journal of Radiology 24 (2023) 155–165. doi:10.3348/kjr.2022.0548.

[96] A. M. Antoniadi, M. Galvin, M. Heverin, L. Wei, O. Hardiman, C. Mooney, A clinical decision support system for the prediction of quality of life in als, Journal of Personalized Medicine 12 (2022). doi:10.3390/jpm12030435.

[97] R. Hendawi, J. Li, S. Roy, A mobile app that addresses interpretability challenges in machine learning–based diabetes predictions: Survey-based user study, JMIR Formative Research 7 (2023). doi:10.2196/50328.

[98] A. J. Barda, C. M. Horvat, H. Hochheiser, A qualitative research framework for the design of user-centered displays of explanations for machine learning model predictions in healthcare, BMC Medical Informatics and Decision Making 20 (2020). doi:10.1186/s12911-020-01276-x.

[99] C. Panigutti, A. Beretta, D. Fadda, F. Giannotti, D. Pedreschi, A. Perotti, S. Rinzivillo, Co-design of human-centered, explainable ai for clinical decision support, ACM Transactions on Interactive Intelligent Systems 13 (2023). doi:10.1145/3587271.

[100] K. Kaczmarek-Majer, G. Casalino, G. Castellano, M. Dominiak, O. Hryniewicz, O. Kamińska, G. Vessio, N. Díaz-Rodríguez, Plenary: Explaining black-box models in natural language through fuzzy linguistic summaries, Information Sciences 614 (2022) 374–399. doi:10.1016/j.ins.2022.10.010.

[101] O. Wysocki, J. K. Davies, M. Vigo, A. C. Armstrong, D. Landers, R. Lee, A. Freitas, Assessing the communication gap between ai models and healthcare professionals: Explainability, utility and trust in ai-driven clinical decision-making, Artificial Intelligence 316 (2023). doi:10.1016/j.artint.2022.103839.

[102] Z. Jin, S. Cui, S. Guo, D. Gotz, J. Sun, N. Cao, Carepre, ACM Transactions on Computing for Healthcare 1 (2020). doi:10.1145/3344258.

[103] F. Gutiérrez, N. N. Htun, V. V. Abeele, R. D. Croon, K. Verbert, Explaining call recommendations in nursing homes: a user-centered design approach for interacting with knowledge-based health decision support systems, in: International Conference on Intelligent User Interfaces, Proceedings IUI, Association for Computing Machinery, 2022, pp. 162–172. doi:10.1145/3490099.3511158.

[104] N. C. Rajashekar, Y. E. Shin, Y. Pu, S. Chung, K. You, M. Giuffre, C. E. Chan, T. Saarinen, A. Hsiao, J. Sekhon, A. H. Wong, L. V. Evans, R. F. Kizilcec, L. Laine, T. McCall, D. Shung, Human-algorithmic interaction using a large language model-augmented artificial intelligence clinical decision support system, in: Conference on Human Factors in Computing Systems - Proceedings, Association for Computing Machinery, 2024. doi:10.1145/3613904.3642024.

[105] S. S. Samuel, N. N. B. Abdullah, A. Raj, Interpretation of svm using data mining technique to extract syllogistic rules: Exploring the notion of explainable ai in diagnosing cad, in: Lecture Notes in Computer Science (including subseries Lecture Notes in Artificial Intelligence and Lecture Notes in Bioinformatics), volume 12279 LNCS, Springer, 2020, pp. 249–266. doi:10.1007/978-3-030-57321-8_14.

[106] W. J. She, C. S. Ang, R. A. Neimeyer, L. A. Burke, Y. Zhang, A. Jatowt, Y. Kawai, J. Hu, M. Rauterberg, H. G. Prigerson, P. Siriaraya, Investigation of a web-based explainable ai screening for prolonged grief disorder, IEEE Access 10 (2022) 41164–41185. doi:10.1109/ACCESS.2022.3163311.

[107] D. Gu, K. Su, H. Zhao, A case-based ensemble learning system for explainable breast cancer recurrence prediction, Artificial Intelligence in Medicine 107 (2020). doi:10.1016/j.artmed.2020.101858.

[108] E. Pisirir, J. M. Wohlgemut, E. Kyrimi, R. S. Stoner, Z. B. Perkins, N. R. Tai, D. W. R. Marsh, A process for evaluating explanations for transparent and trustworthy ai prediction models, in: Proceedings - 2023 IEEE 11th International Conference on Healthcare Informatics, ICHI 2023, Institute of Electrical and Electronics Engineers Inc., 2023, pp. 388–397. doi:10.1109/ICHI57859.2023.00058.

[109] J. Hu, Y. Liang, W. Zhao, K. McAreavey, W. Liu, An interactive xai interface with application in healthcare for non-experts, in: Communications in Computer and Information Science, volume 1901 CCIS, Springer Science and Business Media Deutschland GmbH, 2023, pp. 649–670. doi:10.1007/978-3-031-44064-9_35.

[110] F. L. D. Morais, A. C. B. Garcia, P. S. M. D. Santos, L. A. P. Ribeiro, Do explainable ai techniques effectively explain their rationale? a case study

from the domain expert's perspective, in: Proceedings of the 2023 26th International Conference on Computer Supported Cooperative Work in Design, CSCWD 2023, Institute of Electrical and Electronics Engineers Inc., 2023, pp. 1569–1574. doi:`10.1109/CSCWD57460.2023.10152722`.

[111] K. Aliyeva, N. Mehdiyev, Uncertainty-aware multi-criteria decision analysis for evaluation of explainable artificial intelligence methods: A use case from the healthcare domain, Information Sciences 657 (2024). doi:`10.1016/j.ins.2023.119987`.

[112] Z. Wang, I. Samsten, V. Kougia, P. Papapetrou, Style-transfer counterfactual explanations: An application to mortality prevention of icu patients, Artificial Intelligence in Medicine 135 (2023). doi:`10.1016/j.artmed.2022.102457`.

[113] G. T. Berge, O. C. Granmo, T. O. Tveit, B. E. Munkvold, A. L. Ruthjersen, J. Sharma, Machine learning-driven clinical decision support system for concept-based searching: a field trial in a norwegian hospital, BMC Medical Informatics and Decision Making 23 (2023). doi:`10.1186/s12911-023-02101-x`.

[114] D. Jaber, H. Hajj, F. Maalouf, W. El-Hajj, Medically-oriented design for explainable ai for stress prediction from physiological measurements, BMC Medical Informatics and Decision Making 22 (2022). doi:`10.1186/s12911-022-01772-2`.

[115] M. Nagendran, P. Festor, M. Komorowski, A. C. Gordon, A. A. Faisal, Quantifying the impact of ai recommendations with explanations on prescription decision making, npj Digital Medicine 6 (2023). doi:`10.1038/s41746-023-00955-z`.

[116] L. V. Herm, K. Heinrich, J. Wanner, C. Janiesch, Stop ordering machine learning algorithms by their explainability! a user-centered investigation of performance and explainability, International Journal of Information Management 69 (2023). doi:`10.1016/j.ijinfomgt.2022.102538`.

[117] S. V. Kovalchuk, G. D. Kopanitsa, I. V. Derevitskii, G. A. Matveev, D. A. Savitskaya, Three-stage intelligent support of clinical decision making for higher trust, validity, and explainability, Journal of Biomedical Informatics 127 (2022). doi:`10.1016/j.jbi.2022.104013`.

[118] A. Rind, D. Slijepčević, M. Zeppelzauer, F. Unglaube, A. Kranzl, B. Horsak, Trustworthy visual analytics in clinical gait analysis: A case study for patients with cerebral palsy, in: Proceedings - 2022 IEEE Workshop on TRust and EXpertise in Visual Analytics, TREX 2022, Institute of Electrical and Electronics Engineers Inc., 2022, pp. 8–15. doi:`10.1109/TREX57753.2022.00006`.

[119] F. Cheng, D. Liu, F. Du, Y. Lin, A. Zytek, H. Li, H. Qu, K. Veeramachaneni, Vbridge: Connecting the dots between features and data to explain healthcare models, IEEE Transactions on Visualization and Computer Graphics 28 (2022) 378–388. doi:`10.1109/TVCG.2021.3114836`.

[120] Y. Du, A. M. Antoniadi, C. McNestry, F. M. McAuliffe, C. Mooney, The role of xai in advice-taking from a clinical decision support system: A comparative user study of feature contribution-based and example-based explanations, Applied Sciences (Switzerland) 12 (2022). doi:`10.3390/app122010323`.

[121] C. He, V. Raj, H. Moen, T. Gröhn, C. Wang, L. M. Peltonen, S. Koivusalo, P. Marttinen, G. Jacucci, Vms: Interactive visualization to support the sensemaking and selection of predictive models, in: ACM International Conference Proceeding Series, Association for Computing Machinery, 2024, pp. 229–244. doi:`10.1145/3640543.3645151`.

[122] E. Khodabandehloo, D. Riboni, A. Alimohammadi, Healthxai: Collaborative and explainable ai for supporting early diagnosis of cognitive decline, Future Generation Computer Systems 116 (2021) 168–189. doi:`10.1016/j.future.2020.10.030`.

[123] C. Panigutti, A. Beretta, F. Giannotti, D. Pedreschi, Understanding the impact of explanations on advice-taking: a user study for ai-based clinical decision support systems, in: Conference on Human Factors in Computing Systems - Proceedings, Association for Computing Machinery, 2022. doi:`10.1145/3491102.3502104`.

[124] M. H. Lee, D. P. Siewiorek, A. Smailagic, A human-ai collaborative approach for clinical decision making on rehabilitation assessment, in: Conference on Human Factors in Computing Systems - Proceedings, Association for Computing Machinery, 2021. doi:`10.1145/3411764.3445472`.

[125] W. Jin, X. Li, M. Fatehi, G. Hamarneh, Guidelines and evaluation of clinical explainable ai in medical image analysis, Medical Image Analysis 84 (2023). doi:10.1016/j.media.2022.102684.

[126] M. H. Lee, D. P. Siewiorek, A. Smailagic, A. Bernardino, S. B. I. Badia, Co-design and evaluation of an intelligent decision support system for stroke rehabilitation assessment, Proceedings of the ACM on Human-Computer Interaction 4 (2020). doi:10.1145/3415227.

[127] W. Jin, X. Li, G. Hamarneh, Evaluating explainable ai on a multimodal medical imaging task: Can existing algorithms fulfill clinical requirements?, AAAI Conference on Artificial Intelligence 36 (2022) 11945–11953. doi:10.1609/AAAI.V36I11.21452.

[128] M. H. Lee, D. P. Siewiorek, A. Smailagic, A. Bernardino, S. B. I. Badia, Interactive hybrid approach to combine machine and human intelligence for personalized rehabilitation assessment, in: ACM CHIL 2020 - Proceedings of the 2020 ACM Conference on Health, Inference, and Learning, Association for Computing Machinery, Inc, 2020, pp. 160–169. doi:10.1145/3368555.3384452.

[129] G. Huang, Z. Liu, L. Van Der Maaten, K. Q. Weinberger, Densely Connected Convolutional Networks, in: 2017 IEEE Conference on Computer Vision and Pattern Recognition (CVPR), IEEE, 2017, pp. 2261–2269. URL: https://ieeexplore.ieee.org/document/8099726/. doi:10.1109/CVPR.2017.243.

[130] K. He, X. Zhang, S. Ren, J. Sun, Deep Residual Learning for Image Recognition, in: 2016 IEEE Conference on Computer Vision and Pattern Recognition (CVPR), IEEE, 2016, pp. 770–778. URL: http://ieeexplore.ieee.org/document/7780459/. doi:10.1109/CVPR.2016.90.

[131] P. Geurts, D. Ernst, L. Wehenkel, Extremely randomized trees, Machine Learning 63 (2006) 3–42. URL: https://link.springer.com/article/10.1007/s10994-006-6226-1. doi:10.1007/S10994-006-6226-1/METRICS.

[132] J. H. Friedman, Greedy function approximation: A gradient boosting machine, The Annals of Statistics 29

(2001) 1189–1232. URL: https://projecteuclid.org/journals/annals-of-statistics/volume-29/issue-5/Greedy-function-approximation-A-gradient-boosting-machine/10.1214/aos/1013203451.full. doi:10.1214/aos/1013203451.

[133] S. M. Lundberg, S.-I. Lee, A Unified Approach to Interpreting Model Predictions, in: Advances in Neural Information Processing Systems, 2017, pp. 4766–4775. URL: https://dl.acm.org/doi/pdf/10.5555/3295222.3295230. doi:10.5555/3295222.3295230.

[134] R. R. Selvaraju, M. Cogswell, A. Das, R. Vedantam, D. Parikh, D. Batra, Grad-CAM: Visual Explanations from Deep Networks via Gradient-Based Localization, Proceedings of the IEEE International Conference on Computer Vision 2017-October (2017) 618–626. doi:10.1109/ICCV.2017.74.

[135] V. Venkatesh, M. G. Morris, G. B. Davis, F. D. Davis, User acceptance of information technology: Toward a unified view, MIS Quarterly: Management Information Systems 27 (2003) 425–478. doi:10.2307/30036540.

[136] A. Jeyaraj, Y. K. Dwivedi, V. Venkatesh, Intention in information systems adoption and use: Current state and research directions, International Journal of Information Management 73 (2023) 102680. doi:10.1016/J.IJINFOMGT.2023.102680.

[137] G. Lins de Holanda Coelho, P. H. P. Hanel, L. J. Wolf, The Very Efficient Assessment of Need for Cognition: Developing a Six-Item Version*, Assessment 27 (2020) 1870–1885. URL: https://journals.sagepub.com/doi/10.1177/1073191118793208. doi:10.1177/1073191118793208/ASSET/9542525C-070A-44A1-AA2B-BA602BEDEBE4/ASSETS/IMAGES/LARGE/10.1177{\_}1073191118793208-FIG2.JPG.

[138] O. P. John, S. Srivastava, The Big-Five trait taxonomy: History, measurement, and theoretical perspectives, 1999.

[139] P. Kouki, J. Schaffer, J. Pujara, J. O'Donovan, L. Getoor, Generating and Understanding Personalized Explanations in Hybrid Recommender Systems, ACM Transactions on Interactive Intelligent Systems (TiiS)

10 (2020). URL: https://dl.acm.org/doi/10.1145/3365843. doi:10.1145/3365843.

[140] L. Coroama, A. Groza, Evaluation Metrics in Explainable Artificial Intelligence (XAI), in: Communications in Computer and Information Science, volume 1675 CCIS, Springer Science and Business Media Deutschland GmbH, 2022, pp. 401–413. doi:10.1007/978-3-031-20319-0{\_}30.

[141] R. R. Hoffman, S. T. Mueller, G. Klein, J. Litman, Metrics for Explainable AI: Challenges and Prospects (2018) 1–50. URL: http://arxiv.org/abs/1812.04608.

[142] S. G. Hart, L. E. Staveland, Development of NASA-TLX (Task Load Index): Results of Empirical and Theoretical Research, in: P. A. Hancock, N. Meshkati (Eds.), Human Mental Workload, volume 52 of *Advances in Psychology*, North-Holland, 1988, pp. 139–183. URL: https://www.sciencedirect.com/science/article/pii/S0166411508623869. doi:https://doi.org/10.1016/S0166-4115(08)62386-9.

[143] F. Doshi-Velez, B. Kim, Towards A Rigorous Science of Interpretable Machine Learning, Arxiv (2017) 1–13. URL: http://arxiv.org/abs/1702.08608.

[144] M. Szymanski, M. Millecamp, K. Verbert, Visual, textual or hybrid: The effect of user expertise on different explanations, International Conference on Intelligent User Interfaces, Proceedings IUI (2021) 109–119. URL: /doi/pdf/10.1145/3397481.3450662?download=true. doi:10.1145/3397481.3450662.

[145] O. Asan, A. E. Bayrak, A. Choudhury, Artificial Intelligence and Human Trust in Healthcare: Focus on Clinicians, Journal of Medical Internet Research 22 (2020) e15154. URL: https://www.jmir.org/2020/6/e15154. doi:10.2196/15154.

[146] P. K. Kahr, G. Rooks, M. C. Willemsen, C. C. P. Snijders, Understanding Trust and Reliance Development in AI Advice: Assessing Model Accuracy, Model Explanations, and Experiences from Previous Interactions, ACM Transactions on Interactive Intelligent Systems

14 (2024) 1–30. URL: https://dl.acm.org/doi/10.1145/3686164. doi:10.1145/3686164.

[147] N. Ullah, J. A. Khan, I. De Falco, G. Sannino, Explainable Artificial Intelligence: Importance, Use Domains, Stages, Output Shapes, and Challenges, ACM Computing Surveys 57 (2024). URL: https://dl.acm.org/doi/10.1145/3705724. doi:10.1145/3705724/ASSET/357B3925-47D2-415F-B28F-FD7165CB5168/ASSETS/GRAPHIC/CSUR-2024-0085-F06.JPG.

[148] Satisfaction, in: Merriam-Webster.com, Merriam-Webster, 2025. Retrieved June 6, 2025, from https://www.merriam-webster.com/dictionary/satisfaction.

[149] J. Brooke, SUS – a quick and dirty usability scale, 1996, pp. 189–194.

[150] F. D. Davis, Perceived usefulness, perceived ease of use, and user acceptance of information technology, MIS Quarterly: Management Information Systems 13 (1989) 319–339. doi:10.2307/249008.

[151] M. Velmurugan, C. Ouyang, C. Moreira, R. Sindhgatta, Developing a Fidelity Evaluation Approach for Interpretable Machine Learning (2021). URL: http://arxiv.org/abs/2106.08492.

[152] M. A. Webb, J. P. Tangney, Too Good to Be True: Bots and Bad Data From Mechanical Turk, Perspectives on Psychological Science (2022). URL: https://scholar.google.com/scholar_url?url=https://journals.sagepub.com/doi/pdf/10.1177/17456916221120027&hl=es&sa=T&oi=ucasa&ct=usl&ei=--pCaLrMG6alieoP5f2nkAE&scisig=AAZF9b9KExuffqSdRYIvL28b3eH5. doi:10.1177/17456916221120027/ASSET/F0D2D167-57AC-453B-A3E6-16E5C2B3C2E3/ASSETS/IMAGES/LARGE/10.1177{\_}17456916221120027-FIG1.JPG.

[153] P. Pu, L. Chen, R. Hu, A user-centric evaluation framework for recommender systems, in: Proceedings of the fifth ACM conference on Recommender systems - RecSys '11, ACM Press, New York, New York, USA, 2011, p. 157. URL: http://dl.acm.org/citation.cfm?doid=2043932.2043962. doi:10.1145/2043932.2043962.

[154] Y. Jin, L. Chen, W. Cai, X. Zhao, Y. Jin, L. Chen, W. Cai, X. Zhao, CRS-Que: A User-centric Evaluation Framework for Conversational

Recommender Systems, ACM Transactions on Recommender Systems 2 (2024) 1–34. URL: https://dl.acm.org/doi/pdf/10.1145/3631534. doi:10.1145/3631534.

## Appendix A. Query for each database

| PICO term | Search category | Search Word |
|---|---|---|
| Population | Explanation | transparency<br>explanatory<br>interpretability<br>explainability<br>understanding<br>intelligible |
| | AI systems | Explainable Artificial Intelligence<br>Explainable AI<br>XAI<br>Interpretable Machine Learning<br>Interpretable ML<br>IML<br>Human-centric AI<br>Human-centered AI<br>Human-centred AI<br>XAI method<br>Recommender systems<br>Decision support systems<br>AI model prediction |
| Intervention | Humans | user<br>ux<br>quantitative<br>qualitative<br>human<br>participant<br>empirical |
| | Evaluation | experiment<br>test<br>evaluate<br>asses<br>compare<br>metric<br>measure<br>evidence<br>user-study |

Table A.10: Terms used to query all databases

## Appendix B. Measurements per property

Table B.11: Table of all measurements and their definitions based on the studies. For each quote related to a property, we encoded the type of measurement. These are all reflected in the following table:

| Property | Measurement Type | Paper | Measurement (quote from the paper) |
|---|---|---|---|
| AI model certainty | Closed Questions | [101] | It is important for me to know how uncertain (in %) the model is about its recommendation |
| AI model performance | Metric | [116] | Accuracy |
| | | [79] | Accuracy, Precision, Recall, F1 |
| | | [125] | AUC |
| | | [125] | Accuracy |
| | | [125] | Accuracy |
| | | [59] | Sensitivity, Specificity, AUC |
| | | [128] | F1 |
| | | [121] | Absolute error, Percentage of error |
| | | [91] | Accuracy |
| Alignment with situational context | Closed Questions | [80] | Did you feel that the AI assistance accelerated or slowed the workflow? |
| | Open Questions | [58] | We gathered feedback from the clinicians on the usefulness of information integrated into the conceptual prototype (MR1) and the correspondence to the diag- nostic workflow (MR4). |
| | | [110] | What suggestions would you have for making AI and XAI usable in the medical domain? |
| | | [118] | In these interviews we discussed the usability of the visual interface and how such a visual analytics tool can be integrated into the clinical workflow. |
| | | [50] | Adoption strategy for the tool<br>1. How did you use each explanation strategy during the experiment?<br>2. Was there any notable strategy for adopting the explanations rather than merely accepting the information in the explanations? |

Table B.11 – continued from previous page

| Property | Measurement Type | Paper | Measurement (quote from the paper) |
|---|---|---|---|
| | | [80] | What elements of the tool should be customizable to specific user needs? |
| Case difficulty | Closed Questions | [64] | the mean difficulty feedback by the doctors (1: the task is easy, 2: the task is normal, 3: the task is difficult) |
| | | [49] | Participants reported difficulty |
| | | [77] | Participants reported the perceived degree of difficulty (or complexity) of the case |
| | Metric | [63] | Authors reported difficulty |
| | | [86] | Authors reported difficulty |
| | | [117] | Authors reported difficulty |
| | | [85] | Authors reported difficulty |
| | | [50] | Authors reported difficulty |
| | | [88] | Authors reported difficulty |
| | | [92] | Authors reported difficulty |
| | | [104] | Authors reported difficulty |
| Cognitive load | Closed Questions | [101] | The barplot with feature contribution is easy to interpret |
| | | [101] | The colour bar with the score is easy to interpret |
| | | [101] | The scatterplot with all patients is easy to interpret |
| | | [122] | Explanations are easily understandable. |
| | | [122] | Explanations help reducing the learning time on the system |
| | | [64] | Which XAI solutions(s) were easier to interpret via visual maps? |
| | | [86] | NASA TLX |
| | | [117] | The visualizations are clear and I don't spend much time on their interpretation |
| | | [124] | NASA TLX [Effort and Frustration] |
| | | [126] | NASA TLX [Effort and Frustration] |
| | | [103] | a4) User experienc |

Table B.11 – continued from previous page

| Property | Measurement Type | Paper | Measurement (quote from the paper) |
|---|---|---|---|
| | | [80] | Did the AI assistance make the task less complex, demanding, or exacting? |
| | | [80] | Was the user interface easy or frustrating to use? |
| | | [70] | I was insecure, discouraged, and stressed while using the system |
| | | [70] | The system helped me think through and complete the assessment tasks with less effort |
| | | [74] | Was the AI system overwhelming or not? |
| | | [87] | Using RadiologyAI was very frustrating. Using RadiologyAI, I easily found the information I wanted to help me decide on a diagnosis. (reversed) Using RadiologyAI took too much time. Using RadiologyAI was easy. (reversed) Using RadiologyAI required too much effort. Using RadiologyAI was too complex. |
| | | [81] | NASA TLX |
| | | [100] | I think that most people would learn to understand the explanations very quickly |
| | Open Questions | [117] | Does a reference to a standard scale facilitates critical assessment of a recommendation? |
| | | [80] | What could be changed to further reduce workload? What features could make the process less rushed? What features could minimize feelings of irritation/annoyance |
| | | [80] | What features could make the task less demanding, complex, and exacting? What features could help decrease the mental activity required to carry out the task? What features could improve/minimize fatigue? |
| Completeness | Metric | [125] | The area under PC (AUPC(H)) |

Table B.11 – continued from previous page

| **Property** | **Measurement Type** | **Paper** | **Measurement (quote from the paper)** |
|---|---|---|---|
| Confidence | Closed Questions | [101] | Even when my initial course of action was different to what CORONET recommended, I still had full confidence in my original decision. |
| | | [101] | When my initial decision was the same as CORONET had recommended, I felt reassured |
| | | [63] | Participants reported confidence on their decision |
| | | [85] | Participants reported confidence on their decision |
| | | [76] | Participants reported confidence on their decision |
| | | [123] | Participants reported confidence on their decision |
| | | [49] | Participants reported confidence on their decision |
| | | [65] | Participants reported confidence on their decision |
| | | [54] | Participants reported acceptance of the result |
| | | [68] | Participants reported confidence on their decision |
| | | [68] | XAI increases clinicians' confidence in their own diagnoses |
| | | [77] | his or her confidence in the proposed diagnosis |
| | | [89] | Participants reported confidence on their decision |
| | | [80] | Did the AI assistance increase your level of confidence when grading splenic injuries? |
| | | [80] | How confident you are with your diagnosis |
| | | [94] | Do you feel confident to decide to use the tool? |
| | | [90] | Participants reported confidence on their decision |

Table B.11 – continued from previous page

| Property | Measurement Type | Paper | Measurement (quote from the paper) |
|---|---|---|---|
| | | [97] | I feel very confident about using the system |
| | | [92] | Participants reported confidence on their decision |
| | | [92] | Participants reported confidence on their decision |
| | | [75] | Participants reported confidence on their decision |
| | | [78] | Participants reported confidence on their decision |
| | | [108] | After seeing the explanation of prediction, I am more inclined to change my decision if the system prediction conflicts with my initial belief |
| | | [108] | After seeing the explanation of prediction, I feel more confident to make my decision if the system prediction supports my initial belief |
| | Open Questions | [103] | Do you feel confident with the priority suggestions given by the application? |
| | User Behaviour | [99] | Participants were asked to estimate the patient's chances of developing an acute MI in the near future on a scale from 0 to 100% and their confidence in the estimate on a sliding scale. |
| | | [99] | The confidence shift was measured as the difference of the reported participant's confidence in the estimate before and after receiving the AI advice |
| Contrastivity | Closed Questions | [82] | Participants reported contrastivity |
| Controllability | Closed Questions | [100] | I could change the level of detail on demand |
| | Open Questions | [103] | Do you have a sense of control when you provide feedback to the application? |
| | | [80] | To what degree is an interactive method preferable over a non-interactive one? |

Table B.11 – continued from previous page

| Property | Measurement Type | Paper | Measurement (quote from the paper) |
|---|---|---|---|
| Correctness | Metric | [125] | For the truthfulness assessment, we conducted cumulative feature removal and modality importance (MI) evaluation for the two clinical tasks, and proposed two novel metrics and MI correlation respectively. We also conducted a synthetic data experiment on the glioma grading task |
| | | [125] | G3: Truthfulness Explanations should truthfully reflect the AI model decision process. This is the prerequisite for G4. |
| | | [125] | The area under PC (AUPC(H)) |
| | | [125] | modality Shapley value [defined in study] |
| | | [125] | uthfulness. An explanation should truthfully reflect the model decision pro- cess. |
| Curiosity | User Behaviour | [104] | Notably, when faced with content scenarios, participants asked an average of 1.5 more questions than in risk scenarios. |
| | | [104] | Teams using GutGPT in con- tent scenarios submitted more queries to the chatbot than teams in risk scenarios. There were an average of 3.9 queries in content scenarios compared to 2.4 in risk scenarios. This indicates that the chatbot feature was used more heavily when making decisions regarding a care plan, and less utilized for risk assessment. |
| Efficiency | Closed Questions | [122] | The system with explanations would help me completing the assessment in less time. |
| | | [122] | The system without explanations would help me completing the assessment in less time. |
| | | [64] | Which XAI solutions(s) had the better performance in terms of running-time? |
| | | [83] | I am able to make decisions faster thanks to XAI view of the system. |

Table B.11 – continued from previous page

| Property | Measurement Type | Paper | Measurement (quote from the paper) |
|---|---|---|---|
| | | [80] | Did you feel that the AI assistance accelerated or slowed the workflow? |
| | | [100] | I received the explanations in a timely and efficient manner |
| | User Behaviour | [64] | Average completion rate (depending on the top completion limit for each task) |
| | | [64] | Mean task completion time (in seconds) |
| | | [85] | Completion time when presented with a diagnostic choice for individual and overall questions. |
| | | [76] | Completion time across the four conditions. |
| | | [49] | 5) Completion Time: The average time (in seconds) it took participants to agree or disagree with the system. Similar to yes- rate, completion time does not measure performance. However, it tells us how long it took participants to make decisions, which may be related to participants' level of engagement or effort. |
| | | [91] | The main objective of this is to check the accuracy and the time taken to make the decisions. |
| | | [70] | Duration of Decision Making |
| | | [56] | Column Steps indicates the minimum number of steps (in terms of links to click, overviews to open, or questions to pose) required by each explanatory tool to provide the correct answer. |
| | | [75] | Time The time spent in making decisions. |
| Explanation power | Closed Questions | [82] | The extent the map highlights structures or findings that allow for the correct diagnosis to be reached more easily than if the map was not seen |
| | | [101] | The barplot with feature contribution convinces me to accept or reject the model's recommendation |

Table B.11 – continued from previous page

| **Property** | **Measurement Type** | **Paper** | **Measurement (quote from the paper)** |
|---|---|---|---|
| | | [101] | The colour bar with the score convinces me to accept or reject the model's recommendation |
| | | [101] | The scatterplot with all patients convinces me to accept or reject the model's recommendation |
| | | [100] | The group of sentences provides truthful statements and avoids providing information not supported by evidence (maxim of quality) |
| | Open Questions | [117] | Are you convinced with the interpretations that the system provides? |
| | User Behaviour | [123] | Weight of Advice |
| | | [49] | (4) Yes Rate: the percentage of times participants agreed with the system. Yes-rate does not measure performance. Rather, it pro- vides insights into participants' tendencies to agree with the system across interface conditions. |
| | | [99] | Weight of Advice |
| | | [120] | It measures the percentage shift in judgment after advice, and it quantifies how much the participants follow the advice they receive. |
| | | [120] | Weight of Advice |
| | | [77] | Decision changes |
| | | [55] | Change in mental model (CMM) Difference in perceived feature importance before and after viewing model explanation |
| | | [70] | we analyzed the number of times when AI explanations assisted participants to change and make 'right' or 'wrong' decisions |

Table B.11 – continued from previous page

| **Property** | **Measurement Type** | **Paper** | **Measurement (quote from the paper)** |
|---|---|---|---|
| | | [56] | Carnap's main criteria of explication adequacy [11] are similar- ity, exactness and fruitfulness.5 Similarity means that the explica- tum should be detailed about the explicandum, in the sense that at least many of the intended uses of the explicandum, brought out in the clarification step, are preserved in the explicatum. On the other hand, exactness means that the explication should be embedded in some sufficiently clear and exact linguistic framework, while fruitfulness implies that the explicatum should be useful and usable in a variety of other good explanations (the more, the better). |
| | | [75] | Adherence The adherence of decision-making to recommendations. |
| | | [71] | $D_{i,j}$ as the behavioral distrust measure of annotator i for batch j, quantified as the average absolute discrepancy between the AI decision and the anno- tator's decision (Chu et al., 2020). |
| | | [71] | Finally, the behavioral measure of trust $b_{i,k}$ of annotator i at sample k is defined as the inverse of the absolute dis- crepancy between the human annotator i and ML prediction for sample k. |
| Information correctness | Closed Questions | [73] | Participants reported plausibity |
| | | [112] | Participants reported medical relevance |
| | | [84] | Images in the video look like a chest X-ray |
| | | [59] | Does it look reasonable at a glance? |
| | | [57] | Are the highlighted sentences topically related to the query by using either TF-IDF, BM25 or BioBERT? |
| | | [57] | Are the top sentences (with the best model between TF-IDF, BM25, or BioBERT) correctly supported by scientific evidence (scientific journal articles)? |

Table B.11 – continued from previous page

| **Property** | **Measurement Type** | **Paper** | **Measurement (quote from the paper)** |
|---|---|---|---|
| | | [100] | The group of sentences provides truthful statements and avoids providing information not supported by evidence (maxim of quality) |
| | Metric | [112] | The range of validity is between 0 and 1, and a higher score will be desired. Local outlier factor measures the closeness of the counterfactuals to the training data distribution of the desired class [30]. LOF is calculated as the proportion of testing samples that are considered outliers using the local density of the nearest neigh- bours. |
| | | [112] | Validity is defined as the fraction of the counterfactuals that are valid based on the desired target class [29]. |
| | | [84] | Foreign Object Preservation (FOP) score [defined in study] |
| | | [84] | Frechet Inception Distance (FID) [defined in study] |
| | | [68] | overlap in ontological explanations between the clinicians and the AI. |
| | | [100] | Degree of Truth [defined in study] |
| Information expectedness | Closed Questions | [82] | The extent the map highlights the structures that present a fracture or those in which fractures are excluded according to the machine's suggestion |
| | | [73] | Participants reported whether the fact was new to them or not |
| | | [84] | Images in the video look like the chest X-ray from a given subject |
| | | [59] | Does it make sense? |
| | | [124] | [System - Condition X] generates new insights on patient's performance |
| | | [126] | [Tool - Condition X] generates new insights on patient's performance |

Table B.11 – continued from previous page

| Property | Measurement Type | Paper | Measurement (quote from the paper) |
|---|---|---|---|
| | | [127] | How closely the highlighted areas of the heatmap match with your clinical judgment? |
| | | [67] | How closely does the highlighted area of the color map match with your clinical judgment? |
| | | [94] | Participants reported with the shown information |
| | | [70] | The system provided new insights on patient's performance for assessment |
| | | [97] | The AI predictions align with my medical opinion |
| | | [97] | The explanation of the prediction is clear and reasonable |
| | Metric | [125] | Feature portion (FP), Modality-specific feature importance (MSFI) [defined in study] |
| | | [127] | MSFI [defined in study] |
| | | [68] | We used these explanations to determine the extent to which the XAI system and clinicians detected similar expla- nations on the same lesions. We evaluated our XAI's alignment with the clinicians' ontological explanations and annotated regions of interest (ROI) by assessing their overlap. |
| | Open Questions | [117] | Does a reference to a standard scale facilitates critical assessment of a recommendation? |
| | | [50] | Did the explanations correspond well to your perception of the important waveform patterns? |
| Intention to use | Closed Questions | [62] | Participants reported whether they would use the system |
| | | [107] | Participants reported whether they want to continue using the system |
| | | [117] | I intend to use the tool to have a second opinion on the patient risks I feel like I will use it in the future |

Table B.11 – continued from previous page

| Property | Measurement Type | Paper | Measurement (quote from the paper) |
|---|---|---|---|
| | | [124] | I would use [System - Condition X] to understand and assess patient's performance |
| | | [126] | I would use [Tool - Condition X] to understand and assess patient's performance |
| | | [83] | I want to use this system for auto-decision-making in brain tumor diagnosis. |
| | | [123] | Participants reported whether they would use the system |
| | | [80] | Would you be willing to use a productized tool like this in clinical practice? |
| | | [80] | Would you use a productized tool with these features in the future? |
| | | [67] | How likely will you incorporate this AI's suggestions into your routine clinical practices, such as diagnosis, prognosis, and medical management? |
| | | [115] | Participants reported the likelihood of using an AI system for sepsis prescriptions |
| | | [94] | Will you use it if finally implemented? |
| | | [70] | I would use this system to understand and assess patient's exercise performance in practice |
| | | [74] | Participants reported whether they would use the system in their current work |
| | Open Questions | [80] | What features would make you more likely to use the tool in your future practice? |
| Necessity | Closed Questions | [100] | I found that the data included all relevant known causal factors with sufficient precision and granularity |
| | | [100] | The group of sentences provides all the information we need, and no more (maxim of quantity) |
| | Metric | [84] | Counterfactual Validity (CV) score |

Table B.11 – continued from previous page

| Property | Measurement Type | Paper | Measurement (quote from the paper) |
|---|---|---|---|
| | | [72] | To assess the similarity, we investigated only whether the produced explanation contained all the evidence that clinicians mentioned as significant. As we restricted the set of significant evidence - evidence included in the explanation - with the intent to rapidly produce a concise explanation, and not necessarily the most complete one, our main intention was to investigate whether our produced concise ex- planation does not miss factors that clinicians considered significant and not whether extra variables are included in the explanation. |
| | Open Questions | [110] | What elements you missed in the explanations provided by the XAI method (SHAP, LIME, or PFI)? What information would be valuable to be presented scientific knowledge? |
| | | [114] | Does the report provide the AI explanation needed for psychiatrists and patient |
| Perceived model competence | Closed Questions | [83] | The system distinguishes normal and tumor brain images successfully. |
| | | [83] | This model is not good about showing how it is diagnosing tumors. |
| | | [83] | This system can be used further for alternative tumor detections. |
| | | [83] | This system can be used in further treatment and diagnosis strategies. |
| | | [123] | Participants reported system reliability, predictability, and efficiency |
| | | [99] | Participants reported system reliability, predictability, and efficiency |
| | | [89] | Participants reported confidence on the system classification |
| | | [81] | The system seems capable |

Table B.11 – continued from previous page

| **Property** | **Measurement Type** | **Paper** | **Measurement (quote from the paper)** |
|---|---|---|---|
| | User Behaviour | [121] | Figure 7 shows the ranks of the models by the par- ticipants in boxplots: The smaller the ranks are, the better the models were con- sidered by the participants. A Friedman test revealed a small but statistically sig- nificant difference in the models' ranking (Effect size: 0.18, p = 0.045). A pair-wise comparison using Wilcoxon Signed Rank tests showed large and moderate differ- ences in the ranking of M2-4 compared with M1; the dif- ferences among M2-4 are small. After the Holm correction, the differences are not statistically significant. |
| Perceived user performance | Closed Questions | [122] | The system with explanations would help me providing a more accurate diagnosis than the one I would provide without the use of the tool |
| | | [122] | The system without explanations would help me providing a more accurate diagnosis than the one I would provide without the use of the tool |
| | | [49] | Participants reported satisfaction with their performances |
| | | [80] | Could the tool improve patient decision-making? |
| | | [80] | Did the tool improve your accuracy in grading splenic injuries? |
| | | [70] | I think AI, data-driven tool can improve rehabilitation assessment |

Table B.11 – continued from previous page

| Property | Measurement Type | Paper | Measurement (quote from the paper) |
| --- | --- | --- | --- |
| Performance | User Behaviour | [52] | Pooled results of all evaluators of decision support system (DSS) 2.0. n, number of cases, expressed as the number of cases of the class analysed divided by the total number of cases evaluated (i.e. in the first class, 'Meningioma WHO grade I, 6/38' means 6 of 38 cases with readings in DSS 1.0; '53/354' means 53 evaluations from all seven evaluators on these same cases for a total of 354 read- ings). AUC, area under the curve, followed by the 95% symmetric confidence interval in parentheses. |
| | | [82] | Accuracy of human readers pre and post AI support |
| | | [69] | To further investigate, we plot the receiver operating character- istics (ROC) and precision-recall (PR) for all physicians, as well as the ROC and PR curves for Explainer. |
| | | [86] | Finally, we measured the number of False-Positives and False-Negatives. |
| | | [116] | The dependent variable performance measures the objective performance of the algorithm. T |
| | | [85] | There was no significant difference in agreement between the test and control groups regarding classification of furcation involvement for all questions except question 2 (P < 0.05; Figure 2). |
| | | [76] | We conducted the Repeated Measures ANOVA analysis to compare the average accuracy, |
| | | [128] | While accommodating therapist's additional nine feature-based feedback on each patient, both Knowledge Model (KM) and Hybrid Model (HM) improve their performance on all exercises (Figure 6). |

Table B.11 – continued from previous page

| Property | Measurement Type | Paper | Measurement (quote from the paper) |
|---|---|---|---|
| | | [49] | (1) Accuracy: the percentage of correct agree/disagree decisions. Each sequence of 12 articles included 6 true positive and 6 false positive cases. Thus, we expected accuracy values to be $> 50\%$. (2) Precision: the percentage of "agree" decisions involving a true positive case. Precision captures whether participants rejected false positive cases when they agreed with the system. (3) Recall: the percentage of true positive cases for which participants agreed with the system. Recall captures whether participants agreed with the system for all true positive cases. |
| | | [65] | In P1, the use of XAI improved the mean preferential and differential diagnostic accuracy by |
| | | [65] | The mean preferential accuracy is calculated as the number of cases in which a pulmonologists' preferential diagnosis matched the gold standard, averaged over the entire cohort. |
| | | [50] | First, we considered the accuracy with which the technicians scored the sleep stages under different sleep staging settings. |
| | | [60] | After the risk function selection and model inspection, the EBM model contained 67 1D risk functions. It achieved a PR-AUC of $0.119 \pm 0.020$ and a ROC-AUC of $0.680 \pm 0.025$ (Figure 4). For recall values of 0.4, 0.5, 0.6, and 0.8 the precision values were $0.130 \pm 0.032$, $0.111 \pm 0.019$, $0.105 \pm 0.013$, and $0.082 \pm 0.005$. |

Table B.11 – continued from previous page

| **Property** | **Measurement Type** | **Paper** | **Measurement (quote from the paper)** |
|---|---|---|---|
| | | [77] | accuracy, we observed some interesting differences between residents (N = 6) and specialists (N = 10) (see Fig. 3). When unaided, the residents performed better than the specialists, with an average accuracy of .83 (SD: .06) vs. .76 (SD: .08) and with a large effect size (1.03, NDN = 2), albeit not significantly so (p-value = .5, T = 2.1). This could be due to a greater commitment to the task of the residents (as compared with specialists), who took the opportunity to test their skills and (informally) compete with each other: indeed, on average, residents took 4 min more (15%) to complete the task than specialists. |
| | | [91] | The main objective of this is to check the accuracy and the time taken to make the decisions. |
| | | [70] | 3.5.3 Counts of 'right' and 'wrong' decisions. In addition to the performance and agreement level, we analyzed the counts of 'right' and 'wrong' decisions by participants to further analyze their overreliance on 'wrong' AI outputs. Also, we measured the count of (1) agreeing with 'right' AI outputs, (2) rejecting 'wrong' AI outputs, (3) agreeing with 'wrong' AI outputs (i.e. overreliance), and (4) rejecting 'right' AI outputs for further analys |
| | | [70] | In this study, we utilized the annotations of a therapist from the dataset [40] as ground truths and eval- uate participants' performance on decision-making tasks before/after reviewing AI outputs and explanations |
| | | [56] | More specifically, effectiveness here is defined as "accuracy and completeness with which users achieve specified goals |

Table B.11 – continued from previous page

| Property | Measurement Type | Paper | Measurement (quote from the paper) |
|---|---|---|---|
| | | [95] | Statistical Analysis We assessed the concordance of the above-described five expert-determined standards with the clinical gold standard and compared their sensitivity, specificity, and accuracy. |
| | | [90] | asking then each user to provide their definitive diagnosis |
| | | [92] | ecision performance |
| | | [75] | Accuracy The accuracy of outcomes produced by decision makers. |
| | | [51] | The ratings obtained from the radiographers were used to plot receiver operating characteristic (ROC) curves with and without decision support. These curves are shown in Figure 2. Each point on the curve corresponds to the use of a different position on the five-point rating scale as the threshold for a diagnosis of malignancy. |
| | | [78] | First, we have attached to each participant a score that measures their performance in P1 to test their ability to classify skin lesion images by exploiting the explanation. |
| | | [71] | We first obtain an overall understanding of cases in which the human annotators and the AI system depict differences in performance. |
| Personal characteristics | Closed Questions | [123] | Need for Cognition |
| | | [99] | Need for Cognition |
| | | [47] | Ease of Satisfaction |
| | | [47] | Need for Cognition |
| | | [71] | Big Five Inventory |
| Prediction expectedness | Closed Questions | [101] | I was surprised when CORONET recommended an action different to my own |
| | | [65] | Participants reported agreement with XAI suggestion |
| | | [97] | The predicted result is reasonable |

Table B.11 – continued from previous page

| **Property** | **Measurement Type** | **Paper** | **Measurement (quote from the paper)** |
| --- | --- | --- | --- |
| | Open Questions | [50] | How did you perceive the provided explanations when the automated predictions agreed or disagreed with your decisions? Did it affect your trust in the system? |
| Relevance to the task | Closed Questions | [73] | Participants reported whether the fact was relevant to the domain of biomedicine. |
| | | [62] | Participants reported whether the module is relevant to the task |
| | | [101] | CORONET helps in cases where I am less confident in the decision on how to proceed |
| | | [101] | CORONET helps me in making safe clinical decisions on patient management |
| | | [101] | It is important for me to know how the model makes its recommendation for my individual patient |
| | | [101] | It is important for me to know the features of my patient contribute to the model's recommendation |
| | | [101] | It is important for me to know the mathematics behind the model's recommendations |
| | | [112] | Participants validated the actionability of event sequences |
| | | [117] | Using the system enhances the effectiveness of managing risks of my patients |
| | | [84] | The changes in the video are related to Cardiomegaly |
| | | [124] | [System - Condition X] is useful to understand and assess patient's performance |
| | | [126] | [Tool - Condition X] is useful to understand and assess patient's performance |
| | | [83] | This model can be located in a wider healthcare support system like IoHT. |
| | | [96] | Does the explanation provided help you decide on actionable steps you can undertake? |

Table B.11 – continued from previous page

| Property | Measurement Type | Paper | Measurement (quote from the paper) |
|---|---|---|---|
| | | [96] | Regarding our CDSS, would the provided output and explanation help you justify your clinical decision-making (e.g., to patients and colleagues)? |
| | | [70] | The system provided useful information to understand patient's performance for assessment |
| | | [72] | How useful was the prediction of coagulopathy for confirming your assessment? |
| | | [78] | The participants are asked to quantify how much the exemplars and counter-exemplars helped them to classify skin lesion images in accordance with the AI |
| | | [100] | I understood the explanations within the context of my work |
| | Open Questions | [80] | Did the tool help you think through this multistage diagnostic process? |
| | | [80] | What features could make the system more useful for patient decision-making? |
| | User Behaviour | [112] | For the second task, we defined the relevance score to determine the fraction of suggested edits that are considered medically relevant from the expert validation. |
| Reliance | Closed Questions | [88] | I would likely decrease the frequency of visual field testing for this patient if I had predicted MDs available from this algorithm. |
| Representativeness | Closed Questions | [64] | Which XAI solutions(s) did you find more stable? |
| | | [64] | Which XAI solutions(s) did you find more stable? |
| | | [100] | I did not find inconsistencies between explanations |
| | | [100] | I did not find inconsistencies between explanations |

Table B.11 – continued from previous page

| Property | Measurement Type | Paper | Measurement (quote from the paper) |
|---|---|---|---|
| | Metric | [100] | . The degree of support of a linguistic summary LS indicates how many objects in the dataset are covered by the particular summary, |
| Satisfaction | Closed Questions | [101] | I am satisfied with the output information that CORONET provides |
| | | [122] | In general, I am satisfied with the explanations provided by the system. |
| | | [116] | Perceived Goodness of explanation |
| | | [59] | Does it look good? |
| | | [99] | Explanation Satisfaction Scale |
| | | [74] | Participants reported if they would recommend its use to a colleague. |
| | | [81] | How well the assistant worked? |
| | | [75] | Participants reported satisfaction towards the support received |
| Size | Closed Questions | [122] | Explanations are not lengthy. |
| | | [55] | This explanation is long and uses more words than required |
| | | [57] | Do you think highlighting a single sentence is enough to capture both the topicality and truthfulness of the document? |
| | | [100] | The group of sentences is clear, and as brief and orderly as possible, avoiding obscurity and ambiguity (maxim of manner) |
| | Metric | [55] | The authors refer to unique variables within the rules as cognitive chunks, which contribute to complexity in understanding. In our exper- iment, global explanations naturally contain more cognitive chunks. |
| | | [55] | The hard scenario for both explanation types, consist of larger trees of sim- ilar structures. |
| | | [90] | Features detail level: high, Features detail level: low |

Table B.11 – continued from previous page

| Property | Measurement Type | Paper | Measurement (quote from the paper) |
|---|---|---|---|
| Structure | Closed Questions | [97] | The order of feature importance is logical |
| | | [100] | The group of sentences is clear, and as brief and orderly as possible, avoiding obscurity and ambiguity (maxim of manner) |
| | Open Questions | [114] | What is your opinion concerning the report's display and the attached instructions? |
| Sufficiency | Closed Questions | [122] | Explanations are detailed enough. |
| | | [97] | The explanation for complex medical features is sufficient |
| | | [57] | Do the highlighted sentences (with the best model between TF-IDF, BM25, or BioBERT) provide sufficient information to determine the truthfulness of the document? |
| | | [57] | Does the scientific evidence provide sufficient information to assess the truthfulness of the document? |
| | | [100] | I did not need more references in the explanations (e.g., medical guidelines, regulations) |
| | Open Questions | [110] | What elements you missed in the explanations provided by the XAI method (SHAP, LIME, or PFI)? What information would be valuable to be presented scientific knowledge? |
| Technology Acceptance | Closed Questions | [65] | Paricipants reported enthusiasm about AI |
| | | [120] | Participants reported prior experience, Participants reported their inclination towards CDSS use |
| | | [71] | Propensity to Trust Machines |
| Trust | Closed Questions | [62] | Participants reported trustworthines |

Table B.11 – continued from previous page

| Property | Measurement Type | Paper | Measurement (quote from the paper) |
|---|---|---|---|
| | | [122] | I have the feeling of trust in the system with explanations regarding its prediction of the overall cognitive status. I have the feeling of trust in the system with explanations regarding its prediction about the anomaly level of activities. |
| | | [122] | I have the feeling of trust in the system without explanations regarding its prediction of the overall cognitive status. I have the feeling of trust in the system without explanations regarding its prediction of the anomaly level of activities. |
| | | [117] | I trust the system as it provides interpretations of the results I trust the system because it provides references to the standard scales |
| | | [124] | I can trust information from [System - Condition X] |
| | | [126] | I can trust information from [Tool - Condition X] |
| | | [83] | This system is trustworthy in terms of diagnosis of brain tumors. |
| | | [46] | Participants reported trustworthines |
| | | [96] | Does the explanation provided add towards your trust of model predictions? |
| | | [68] | Participants reported trust in AI decision |
| | | [68] | XAI increases clinicians' trust in machine decisions |
| | | [89] | Participants reported trustworthines |
| | | [80] | Did you trust the AI assistance to correctly grade the injuries? |
| | | [80] | Do you trust the AI assistance? |
| | | [67] | What is your trust level in this AI model? |
| | | [70] | I can trust the provided assessment scores or/and analysis from the system |
| | | [88] | Itrust the predicted MD enough to incorporate it into my decision-making |

Table B.11 – continued from previous page

| Property | Measurement Type | Paper | Measurement (quote from the paper) |
|---|---|---|---|
| | | [97] | The application enhances my trust in AI predictions |
| | | [72] | How much would you say that you trust the prediction of the model? |
| | | [78] | Participants reported trustworthines |
| | | [71] | Participants reported trustworthines |
| | Open Questions | [110] | Do you trust in AI? |
| | | [118] | We also discussed how the visual interface affects their trust in the automated classifications. |
| | | [50] | How did you perceive the provided explanations when the automated predictions agreed or disagreed with your decisions? Did it affect your trust in the system? |
| | | [80] | To what degree does your ability to verify or reject the algorithm's conclusions improve your trust in the method? |
| | User Behaviour | [71] | We further explore how annotators change their trust in AI over time, depending on the error of the AI system. We build a LME model with random intercept that estimates the behavioral measure of trust in AI as a function of time and AI error, as follows:<br>bi, k ¼ b þ |
| Understanding | Closed Questions | [70] | The system provided useful information to understand patient's performance for assessment |
| | | [74] | Was the AI system confusing or not confusing |
| | | [74] | Was the AI system misleading or clear? |
| | User Behaviour | [53] | Its purpose was to test (see Section 1) the information recall and understanding gain in the participants with respect to the priming method they used and their interaction with an aligning or misaligning agent, or no interaction at all. |

Table B.11 – continued from previous page

| **Property** | **Measurement Type** | **Paper** | **Measurement (quote from the paper)** |
|---|---|---|---|
| Understanding explanation | Closed Questions | [62] | Participants reported whether they understood the information |
| | | [101] | The barplot with feature contribution is easy to interpret |
| | | [101] | The colour bar with the score is easy to interpret |
| | | [101] | The scatterplot with all patients is easy to interpret |
| | | [122] | Explanations are easily understandable. |
| | | [64] | Which XAI solutions(s) were easier to interpret via visual maps? |
| | | [117] | The interpretations are clear, and I understand the reasoning |
| | | [117] | The visualizations are clear and I don't spend much time on their interpretation |
| | | [84] | I understand how the AI system made the above assessment for Cardiomegaly |
| | | [83] | I am able to understand well about detection, by looking at the heat maps. |
| | | [46] | Participants reported understanding |
| | | [97] | The explanation of the prediction is clear and reasonable |
| | | [109] | How useful do you find the counterfactual examples it generates to your understanding of counterfactual explanations? |
| | | [109] | tree to your understanding the global & local explanations? |
| | | [72] | How clear was the explanation of the prediction of coagulopathy? |
| | | [100] | I did not need support to understand the explanations |
| | | [100] | The group of sentences is clear, and as brief and orderly as possible, avoiding obscurity and ambiguity (maxim of manner) |

Table B.11 – continued from previous page

| Property | Measurement Type | Paper | Measurement (quote from the paper) |
|---|---|---|---|
| | User Behaviour | [110] | According to the results, what are the three factors that most influenced the cancer diagnosis? |
| Understanding model behaviour | Closed Questions | [62] | Participants reported trustworthines |
| | | [101] | I understand when and why CORONET may provide the wrong recommendation in some cases |
| | | [122] | Explanations are useful to understand the reason for the system's prediction. |
| | | [117] | It explains me why a certain risk assessment is done |
| | | [117] | The model outcomes are clear and understandable |
| | | [84] | Which explanation helped you the most in understanding the assessment made by the AI system? |
| | | [83] | It is possible to predict limits of the system, thanks to heat map results. |
| | | [96] | Does the visual representation of the CDSS output help you rationalise the predictions? |
| | | [96] | Does the visual representation of the CDSS output help you understand the predictions? |
| | | [55] | Based on the explanation I understand how the model would behave for another patient |
| | | [55] | This explanation helps me completely understand why the AI system made the prediction |
| | | [70] | The system was transparent about why it provided a particular assessment score |
| | | [97] | The data set explanation assists me in understanding the data set used by the machine learning model for prediction |
| | | [97] | The explanation helps me comprehend how machine learning generates the prediction |
| | | [57] | Are topicality and truthfulness scores useful to understand the ranking? |

Table B.11 – continued from previous page

| **Property** | **Measurement Type** | **Paper** | **Measurement (quote from the paper)** |
|---|---|---|---|
| | | [87] | RadiologyAI made its reasoning process clear to me. It was readily apparent to me how RadiologyAI generates its predictions. I could not understand how RadiologyAI performs its job. (reversed) I could easily understand the reasoning process of RadiologyAI. It was easy for me to understand the inner workings of RadiologyAI. I could understand why and how RadiologyAI recommends the diagnoses to me. The logic of RadiologyAI was clear to me. |
| | | [81] | I understand what the system is thinking |
| | | [108] | Seeing the explanation of prediction makes me understand the reasoning of a system's prediction |
| | | [108] | Seeing the explanation of prediction makes the system more transparent |
| | | [100] | I found the explanations helped me to understand causality |
| | Open Questions | [117] | Can you understand a model output without interpretations? |
| | | [110] | The explanation provided by the XAI method (SHAP, LIME, or PFI) is comprehensible and helped me understanding how the 'Dr. Artificial Intelligence' system came to the diagnosis of cancer. |
| | User Behaviour | [110] | Based on what you saw on the system, how do you explain the result provided by the XAI method (SHAP, LIME, or PFI)? |
| | | [55] | Error in Understanding (EU) Difference between model feature importance and perceived feature importance after viewing explanation |
| Usability | Open Questions | [118] | In these inter- views we discussed the usability of the visual interface and how such a visual analytics tool can be integrated into the clinical workflow. |

Table B.11 – continued from previous page

| Property | Measurement Type | Paper | Measurement (quote from the paper) |
|---|---|---|---|
| Usefulness | Closed Questions | [107] | Participants reported whether the system could help them improve their decision-making |
| | | [122] | Explanations are necessary to understand the reason for the system's prediction regarding the overall cognitive status Explanations are necessary to understand the reason for the system's prediction regarding the overall anomaly level of activities |
| | | [122] | Explanations are useful to understand the reason for the system's prediction. |
| | | [122] | The explanations would help me deciding whether the prediction of the system is correct. |
| | | [117] | The system helps me to make more informed decisions |
| | | [84] | Which explanation helped you the most in understanding the assessment made by the AI system? |
| | | [85] | Participants reported usefulness |
| | | [46] | Participants reported usefulness |
| | | [96] | Does the visual representation of the CDSS output help you understand the predictions? |
| | | [77] | Participants reported usefulness |
| | | [89] | Participants reported usefulness |
| | | [80] | Did the AI assistance help you think through the grading process? |
| | | [80] | Did the AI assistance help you think through the problem? |
| | | [115] | Participants reported usefulness |
| | | [74] | Was the AI system not useful or useful? |
| | | [88] | The predicted MD provides additional useful information beyond the existing clinical information available |
| | | [90] | Participants reported perceived utility |

Table B.11 – continued from previous page

| **Property** | **Measurement Type** | **Paper** | **Measurement (quote from the paper)** |
|---|---|---|---|
| | | [109] | How useful do you find the counterfactual examples it generates to your understanding of counterfactual explanations? |
| | | [109] | How useful do you find the decision tree to your understanding the global & local explanations? |
| | | [92] | Participants reported perceived utility |
| | | [57] | Are topicality and truthfulness scores useful to understand the ranking? |
| | | [100] | I found the explanations helped me to understand causality |
| | | [100] | The group of sentences is relevant to the discussion objective of explaining the model (maxim of relation) |
| | Open Questions | [58] | We gathered feedback from the clinicians on the usefulness of information integrated into the conceptual prototype |
| | | [117] | Do you require an interpretation to be able to critically assess a model output? |
| | | [114] | Can the explainable reports be useful for addi- tional medical applications? |
| | | [114] | How useful are the report parameters for the physi- cians and patients in understanding how the model is making its decision? |
| | | [50] | Were the automated predictions and explanations provided in the clinical decision support systems helpful during the experiment? If not, which aspects were unhelpful? |
| | | [66] | Here, an average of 3.38 (SD0.49) was given to the usefulness of explanation types where 1 was considered as "Strongly disagree" and 5 |

Table B.11 – continued from previous page

| Property | Measurement Type | Paper | Measurement (quote from the paper) |
| --- | --- | --- | --- |
| | User Behaviour | [78] | This point aims to understand if the explanations returned by abele significantly help in separating different images, even for non-expert users. From another perspective, this can be considered the human evaluation of the usefulness |
| Variance user decision | User Behaviour | [126] | Thus, this study utilizes the level of agreement of therapists' assessment (F1-score) as a metric to analyze the effect of a decision support system. Therapists generate assessment on patient's exercise performances while using each interface. We utilize this therapists' assessment to compute the agreement level of therapists, and explore whether our proposed interface with user-specific analysis supports therapists more consistent assessment. |
| | | [65] | their inter-rater agreement on the preferential diagnosis. |
| | | [50] | Interrater reliability between polysomnography technicians has been a critical issue in sleep staging because of the variability in interpreting polysomnography recordings among technicians [38]. Following previous work in sleep medicine, which demonstrated that an adequate information system could reduce interrater reliability [19], we investigated whether the information from our CDSS could enhance this property. With this objective, interrater reliability was measured using the Cohen score [39]. |
| | | [80] | Inter-rater agreement and prediction accuracy are both important parameters for establishing internal validity of a grading system [45]. The level of inter-rater agreement determined by Cohen's kappa increased from "moderate" to "substantial" with AI/ML assistance [46] |

Table B.11 – continued from previous page

| **Property** | **Measurement Type** | **Paper** | **Measurement (quote from the paper)** |
|---|---|---|---|
| | | [94] | inter-observer agreement was good (kappa=0.7) and the median agreement between residents and WHO panel experts was (kappa=0.7) before using the SDT and median kappa=0.8 after using the SDT (Fig. 5). |
| | | [70] | 3.5.2 Agreement Level. Most medical diagnoses rely on standardized guidelines [23, 26, 60]. However, clinicians can be biased in their decision making and expert disagreement is prevalent in medical decision-making tasks [7, 33, 34]. Thus, we also analyzed the agreement level of participants' decisions before/after reviewing AI outputs and explanations. |
| | | [81] | n order to express the clinicians' variability [118] of both Clinician- Only and Clinician-AI setups, we report (Table 3) two measures of the Coefficient of Variability (CV) as: (a) Inter-Variability; and (b) Intra-Variability. T |
| | | [75] | Variation The variation of outcomes produced by decision makers between two or more similar patient cases (which should be consistent). |
| | | [71] | To evaluate the attention of the annotators during the decision making task, we compute the average difference in the annotation scores for the four duplicate samples that were provided in the data. |